\documentclass[trackchanges]{aastex61}

\usepackage{bm} % for bolding vectors
\usepackage{amsmath}
\newcommand*\diff{\mathop{}\!\mathrm{d}}
\newcommand{\Secref}[1]{\hyperref[#1]{Section~\ref*{#1}}}
\newcommand{\Appref}[1]{\hyperref[#1]{Appendix~\ref*{#1}}}

\graphicspath{{./gfx/}{./}}

\shorttitle{Improving and Assessing Planet Sensitivity}
\shortauthors{Ruffio et al.}

\begin{document}

\title{Improving and Assessing Planet Sensitivity of the GPI Exoplanet Survey \\
with a Forward Model Matched Filter}

\correspondingauthor{Jean-Baptiste Ruffio}
\email{jruffio@stanford.edu}
\author[0000-0003-2233-4821]{Jean-Baptiste Ruffio}
\affiliation{Kavli Institute for Particle Astrophysics and Cosmology, Stanford University, Stanford, CA, USA 94305}

\author[0000-0003-1212-7538]{Bruce Macintosh}
\affiliation{Kavli Institute for Particle Astrophysics and Cosmology, Stanford University, Stanford, CA, USA 94305}

\author[0000-0003-0774-6502]{Jason J. Wang}
\affiliation{Astronomy Department, University of California, Berkeley; Berkeley CA, USA 94720}

\author{Laurent Pueyo}
\affiliation{Space Telescope Science Institute, Baltimore, MD, USA 21218}

\author[0000-0001-6975-9056]{Eric L. Nielsen}
\affiliation{SETI Institute, Carl Sagan Center, 189 Bernardo Avenue,  Mountain View, CA, USA 94043}
\affiliation{Kavli Institute for Particle Astrophysics and Cosmology, Stanford University, Stanford, CA, USA 94305}

\author[0000-0002-4918-0247]{Robert J. De Rosa}
\affiliation{Astronomy Department, University of California, Berkeley; Berkeley CA, USA 94720}

\author[0000-0002-1483-8811]{Ian Czekala}
\affiliation{Kavli Institute for Particle Astrophysics and Cosmology, Stanford University, Stanford, CA, USA 94305}

\author[0000-0002-5251-2943]{Mark S. Marley}
\affiliation{NASA Ames Research Center,  Mountain View, CA, USA 94035}

\author[0000-0001-6364-2834]{Pauline Arriaga}
\affiliation{Department of Physics \& Astronomy, University of California, Los Angeles, CA, USA 90095}

\author[0000-0002-5407-2806]{Vanessa P. Bailey}
\affiliation{Kavli Institute for Particle Astrophysics and Cosmology, Stanford University, Stanford, CA, USA 94305}

\author[0000-0002-7129-3002]{Travis Barman}
\affiliation{Lunar and Planetary Laboratory, University of Arizona, Tucson AZ, USA 85721}

\author{Joanna Bulger}
\affiliation{Subaru Telescope, NAOJ, 650 North A{'o}hoku Place, Hilo, HI 96720, USA}

\author[0000-0001-6305-7272]{Jeffrey Chilcote}
\affiliation{Dunlap Institute for Astronomy \& Astrophysics, University of Toronto, Toronto, ON, Canada M5S 3H4}

\author[0000-0003-0156-3019]{Tara Cotten}
\affiliation{Department of Physics and Astronomy, University of Georgia, Athens, GA, USA 30602}

\author{Rene Doyon}
\affiliation{Institut de Recherche sur les Exoplan{\`e}tes, D{\'e}partment de Physique, Universit{\'e} de Montr{\'e}al, Montr{\'e}al QC, Canada H3C 3J7}

\author[0000-0002-5092-6464]{Gaspard Duch\^ene}
\affiliation{Astronomy Department, University of California, Berkeley; Berkeley CA, USA 94720}
\affiliation{Univ. Grenoble Alpes/CNRS, IPAG, F-38000 Grenoble, France}

\author[0000-0002-0176-8973]{Michael P. Fitzgerald}
\affiliation{Department of Physics \& Astronomy, University of California, Los Angeles, CA, USA 90095}

\author[0000-0002-7821-0695]{Katherine B. Follette}
\affiliation{Kavli Institute for Particle Astrophysics and Cosmology, Stanford University, Stanford, CA, USA 94305}

\author{Benjamin L. Gerard}
\affiliation{University of Victoria, 3800 Finnerty Rd, Victoria, BC, Canada V8P 5C2}
\affiliation{National Research Council of Canada Herzberg, 5071 West Saanich Rd, Victoria, BC, Canada V9E 2E7}

\author[0000-0002-4144-5116]{Stephen J. Goodsell}
\affiliation{Gemini Observatory, 670 N. A'ohoku Place, Hilo, HI, USA 96720}

\author{James R. Graham}
\affiliation{Astronomy Department, University of California, Berkeley; Berkeley CA, USA 94720}

\author[0000-0002-7162-8036]{Alexandra Z. Greenbaum}
\affiliation{Department of Astronomy, University of Michigan, Ann Arbor MI, USA 48109}

\author[0000-0003-3726-5494]{Pascale Hibon}
\affiliation{Gemini Observatory, Casilla 603, La Serena, Chile}

\author[0000-0003-1498-6088]{Li-Wei Hung}
\affiliation{Department of Physics \& Astronomy, University of California, Los Angeles, CA, USA 90095}

\author{Patrick Ingraham}
\affiliation{Large Synoptic Survey Telescope, 950N Cherry Av, Tucson, AZ, USA 85719}

\author{Paul Kalas}
\affiliation{Astronomy Department, University of California, Berkeley; Berkeley CA, USA 94720}
\affiliation{SETI Institute, Carl Sagan Center, 189 Bernardo Avenue,  Mountain View, CA, USA 94043}

\author[0000-0002-9936-6285]{Quinn Konopacky}
\affiliation{Center for Astrophysics and Space Science, University of California San Diego, La Jolla, CA, USA 92093}

\author{James E. Larkin}
\affiliation{Department of Physics \& Astronomy, University of California, Los Angeles, CA, USA 90095}

\author{J\'er\^ome Maire}
\affiliation{Center for Astrophysics and Space Science, University of California San Diego, La Jolla, CA, USA 92093}

\author[0000-0001-7016-7277]{Franck Marchis}
\affiliation{SETI Institute, Carl Sagan Center, 189 Bernardo Avenue,  Mountain View, CA, USA 94043}

\author[0000-0002-4164-4182]{Christian Marois}
\affiliation{National Research Council of Canada Herzberg, 5071 West Saanich Rd, Victoria, BC, Canada V9E 2E7}
\affiliation{University of Victoria, 3800 Finnerty Rd, Victoria, BC, Canada V8P 5C2}

\author[0000-0003-3050-8203]{Stanimir Metchev}
\affiliation{Department of Physics and Astronomy, Centre for Planetary Science and Exploration, The University of Western Ontario, London, ON N6A 3K7, Canada}
\affiliation{Department of Physics and Astronomy, Stony Brook University, Stony Brook, NY 11794-3800, USA}

\author[0000-0001-6205-9233]{Maxwell A. Millar-Blanchaer}
\affiliation{Jet Propulsion Laboratory, California Institute of Technology, Pasadena, CA, USA 91125}
\affiliation{NASA Hubble Fellow}

\author[0000-0002-1384-0063]{Katie M. Morzinski}
\affiliation{Steward Observatory, University of Arizona, Tucson AZ, USA 85721}

\author[0000-0001-7130-7681]{Rebecca Oppenheimer}
\affiliation{Department of Astrophysics, American Museum of Natural History, New York, NY, USA 10024}

\author{David Palmer}
\affiliation{Lawrence Livermore National Laboratory, Livermore, CA, USA 94551}

\author{Jennifer Patience}
\affiliation{School of Earth and Space Exploration, Arizona State University, PO Box 871404, Tempe, AZ, USA 85287}

\author[0000-0002-3191-8151]{Marshall Perrin}
\affiliation{Space Telescope Science Institute, Baltimore, MD, USA 21218}

\author{Lisa Poyneer}
\affiliation{Lawrence Livermore National Laboratory, Livermore, CA, USA 94551}

\author[0000-0002-9246-5467]{Abhijith Rajan}
\affiliation{School of Earth and Space Exploration, Arizona State University, PO Box 871404, Tempe, AZ, USA 85287}

\author[0000-0003-0029-0258]{Julien Rameau}
\affiliation{Institut de Recherche sur les Exoplan{\`e}tes, D{\'e}partment de Physique, Universit{\'e} de Montr{\'e}al, Montr{\'e}al QC, Canada H3C 3J7}

\author[0000-0002-9667-2244]{Fredrik T. Rantakyr\"o}
\affiliation{Gemini Observatory, Casilla 603, La Serena, Chile}

\author[0000-0002-8711-7206]{Dmitry Savransky}
\affiliation{Sibley School of Mechanical and Aerospace Engineering, Cornell University, Ithaca, NY, USA 14853}

\author{Adam C. Schneider}
\affiliation{School of Earth and Space Exploration, Arizona State University, PO Box 871404, Tempe, AZ, USA 85287}

\author[0000-0003-1251-4124]{Anand Sivaramakrishnan}
\affiliation{Space Telescope Science Institute, Baltimore, MD, USA 21218}

\author[0000-0002-5815-7372]{Inseok Song}
\affiliation{Department of Physics and Astronomy, University of Georgia, Athens, GA, USA 30602}

\author[0000-0003-2753-2819]{Remi Soummer}
\affiliation{Space Telescope Science Institute, Baltimore, MD, USA 21218}

\author{Sandrine Thomas}
\affiliation{Large Synoptic Survey Telescope, 950N Cherry Av, Tucson, AZ, USA 85719}

\author{J. Kent Wallace}
\affiliation{Jet Propulsion Laboratory, California Institute of Technology, Pasadena, CA, USA 91125}

\author[0000-0002-4479-8291]{Kimberly Ward-Duong}
\affiliation{School of Earth and Space Exploration, Arizona State University, PO Box 871404, Tempe, AZ, USA 85287}

\author{Sloane Wiktorowicz}
\affiliation{Department of Astronomy, UC Santa Cruz, 1156 High Street, Santa Cruz, CA, USA 95064}

\author[0000-0002-9977-8255]{Schuyler Wolff}
\affiliation{Department of Physics and Astronomy, Johns Hopkins University, Baltimore, MD, USA 21218}



\begin{abstract}

We present a new matched filter algorithm for direct detection of point sources in the immediate vicinity of bright stars.
The stellar Point Spread Function (PSF) is first subtracted using a Karhunen-Lo\'eve Image Processing (KLIP) algorithm with Angular and Spectral Differential Imaging (ADI and SDI). The KLIP-induced distortion of the astrophysical signal is included in the matched filter template by computing a forward model of the PSF at every position in the image.
To optimize the performance of the algorithm, we conduct extensive planet injection and recovery tests and tune the exoplanet spectra template and KLIP reduction aggressiveness to maximize the Signal-to-Noise Ratio (SNR) of the recovered planets. We show that only two spectral templates are necessary to recover any young Jovian exoplanets with minimal SNR loss.
We also developed a complete pipeline for the automated detection of point source candidates, the calculation of Receiver Operating Characteristics (ROC), false positives based contrast curves, and completeness contours. 
We process in a uniform manner more than 330 datasets from the Gemini Planet Imager Exoplanet Survey (GPIES) and assess GPI typical sensitivity as a function of the star and the hypothetical companion spectral type. This work allows for the first time a comparison of different detection algorithms at a survey scale accounting for both planet completeness and false positive rate. We show that the new forward model matched filter allows the detection of $50\%$ fainter objects than a conventional cross-correlation technique with a Gaussian PSF template for the same false positive rate.

\end{abstract}




\section{Introduction}

Direct imaging techniques spatially resolve exoplanets from their host star by using high-contrast imaging instruments usually combined with the power of large telescopes, adaptive optics, coronagraphs and sophisticated data processing. This technique currently allows the detection of young ($< 300\,$Myr), massive ($>2\,M_\mathrm{Jup}$), self-luminous exoplanets at host-star separations not yet covered by indirect methods ($a >5\,$au) and therefore helps to constrain planet population statistics. The Gemini Planet Imager (GPI) \citep{Macintosh2014}, operating on the Gemini South telescope, is one of the latest generation of high contrast instruments with extreme adaptive optics. The GPI Exoplanet Survey (GPIES) is targeting 600 young stars and has to date observed more than half of them. As part of the survey, it imaged several known systems and discovered the exoplanet 51 Eridani~b \citep{Macintosh2015}.

High contrast images suffer from spatially correlated noise, called speckles, which originate from optical aberrations in the instrument as well as a diffuse light component resulting from the time averaged uncorrected atmospheric turbulence. The correlation length of the speckles in a raw image is equal to the size of the unocculted Point Spread Function (PSF).
Speckles are often described as quasi-static, since they are correlated across the observed spectral band and across the exposures of a typical observation \citep{Perrin2003, Bloemhof2001, Sivaramakrishnan2002}.  
It has been shown that the Probability Density Function (PDF) of the speckle noise is not Gaussian, but rather better described by a modified Rician distribution \citep{Soummer2004, Bloemhof2004, Fitzgerald2006, Soummer2007, Hinkley2007, Marois2008, Mawet2014}. 
The Rician PDF has a larger positive tail than a Gaussian distribution yielding a comparatively higher number of false positives at constant Signal-to-Noise Ratio (SNR).

In a single image, disentangling the signal of a faint planet buried under speckle noise is challenging because the spatial scale of speckle noise and the signal from a planet are similar, corresponding to the size of the Point Spread Function (PSF).
However, speckles and astrophysical signals behave differently with time and wavelength. This diversity can be used to build a model of the speckle pattern and subtract it from all images.
Two observing strategies are commonly used for speckle subtraction with instruments like GPI; Angular Differential Imaging (ADI) \citep{Marois2006} and Spectral Differential Imaging (SDI) \citep{Marois2000,Sparks2002}.
The observing setup for ADI is different from traditional imaging with altitude/azimuth telescopes, as the instrument field derotator is switched off or adjusted to keep the telescope pupil fixed with respect to sky rotation. As a consequence, the astrophysical signal rotates on the detector, following the sky rotation with respect to the telescope (\textit{i.e.} parallactic angle evolution), while the speckle pattern remains relatively stable. In a similar tactic, SDI exploits the radial linear dependence of the speckle pattern with wavelength to separate it from the planet signal, which remains at the same position at all wavelengths. By definition, an observation with an Integral Field Spectrograph (IFS), like GPI, provides both temporal and spectral diversity of the speckle pattern for ADI or SDI to be used separately or in tandem.

The most common spectral subtraction algorithms are Locally Optimized Combination of Images (LOCI) \cite{Lafreniere2007a}, which uses a least square approach to optimally subtract the speckle noise, and Karhunen-Lo\'eve Image Processing (KLIP) \cite{Soummer2012}, which regularizes the least square problem by filtering out the high order singular modes. Following speckle subtraction, point sources can be searched for by using a matched filter or a more general Bayesian model comparison framework as shown in \cite{Kasdin2006}. However, the distortion of the planet PSF caused by the speckle subtraction algorithm, referred to as self-subtraction, can make it difficult to define an accurate matched filter template. The self-subtraction can be accurately modeled in simplified cases, as shown in the matched filter approach by \cite{Cantalloube2015} using ADI subtracted pair images. 
In the context of planet characterization, the self-subtraction also biases the photometry and the astrometry of the object. The inverse problem is usually solved by injecting negative planets in the raw images and iteratively minimizing the image residuals \citep{Marois2010,Morzinski2015}.
Recently, \cite{Pueyo2016} derived a closed-form approximation of the self-subtraction in KLIP but without applying it in the context of a matched filter. \cite{Wang2016} used this new forward model in a Bayesian framework to estimate the astrometry of $\beta$ Pictoris b.

The main challenge with uniformly and systematically characterizing the detections of a high-contrast imaging exoplanet survey is the high number of false positives even at relatively high SNR. The detection threshold is hard to define because the PDF of the residual noise is generally unknown and depends on the instrument, the choice of data processing, and the dataset itself. The lack of well-known false positive rates makes it very difficult to evaluate the performance of algorithms relative to one another. Currently, candidates are discarded as false positives by visual inspection, which does not permit a rigorous calculation of planet completeness. In order to accurately characterize planet detection statistics, and ultimately constrain the underlying planet population in a uniform and unbiased manner, it is important to improve and systematize exoplanet detection methods.

\par
The goal of this paper is to define a systematic and rigorous approach for exoplanet detection in large direct imaging surveys such that the long-period exoplanet population can be inferred in a meaningful manner. Using the latest KLIP framework, we develop an automated matched-filter based detection algorithm that includes a forward model of the planet self-subtraction \citep[e.g.,][]{Pueyo2016} and accounts for the noise variations in the spatial, temporal and spectral dimensions of a dataset. As of the end of 2016, GPIES has already observed 330 stars, which allows us to precisely estimate the false positive rate and define meaningful detection thresholds for the entire survey.
We conduct a rigorous set of tests to characterize state-of-the-art detection algorithms and demonstrate that a Forward Model Matched Filter (FMMF) most effectively recovers a planet signal while reducing the number of false positive detections. The paper is structured as follow:

\begin{itemize}
\item GPIES observations and data reduction are presented in \Secref{sec:obsRed},
\item The matched filter is described in \Secref{sec:MFs},
\item The optimization of the reduction parameters for GPIES is presented in \Secref{sec:optimization},
\item The residual noise is characterized for the different algorithms in \Secref{sec:noiseCharac} including the calculation of Receiver Operating Characteristics (ROC),
\item The detection sensitivity as a function of separation, referred to as contrast curve, is calculated in \Secref{sec:contCurve} where the contrast is defined as the companion to host star brightness ratio in a spectral band,
\item The follow-up strategy and the vetting of point source candidates is discussed in \Secref{sec:candidateDetec},
\item The contours of the planet completeness, which is the fraction of planets that could have been detected, are then derived in \Secref{sec:completeness},
\item We conclude in \Secref{sec:conclusion}.
\end{itemize}

We refer any reader who is not familiar with the data processing of high contrast images to \Appref{sec:appmethod}, which includes a detailed description of KLIP, the matched filter and the planet PSF forward model.
The mathematical notations are summarized in \Appref{sec:notations}.

\section{Observations and Data Reduction}
\label{sec:obsRed}

\subsection{Observations}
\label{sec:observations}
In this paper, we use 330 observations from the GPI Exoplanet Survey (GPIES) (Gemini programs GS-2014B-Q-500, GS-2015A-Q-500, GS-2015B-Q-500; PI: B. Macintosh) to construct and test our matched filter. A typical GPIES epoch consists in $38 \times 1$ minute exposures in \textit{H} band ($1.5-1.8 \mu m$). The number of usable raw spectral cubes can be lower, due to degrading weather condition or isolated star tracking failures. We arbitrarily consider any dataset with more than 20 usable exposures as complete. In some cases, a few exposures were added to the observing sequence in order to compensate for bad conditions, which resulted in a number of datasets with more than 38 exposures. For each star, we only consider the first complete epoch and ignore any follow-up observations that may have been made. We have not considered datasets with visible debris disks in order to avoid biasing the contrast curves.
In \textit{H} band, a GPI spectral cube has 37 wavelength channels, and $281\times281$ pixels in the spatial dimensions, half of which are however not filled with data due to the tilted IFS field of view. Therefore, a typical GPIES dataset includes approximately $1400$ images at different position angles and wavelengths.

\subsection{Raw Data Reduction}
\label{sec:pipeline}

Spectral data cubes are built from raw IFS detector images using standard recipes from the GPI Data Reduction Pipeline\footnote{Documentation available at \url{http://docs.planetimager.org/pipeline/}.} version 1.3 and 1.4 \citep{Perrin2016}. The process includes correction for dark current, bad pixels, correlated read noise, and cryocooler vibration induced microphonics \citep{Ingraham2014}. Flexure in the instrument slightly shifts the position of the lenslet micro-spectra on the detector. The offset is calibrated using argon arc lamp images at the target elevation, which are then compared with wavelength solutions references taken at zenith \citep{Wolff2014}. Finally, each spectral cube is also corrected for optical distortion according to \cite{Konopacky2014}.
GPI images also contain four fainter copies of the unblocked PSF called satellite spots \citep{Marois2006,Sivaramakrishnan2006}. The spots are used to estimate the location of the star behind the focal plane mask and their flux allows the photometric calibration of the images. The GPI satellite spot to star flux ratio used here is $2.035\times 10^{-4}$ for \textit{H} band \citep{Maire2014}. In the following, the satellite spots are also median combined to estimate an empirical planet PSF that is wavelength dependent.

\subsection{Use of Simulated Planets}
\label{sec:fakes}

Simulated planets are necessary to optimize and characterize a detection algorithm due to the scarcity of real point sources in high-contrast images. We decided to neglect for now the PSF smearing due to the sky rotation in a single exposure.
The planets' spectra are selected from atmospheric models described in Marley et al. (2017, in preparation) and \cite{Saumon2012}. For a given cloud-coverage, the only model parameter having a significant effect on the shape of the spectrum in a single band is the temperature. However, this work is not about atmospheric characterization, so physically unrealistic model temperatures for a given object are not an issue as long as the shape of spectrum is matching. In the following, the references to T-type and L-type planets correspond to the analagous spectra of brown dwarfs. T-type spectra have strong methane absorption features while L-type are cloudy objects whose spectra are dominated by H$_2$O and CO. Of the known extrasolar planets, 51 Eridani b is an example of the T-type \citep{Macintosh2015} and $\beta$ Pictoris b of the L-type \citep{Morzinski2015}. The transmission spectrum of the Earth's atmosphere combined with that of the instrument is estimated by dividing the satellite spots spectrum with a stellar spectrum, which is then used to translate the spectrum from physical units to raw pixel values from the detector. The spectrum of the star is interpolated from the Pickles atlas \cite{Pickles1998} based on the star's spectral type.

\subsection{Speckle Subtraction with KLIP}
\label{sec:PyKLIP}

Each individual image in the dataset is speckle subtracted using a Python implementation of KLIP called PyKLIP\footnote{Available under open-source license at \url{https://bitbucket.org/pyKLIP/pyklip}.} \citep{Wang2015}. The KLIP algorithm consists of building and subtracting a model of the speckle pattern in an image, called science image, from a set of reference images which can be selected from the same dataset or from completely different observations. In this paper, we use a combination of ADI and SDI strategies. The KLIP mathematical framework is summarized in \Appref{sec:appklipForma}.
A side-effect of the speckle subtraction is the distortion of the planet PSF, referred to as self-subtraction. The self-subtraction is fully characterized in \Appref{sec:appselfsubtraction}.

\begin{figure}
\centering
\includegraphics[width=0.4\linewidth]{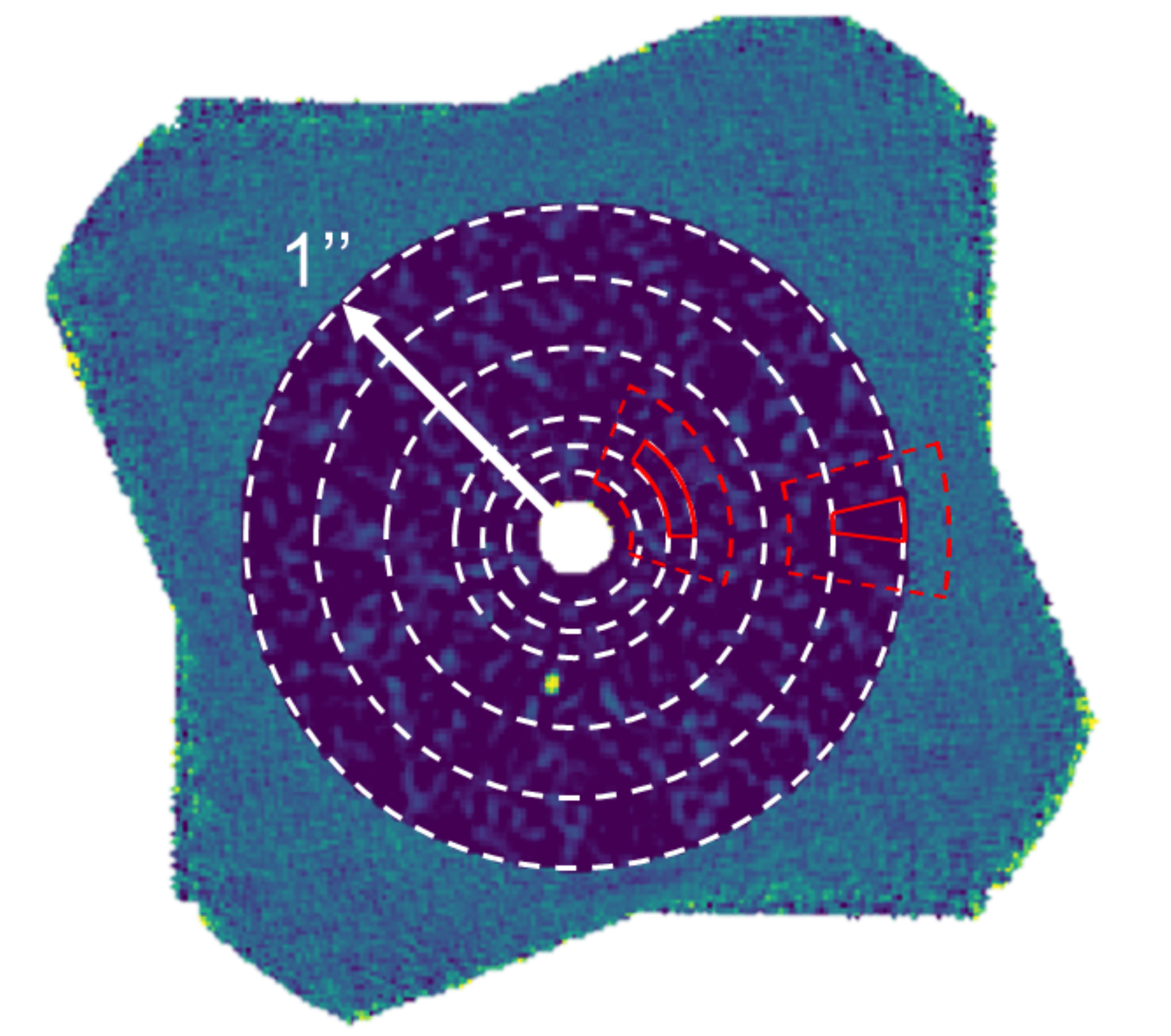}
\caption{Definition of the padded sectors dividing the image. The annuli boundaries used for the unpadded sectors are shown as white dashed circle. The first three annuli are thinner to account for the rapidly varying noise standard deviation close to the focal plane mask. Examples of sectors are drawn in red with the solid line containing 100 pixels and the dashed line delimiting the padded area. The outer working angle has been set to $1\arcsec$.\label{fig:sectors}}
\end{figure}

All individual images are first high-pass filtered by subtracting a Gaussian convolved image with a Full-Width-Half-Maximum (FWHM) of 12 pixels. Then, they are aligned and the reference images are scaled to the same wavelength as the science image. In order to account for spatial variation of the speckle behavior, KLIP is independently applied on small subsections of the field of view. Each image is therefore divided in small $100\ \text{pixel}$ arcs to which a $10\ \text{pixel}$ wide padding is added as illustrated in \autoref{fig:sectors}. For each sector, the reference library is built according to an exclusion criterion based on the displacement and the flux overlap \citep{Marois2014} of the planet PSF between the science image and its reference images. The exclusion criterion and the reference library selection is described in \Appref{sec:exclusionCrit}. The assumed spectrum for the companion used in the exclusion parameter will be referred to as the reduction spectrum. In order to speed up the reduction, we only include the $N_R = 150$ most correlated images. This means that most of the images satisfying the exclusion criterion are in practice not used. In \Secref{sec:aggressOpti}, it is shown that the exclusion criterion has a soft maximum around $0.7$ for T-type planets, which is the value used in the following. 
The number of Karhunen-Lo\`eve modes of the reference library kept for the speckle subtraction is set to $K=30$. This value has been chosen as a reasonable guess based on our experience but it should be rigorously optimized in the future.
In order to limit computation time, we have arbitrarily defined the outer working angle of the algorithm to $1\arcsec$ ($\approx 71$ pixels).

\section{Matched Filter}
\label{sec:MFs}

\subsection{Concept}
\label{sec:matchedFilterConcept}

In the field of signal processing, a matched filter is the linear filter maximizing the SNR of a known signal in the presence of additive noise \citep{Kasdin2006,Rover2011, Cantalloube2015}. A detailed description of the matched filter is presented in appendix \Appref{sec:appmatchedFilter} and we only summarize the key results here.
If the noise samples are independent and identically distributed, the matched filter corresponds to the cross-correlation of a template with the noisy data. In the context of high-contrast imaging, the pixels are neither independent nor identically distributed (i.e., heteroskedastic), which introduces a local noise normalization in the expression of the matched filter.

In a dataset, each image is indexed by its exposure number $\tau$ and its wavelength $\lambda$. We define the vector ${\bm p}_{l}$, with $l=(\tau,\lambda)$, as a specific speckle subtracted image. Similarly, we define the matched filter template ${\bm m}_{l}$ as the model of the planet signal in the corresponding processed image normalized such that it has the same broadband flux as the star. The whitening effect of the speckle subtraction allows one to assume uncorrelated residual noise, which simplifies significantly the matched filter. However, this assumption is not perfectly verified and its consequence is discussed in \Secref{sec:SNR}. The maximum likelihood estimate of the planet contrast at separation $\rho$ and position angle $\theta$ is then given by
\begin{equation}
\tilde{\epsilon}(\rho,\theta)  = \sum\limits_{l}  {\bm p}_{l}^{\top} {\bm m}_{l}/\sigma_{l}^{2} \Bigg / \sum\limits_{l} {\bm m}_{l}^{\top} {\bm m}_{l}/\sigma_{l}^{2},
\label{eq:datasetAmpl}
\end{equation}
where $\sigma_{l}$ is the local standard deviation at the position $(\rho,\theta)$, assuming that it is constant in the neighborhood of the planet. Note that the planet model ${\bm m}_{l}$ approaches zero rapidly when moving away from its center $(\rho,\theta)$ allowing one to only consider postage-stamp sized images containing the putative planet instead of the full images in \autoref{eq:datasetAmpl}.
Then, the theoretical SNR of the planet can be written
\begin{equation}
\mathcal{S}(\rho,\theta)  =  \sum\limits_{l} {\bm p}_{l}^{\top} {\bm m}_{l}/\sigma_{l}^{2} \Bigg / \sqrt{ \sum\limits_{l} {\bm m}_{l}^{\top} {\bm m}_{l}/\sigma_{l}^{2}}.
\label{eq:datasetSNR}
\end{equation}
A detection can be claimed when the SNR is such that the observation cannot be explained by the null-hypothesis.

\subsection{Matched Filter Template and Forward Model}
\label{sec:forwardModel}

\begin{figure}
\centering
\includegraphics[width=1.0\linewidth]{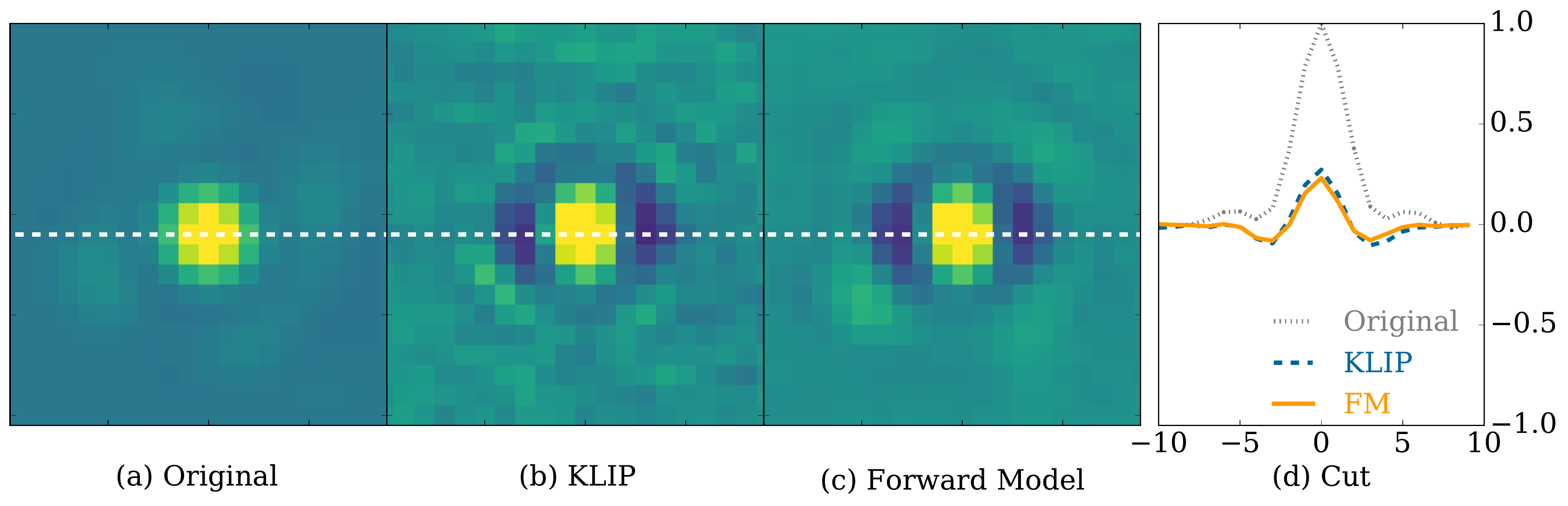}
\caption{KLIP self-subtraction and forward modeled Point Spread Function (PSF). The three panels from left to right represent (a) the original broadband PSF calculated from the GPI satellite spots, (b) the speckle subtracted image of a simulated planet using KLIP, and (c) the KLIP forward model of the PSF calculated at the position of the simulated planet. All three images are collapsed in time and wavelength and have been scaled to their peak value. The last panel (d) includes horizontal cuts of the different PSF in (a), (b) and (c). The negative ring and negative lobes around the central peak are characteristic of the self-subtraction. Both the shape and the final amplitude of the PSF are successfully recovered by the forward model. The simulated planet was injected in 38 cubes of the 51 Eridani b GPIES discovery epoch on 2014 December 18, which is characterized by remarkably stable observing conditions. The star is located approximatively $30$ pixels ($0.4 \arcsec$) above the planet and is not visible here.}
\label{fig:forwardModelExample}
\end{figure}

The calculation of the matched filter template ${\bm m}_{l}$ is complicated by the distortion of the planet PSF after speckle subtraction. The characteristic effect of the self-subtraction manifests as negative wings on each side of the planet PSF and a narrowing of its central peak. It is common practice to define the matched filter template as a 2D Gaussian, but this model fails to include most of the information about the distortion. In this paper, we present a novel matched filter implementation using KLIP and a forward model of the planet PSF from \cite{Pueyo2016} illustrated in \autoref{fig:forwardModelExample}. The forward model is a linearized closed form estimate of the distorted PSF, which is detailed in \Secref{sec:KLIPFM}. It is built from the same satellite spot based PSF as the simulated planet injection in \Secref{sec:fakes} and from the reduction spectrum. The forward model is a function of the PSF's location on the image and therefore needs to be calculated for each pixel.

\begin{figure}
\gridline{\fig{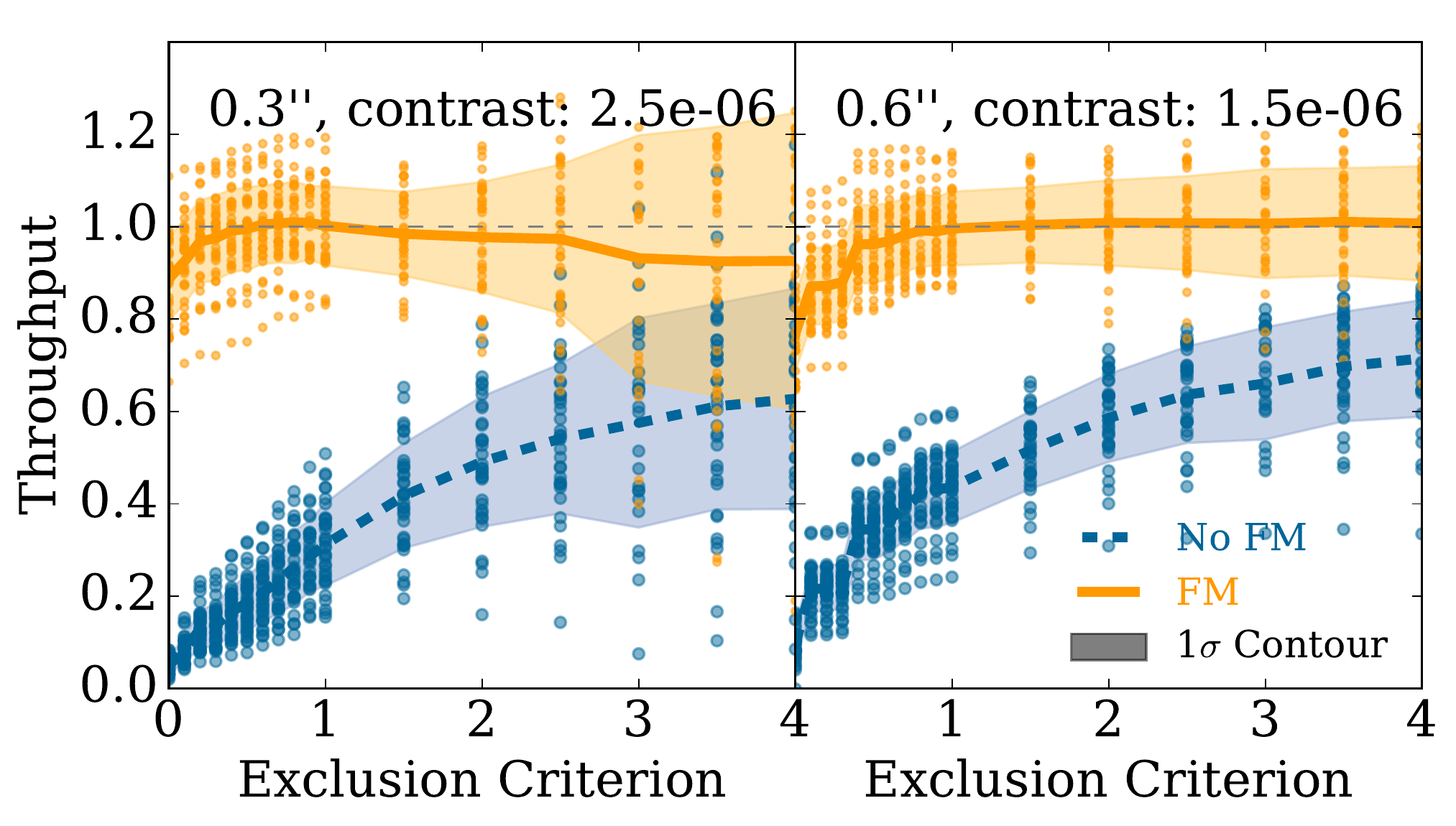}{0.5\textwidth}{(a) HR 7012}
		\fig{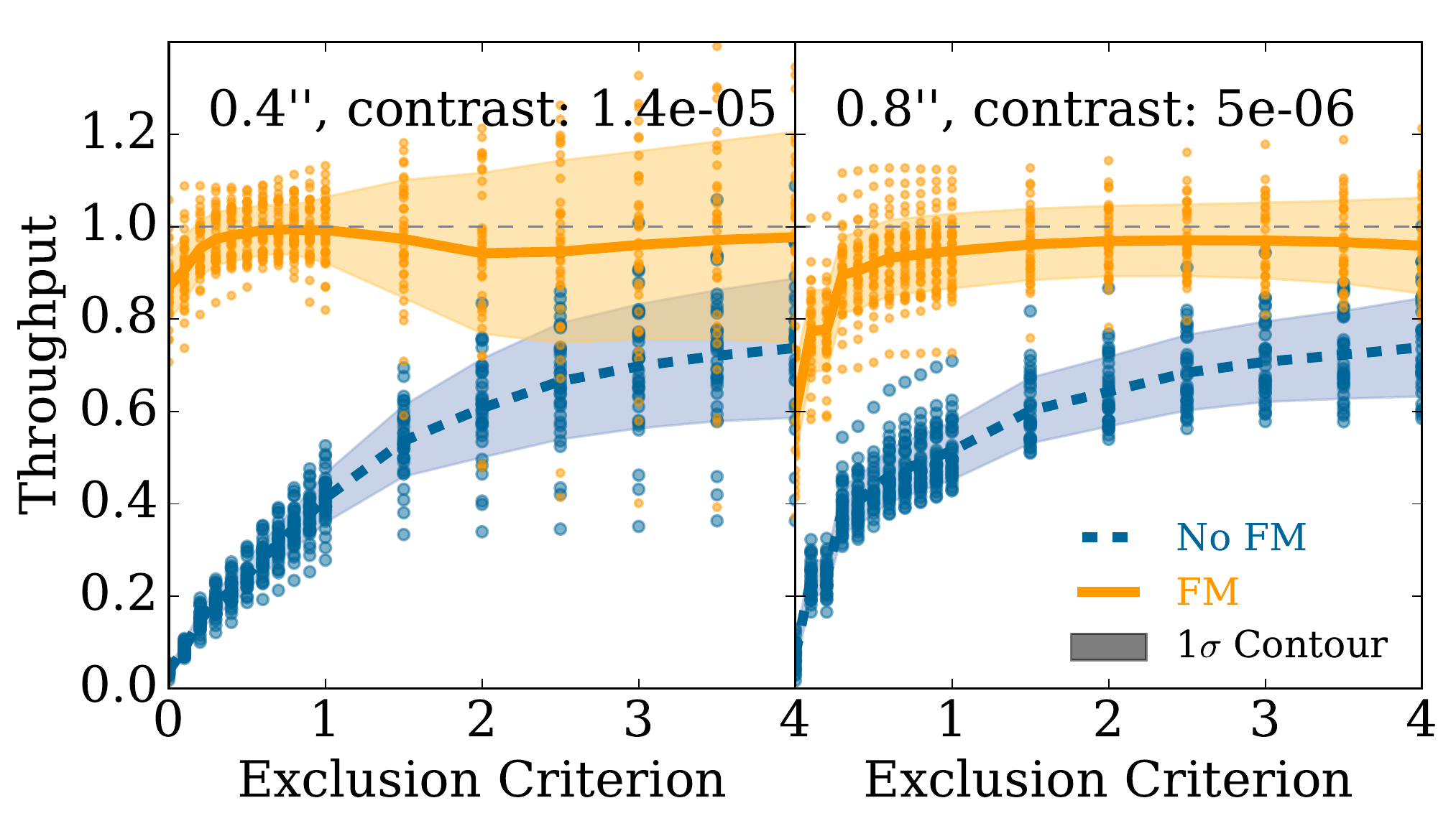}{0.5\textwidth}{(b) HD 131435}
          }
\caption{Algorithm throughput as a function of the exclusion criterion and the planet model. The throughput is estimated for both the Forward Model (FM, Solid Orange) and the original PSF (No FM, Dashed Blue). The conversion factor has no physical units although it matches to pixels when the reduction spectrum is constant. The test was performed with two representative datasets at two different separations: (a) HR 7012 at $0.3\arcsec$ and $0.6\arcsec$ and (b) HD 131435 at $0.4\arcsec$ and $0.8\arcsec$. The datasets were chosen to represent different regimes of data quality; HR 7012 is an example of good dataset while HD 131435 is of median quality. (Dots) Throughput for the $35$ simulated exoplanets injected at different position angles in seven different copies of the dataset. (Shaded) $1\sigma$ spread of the throughput. The original contrast of the simulated planets is indicated at the top of the plot. A spectrum with a methane signature is assumed for the planets and for the reference library selection. \label{fig:thoughvsMvt}}
\end{figure}

In this paper, we define the throughput as the ratio of the estimated contrast of a planet after speckle subtraction over its original contrast. This definition might differ from the conventional idea of throughput in the data processing community for direct imaging.
It is a measure of the ability of an algorithm to recover an unbiased contrast estimate of the point source. This can then be used to validate our implementation of the forward model. A low throughput indicates that the planet signal has been distorted during the speckle subtraction.
The flux of each point source is estimated with \autoref{eq:datasetAmpl} and then compared to the known true contrast of the simulated planet. \autoref{fig:thoughvsMvt} compares the throughput as a function the exclusion criterion, when using either the forward model or only the original PSF as a template. The estimation algorithm is otherwise identical in both cases. \autoref{fig:thoughvsMvt} demonstrates that the forward model accurately models the self-subtraction because the throughput is close to one even for very aggressive reduction, while it drops very quickly to zero with the original PSF. However, as the exclusion parameter becomes too small, the linear approximation of the forward model breaks down and the throughput drops. As expected, the scatter in \autoref{fig:thoughvsMvt} decreases with smaller values of the exclusion criterion.
The test was performed with two representative datasets at two different separations: (a) HR 7012 at $0.3\arcsec$ and $0.6\arcsec$ and (b) HD 131435 at $0.4\arcsec$ and $0.8\arcsec$. The datasets were chosen to represent different regimes of data quality as measured by their exoplanet contrast sensitivity in the fully reduced image”; HR 7012 is an example of good dataset while HD 131435 is of median quality.
A total of 35 simulated exoplanets were injected at each separation in seven copies of the same dataset. For each copy of the dataset, the five simulated planets, that are $72^\circ$ apart, were rotated by $10^\circ$ in position angle to better sample the image.

\begin{figure}
\centering
\includegraphics[trim={0 3.5cm 0 0},clip,width=1\linewidth]{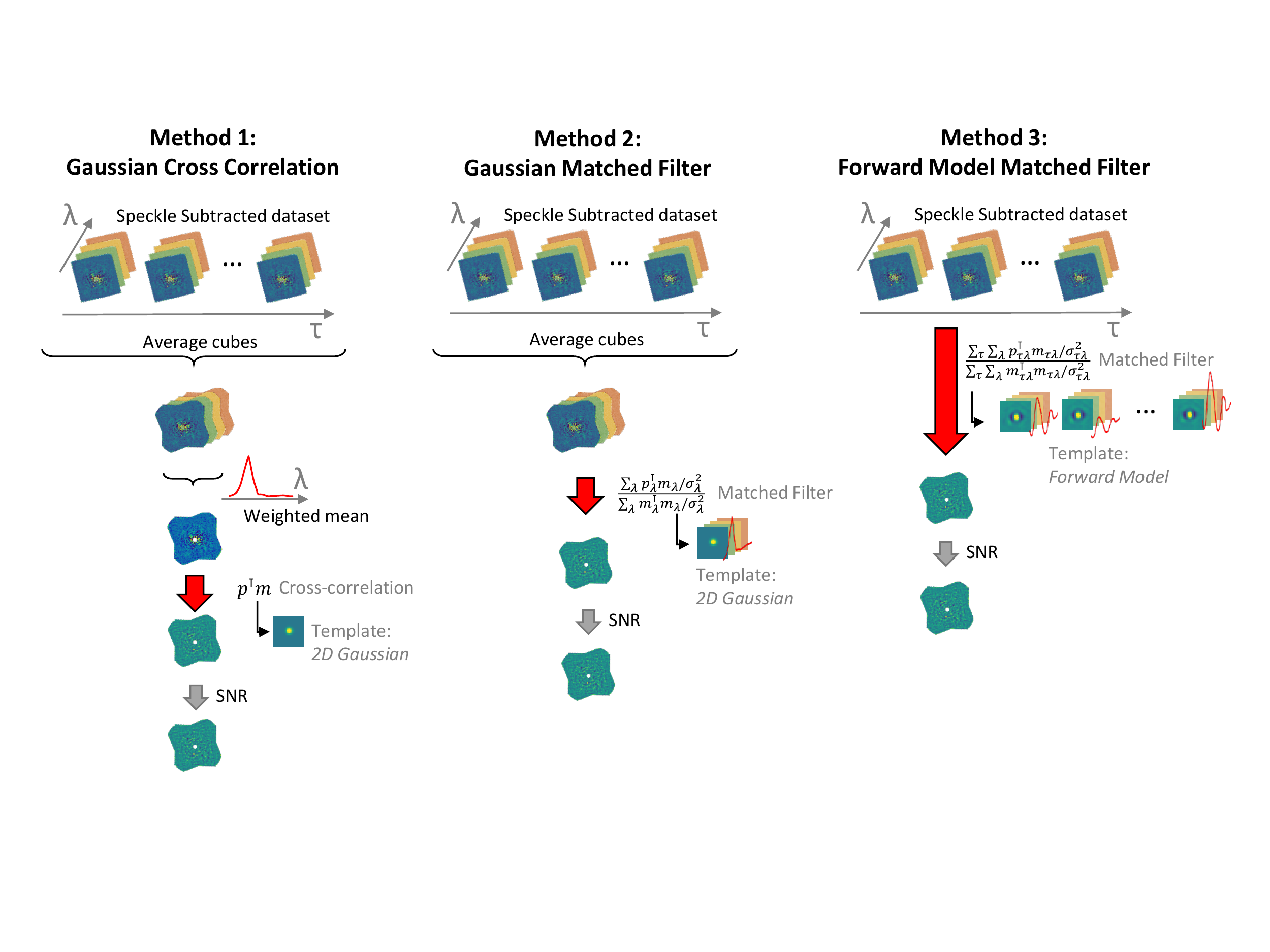} %trim={0 3.5cm 0 0},clip,
\caption{Illustration of three different matched filters. They differ by their template, Gaussian or forward model, and by the way the dataset is combined before the matched filter.}
\label{fig:matchedFilters}
\end{figure}

The new algorithm is compared to two simpler matched filters, all using the same KLIP implementation, but involving a simple Gaussian PSF and no modeling of the planet self-subtraction. The three algorithms will be referred in this paper as Gaussian Cross correlation (GCC), Gaussian Matched filter (GMF), and Forward Model Matched Filter (FMMF). The three methods are detailed in this section and illustrated in \autoref{fig:matchedFilters}. 
The different methods also differ in the way the dataset is collapsed before the matched filter is performed. It is outside the scope of this paper to evaluate the effect of each particular difference. All the algorithms presented in this paper are available in PyKLIP \citep{Wang2015}. A significant fraction of the code is shared with the Bayesian KLIP-FM Astrometry (BKA) method developed in \citep{Wang2016}.

\subsection{Gaussian Cross Correlation}

The Gaussian Cross Correlation (GCC) is the baseline algorithm because it is commonly used in high contrast imaging. The overlapping sectors are mean combined after speckle subtraction in order to limit edge effects. Then, the processed dataset is first derotated, coadded and then collapsed using the reduction spectrum. 
The resulting image is cross correlated with a 2-D Gaussian kernel with a Full-Width-Half-Maximum (FWHM) of $2.4$ pixels. The size of the Gaussian has been chosen to be significantly smaller than the FWHM of the original PSF ($\approx 3.5\ \text{pixels}$) to account for the self-subtraction. The cross correlation is nothing more than a matched filter where the noise is spatially identically distributed. This assumption is not verified for GPI images as discussed in \Secref{sec:noiseDistrib}. In practice, the noise properties only need to be azimuthally uniform; the theoretical SNR from the matched filter is always rescaled as a function of separation by estimating the standard deviation in concentric annuli, as discussed in \Secref{sec:SNR}. To summarize the Gaussian Cross Correlation for GPIES, the template is a $20\times20$ pixels stamp projected on a $281\times281$ pixels image.

\subsection{Gaussian Matched Filter}

The Gaussian Matched Filter (GMF) projects the PSF template on the coadded cubes directly as illustrated in \autoref{fig:matchedFilters}. Therefore, the template is a stamp cube with both spatial and wavelength dimensions. It also uses a Gaussian PSF with a 2.4 pixels FWHM and the reduction spectrum is used to scale the template as a function of wavelength. The matched filter is calculated according to \autoref{eq:datasetSNR} considering a single combined cube. As a consequence, this implementation does not assume identically distributed noise. The local noise standard deviation is estimated at each position in a $20\times20$ pixels stamp from which a central disk with a $2.5\ \text{pixel}$ radius is removed. To summarize the Gaussian Matched Filter for GPIES, the template is a $20\times20\times37$ pixels cube stamp, which includes the spectral dimension, projected on a $281\times281\times37$ pixels datacube.

\subsection{Forward Model Matched Filter}

The Forward Model Matched Filter (FMMF) is performed on uncollapsed dataset according to \autoref{eq:datasetSNR} as illustrated in \autoref{fig:matchedFilters}. It uses the forward model as a template. The sector padding provides the necessary margin to perform the projection of the template anywhere in the image.
The local standard deviation in each image is estimated at the position $(\rho, \theta)$ in a local arc defined by $[\rho-\Delta \rho, \rho+ \Delta \rho]\times[\theta-\Delta \rho/\rho, \theta+\Delta \rho/\rho]$ and $\Delta \rho = 10\ \text{px}$. In this case, the center of the arc is not masked, which results in an overestimation the standard deviation in the presence of a planet signal. This doesn't significantly change the detectability of real objects, because the overestimated local standard deviation plays against both real objects and false positives. 

In practice, only the values of the inner products ${\bm p}_{l}^{\top} {\bm m}_{l}$, ${\bm m}_{l}^{\top} {\bm m}_{l}$ and $\sigma_{l}(\rho,\theta)$ are saved while the sectors are speckle subtracted in order to limit computer memory usage. The final sum of \autoref{eq:datasetSNR} is performed at the very end. One advantage of not combining the data is that it is not necessary to derotate the speckle subtracted images, as long as one accounts for the movement of the model in the data as a function of time and wavelength. This removes the image interpolation associated with derotation and therefore limits interpolation errors. The matched filter calculation is run on a discretized grid, in this case, centered on each pixel in the final image. The matched filter is therefore slightly less sensitive to planets that are not also centered on a pixel. Assuming spatially randomly distributed planets, we estimate that the discretization results in an average loss of SNR of a few percentage points, compared to the case where every planet is centered on a pixel.” To summarize the FMMF for GPIES, the template is a $20\times20\times37\times38$ pixels multidimensional stamp, therefore including both spectral and time dimensions, projected on an uncombined $281\times281\times37\times38$ pixels dataset.

The FMMF reduction of a typical GPI 38 exposure dataset requires around $30$ wall clock hours on a computer equipped with $32$ $2.3$ GHz cores and using $\sim 20$ GB of Random-Access Memory (RAM). As a consequence, a super computer is necessary to process an entire survey. Each dataset is independent and can be run on separate nodes without sharing memory. In this paper, we have used the SLAC bullet cluster to process the entire campaign using of the order of $10^6$ CPU hours including the data processing necessary for the work presented in the remainder of the paper.

\subsection{SNR Calculation}
\label{sec:SNR}

\begin{figure}
\centering
\includegraphics[trim={10cm 3.5cm 10cm 3.5cm},clip,width=0.3\linewidth]{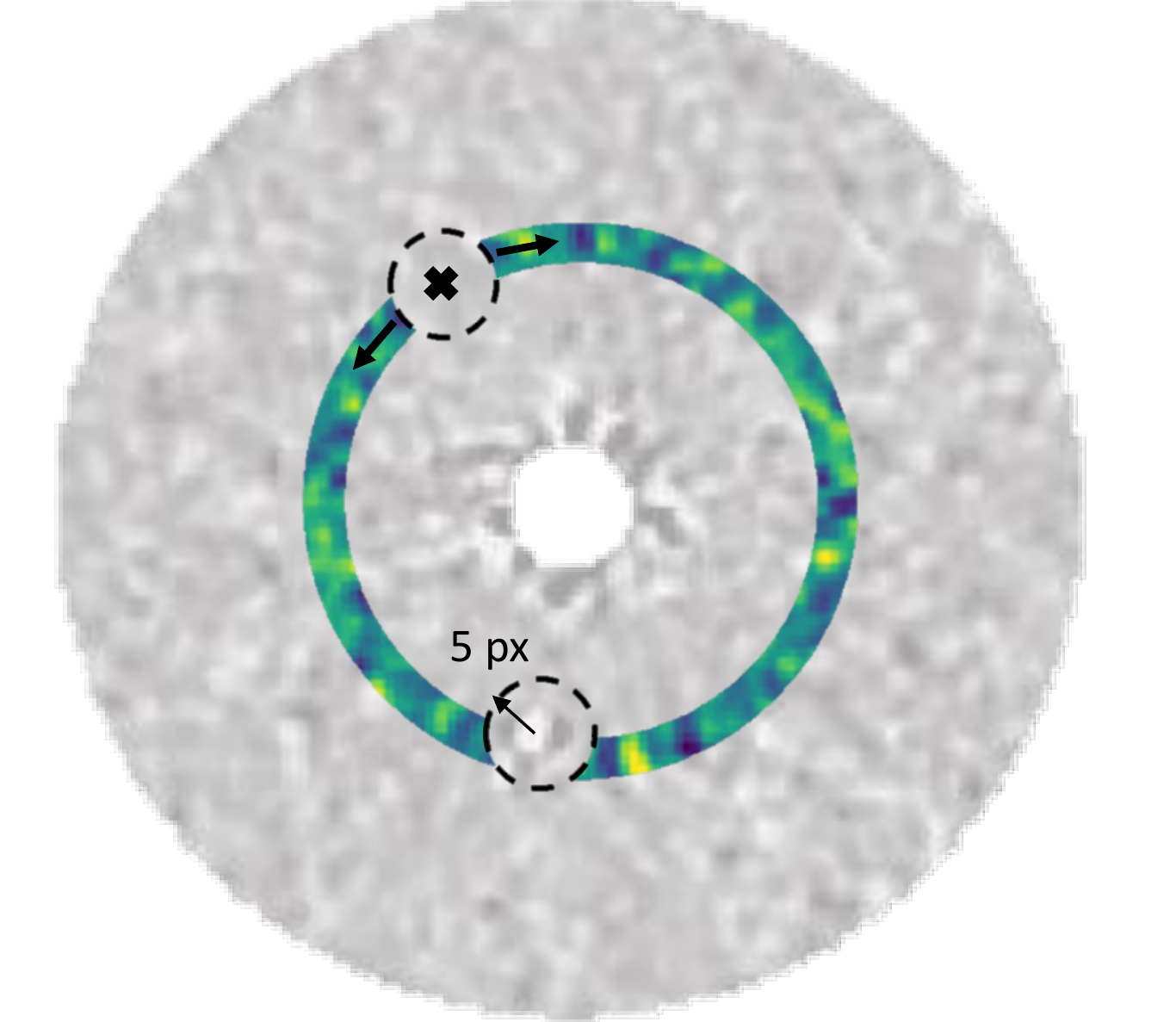}
\caption{Illustration of the estimation of the standard deviation. For each pixel (cross), the standard deviation is empirically calculated in a $4$ pixel wide annulus from which the surroundings of the current pixel as well as any known astrophysical signal has been masked.}
\label{fig:annulus}
\end{figure}

In practice, the theoretical SNR defined in \autoref{eq:datasetSNR} needs to be empirically calibrated by estimating the standard deviation from the matched filter map itself \citep{Cantalloube2015}. Indeed, the SNR is overestimated due to overly optimistic assumptions on the noise distribution.
The residual noise is mostly white and Gaussian in areas not dominated by speckle noise but both assumptions break down close to the mask due to speckle noise, causing the matched filter to lose some validity although it remains nonetheless relatively effective. It is also hard to estimate the local noise accurately, because the noise properties vary from pixel to pixel, adding a layer of uncertainty on the theoretical SNR. This is why an empirical standard deviation is estimated in the matched filter map as a function of separation using a $4$ pixels wide annulus as illustrated in \autoref{fig:annulus}. In order to prevent a planet from biasing its own SNR, the standard deviation is calculated at each pixel while masking a disk with a $5$ pixel radius centered on that pixel from the annulus. In addition, all the known astrophysical objects are also masked. 
In the particular case of injected simulated planets, the standard deviation is estimated from the planet free reduction.
In the following, unless specified otherwise, any reference to a SNR relates to the calibrated SNR.

For a centered Gaussian noise, the SNR is related to the False Positive Rate (FPR) following
\begin{equation}
\mathrm{FPR} = \frac{1}{2}\left(1-\mathrm{erf}\left(\frac{\mathrm{SNR}}{\sqrt{2}}\right)\right).
\end{equation}
It is common practice to choose a $5\sigma$ detection threshold or higher as discussed in \cite{Marois2008}. For a Gaussian distribution, a $5\sigma$ threshold corresponds to a False Positive Rate equal to $\mathrm{FPR} = 2.9 \times 10^{-7}$, which represents a false detection every $3.4 \times 10^6$ independent samples or equivalently of the order of $1000$ GPI epochs. 
The deviation from Gaussianity of the residual noise in the final image will dramatically increase this false positive fraction as shown in \Secref{sec:ROC}. Small sample statistics \citep{Mawet2014} also increases the false positive fraction but it is not expected to be a dominant term in this work and has therefore been neglected. Indeed, the inner working angle in the code is larger than three resolution elements at $1.6 \,\mu m$.

\subsection{Example Reduction}

\begin{figure}
\centering
\includegraphics[width=1\linewidth]{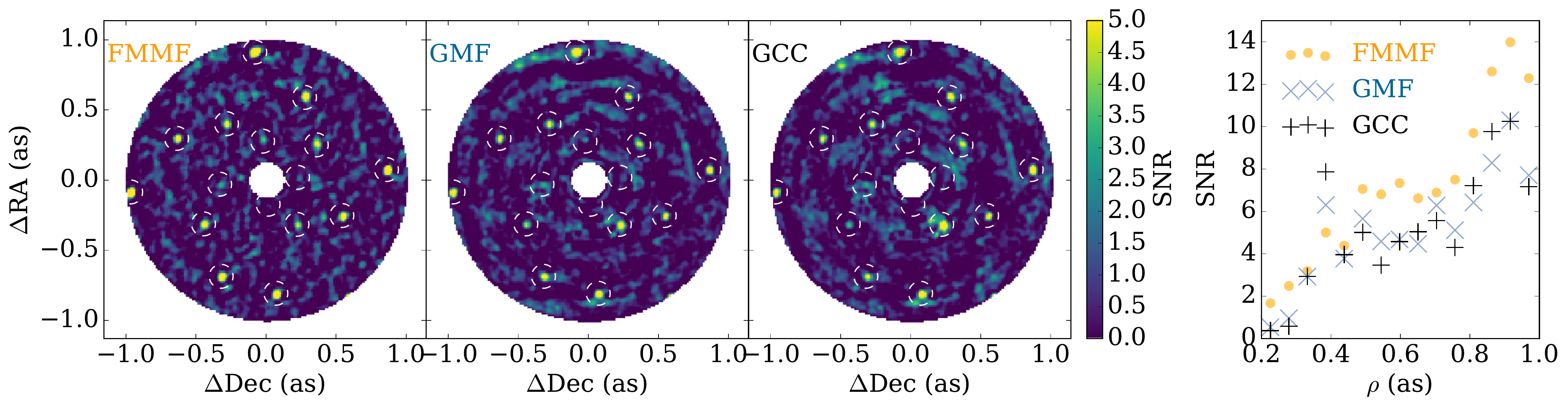}
\caption{Reduction of the HR 7012 epoch including 16 simulated T-type planets with three different algorithms, from left to right: Forward Model Matched Filter (FMMF), Gaussian Matched Filter (GMF) and Gaussian Cross Correlation (GCC). Each image corresponds to a SNR map where the simulated planets have been circled. The right most figure shows the SNR of the simulated planets as a function of separation for the three algorithms. A T-type spectrum similar to 51 Eridani b was used for the simulated planets and a spectrum with a sharper methane signature was used for the reduction and the matched filter.
\label{fig:FMMFShowOff}}
\end{figure}

\begin{figure}
\centering
\includegraphics[width=1\linewidth]{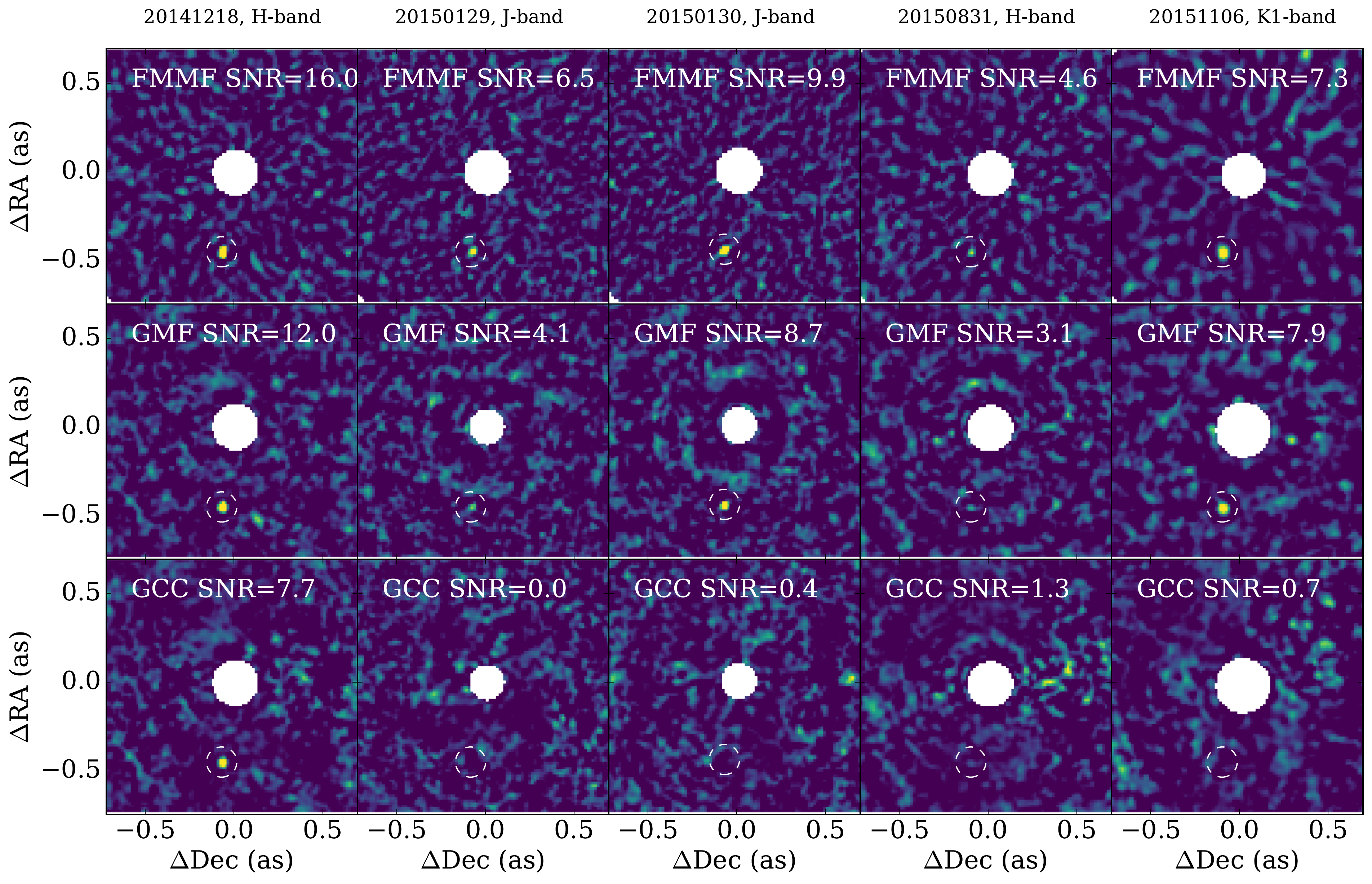}
\caption{Reduction of several 51 Eridani epochs with three different algorithms, from the top to the bottom row: Forward Model Matched Filter (FMMF), Gaussian Matched Filter (GMF) and Gaussian Cross Correlation (GCC). Each column corresponds to a specific dataset. The images correspond to SNR maps and the local SNR maximum at the location of the planet (dashed circle) is printed in each stamp.
\label{fig:cEriShowOff}}
\end{figure}

\autoref{fig:FMMFShowOff} gives an example of reduction using the three different algorithms on the HR 7012 GPIES dataset observed on 2015 April 08. This dataset doesn't contain any visible astrophysical signal. Simulated planets were injected according to \Secref{sec:fakes}. With the exception of the anomaly at $0.4\arcsec$, the FMMF consistently yields a higher SNR than the two other methods and the final image shows fewer residual features. In the following, the simulated datasets are only reduced at the position of the simulated planets in order to speed up the algorithm.
\autoref{fig:cEriShowOff} shows the processed data for several follow-up epochs of 51 Eridani also using the three algorithms.
We show that FMMF would have marginally detected 51 Eridani b in all epochs, while it is not true with the Gaussian Matched Filter and the Gaussian Cross Correlation.

\section{Optimization}
\label{sec:optimization}

The free parameters for KLIP and the matched filter are the exclusion criterion, the reduction spectrum, the number of Karhunen-Lo\'eve modes, the number of reference images, and the shape of the sectors.
In this paper, we are only optimizing with respect to the exclusion criterion and the spectrum of the template. The exclusion criterion is only optimized for T-type objects but we do not expect a significant difference for L-type objects. Preliminary tests show that the performance of the algorithm is less sensitive to the choice of the other parameters.

\subsection{Exclusion Criterion}
\label{sec:aggressOpti}
The exclusion criterion is defined in \Secref{sec:exclusionCrit} and its optimal value is a trade-off between achieving a better speckle subtraction and maintaining a stronger planet signal. The forward model helps to keep the throughput close to unity for more aggressive reductions, therefore improving the overall SNR.

\begin{figure}
\gridline{\fig{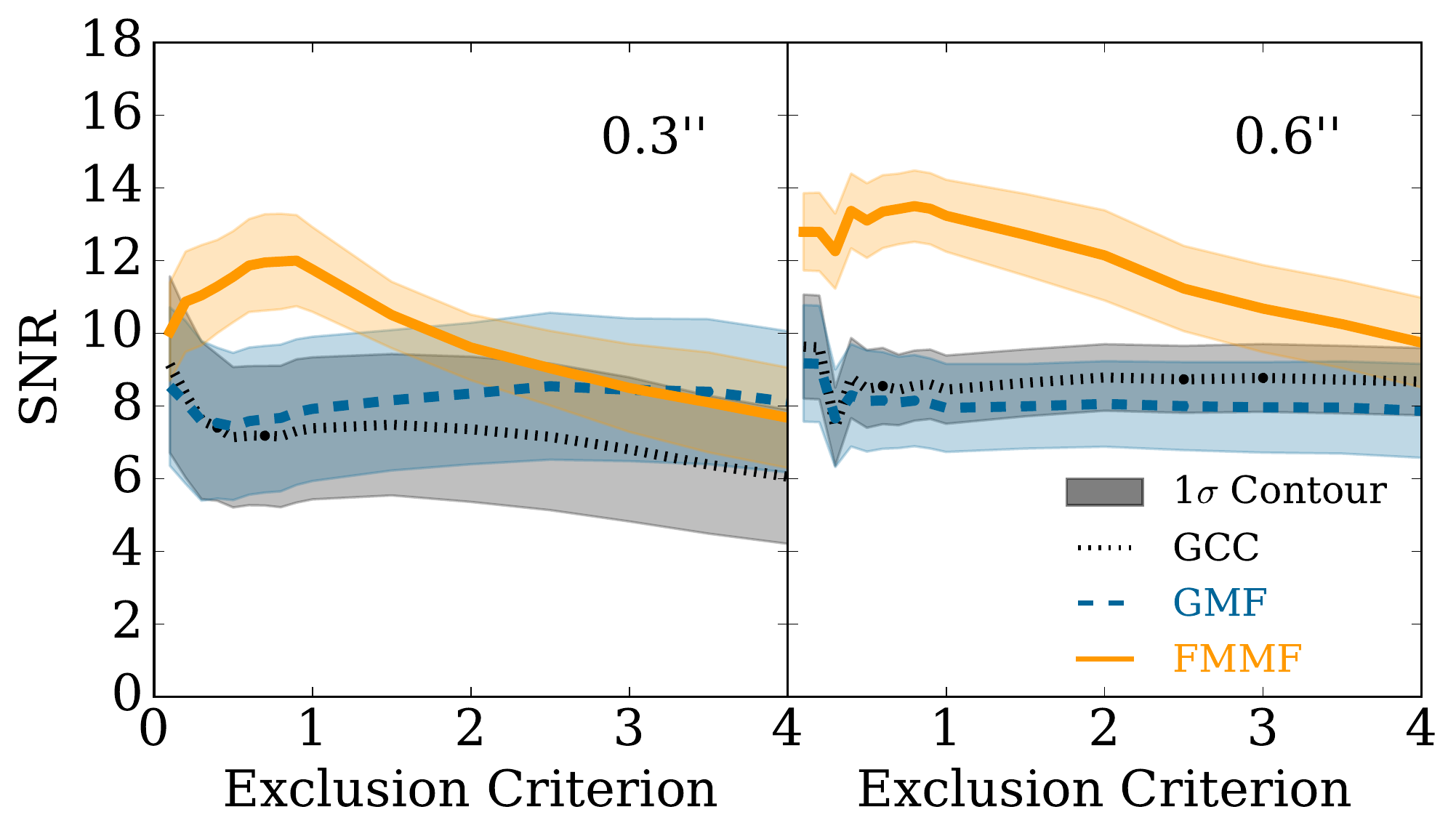}{0.5\textwidth}{(a) HR 7012}
		\fig{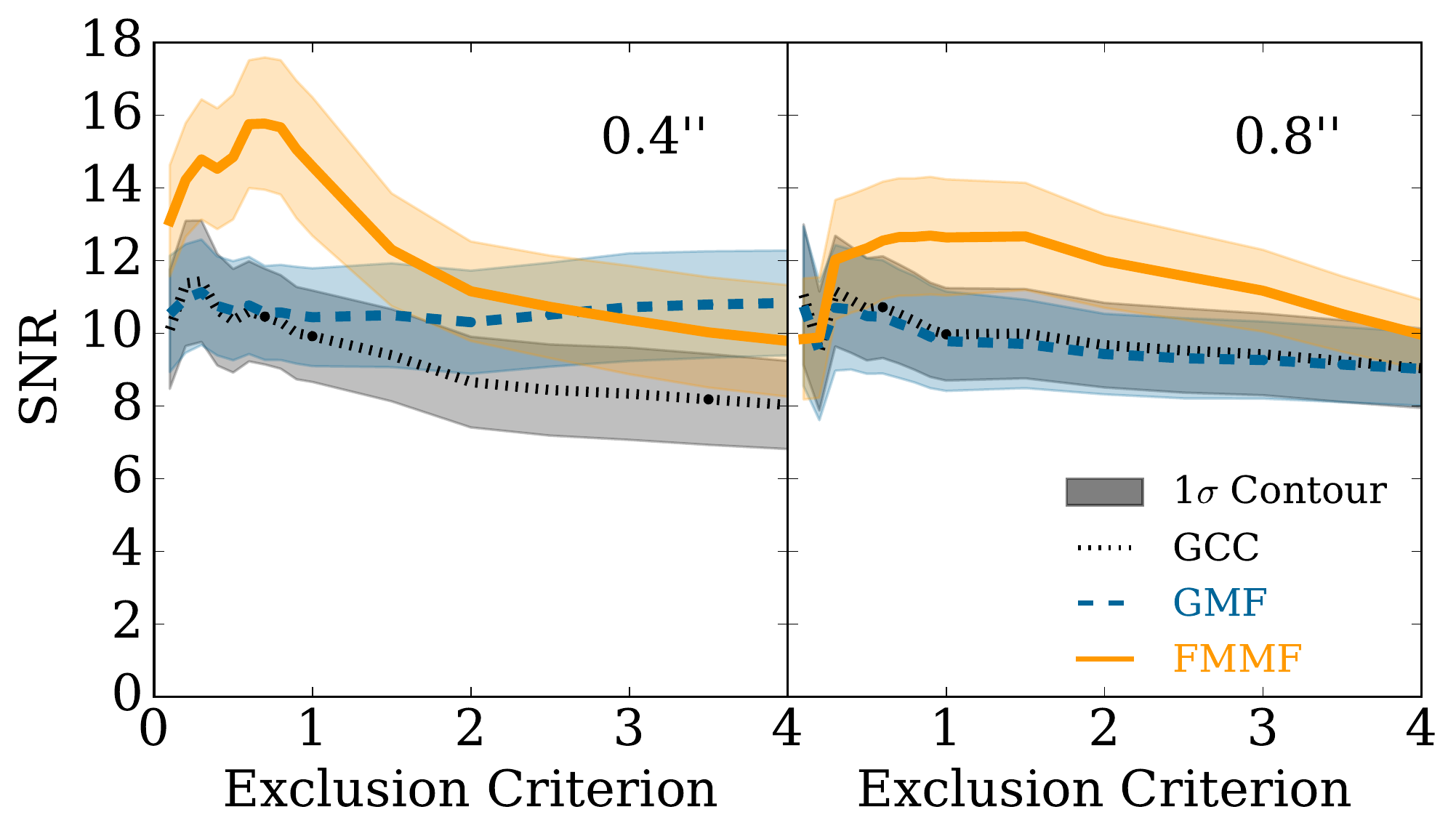}{0.5\textwidth}{(b) HD 131435}
          }
\caption{Signal-to-Noise Ratio (SNR) of simulated planets as a function of the exclusion parameter for the three algorithms: Forward Model Matched Filter (FMMF, Solid Orange), Gaussian Matched Filter (GMF, Dashed Blue) and Gaussian Cross Correlation (GCC, Dotted Black). The test was performed with two representative datasets at two different separations: (a) HR 7012 at $0.3\arcsec$ and $0.6\arcsec$ and (b) HD 131435 at $0.4\arcsec$ and $0.8\arcsec$. The datasets were chosen to represent different regimes of data quality; HR 7012 is an example of good dataset while HD 131435 is of median quality. The SNR was calculated for $35$ simulated exoplanets injected at different position angles in 7 different copies of the original dataset. The shaded region is the $1\sigma$ spread of the SNR. A T-type spectrum similar to 51 Eridani b was used for the simulated planets and a spectrum with a sharper methane signature was used for the reduction. \label{fig:SNRvsMvt}}
\end{figure}

The optimal value of the exclusion criterion is found by calculating the SNR of simulated planets for different values of the parameter. We used similar simulated planets as for \autoref{fig:thoughvsMvt} with a $1000\,\textrm{K}$ cloud-free model spectrum and a $600\,\textrm{K}$ cloud-free spectrum for the reduction. The choice of spectra is discussed in \Secref{sec:specMismatch}. \autoref{fig:SNRvsMvt} shows that the exclusion parameter has a soft optimimum around $0.7$.  The optimal exclusion criterion doesn't seem to significantly depend on separation or dataset quality. This apparent stability of the FMMF optimum suggests that a single value of the exclusion criterion can be used for the entire survey. Consequentially, all following reductions in \textit{H} band will be performed with an exclusion criterion of $0.7$. The optimal exclusion criterion is expected to change depending on the filter used for the observation and future work should involve separate optimizations in the different spectral bands.
The FMMF almost always yields a better SNR than the two other algorithm but this does not necessarily mean it has better detection efficiency.
Interestingly, neither the GCC nor GMF have a consistent or even a well-defined optimum. It is very common in the field to reduce a dataset with different sets of parameters and select the best one a posteriori but we believe that the FMMF limits the need to fine tune the parameters for each dataset.

\subsection{Spectral Mismatch}
\label{sec:specMismatch}

In this section, we discuss the optimization of the reduction spectrum used in the forward model and the reference library selection. The goal is to estimate the number of reduction spectra that should be used to recover the widest variety of planets. Currently detectable exoplanets and brown dwarfs are expected to feature spectra ranging from the T to the L spectral types. T-type objects are characterized by methane absorption bands visible in \textit{H} band and an energy peak around $1.6\,\mu m$ while L-type objects feature cloudy atmosphere with a flatter spectrum that peaks in the second half of the band.

\begin{figure}
\gridline{\fig{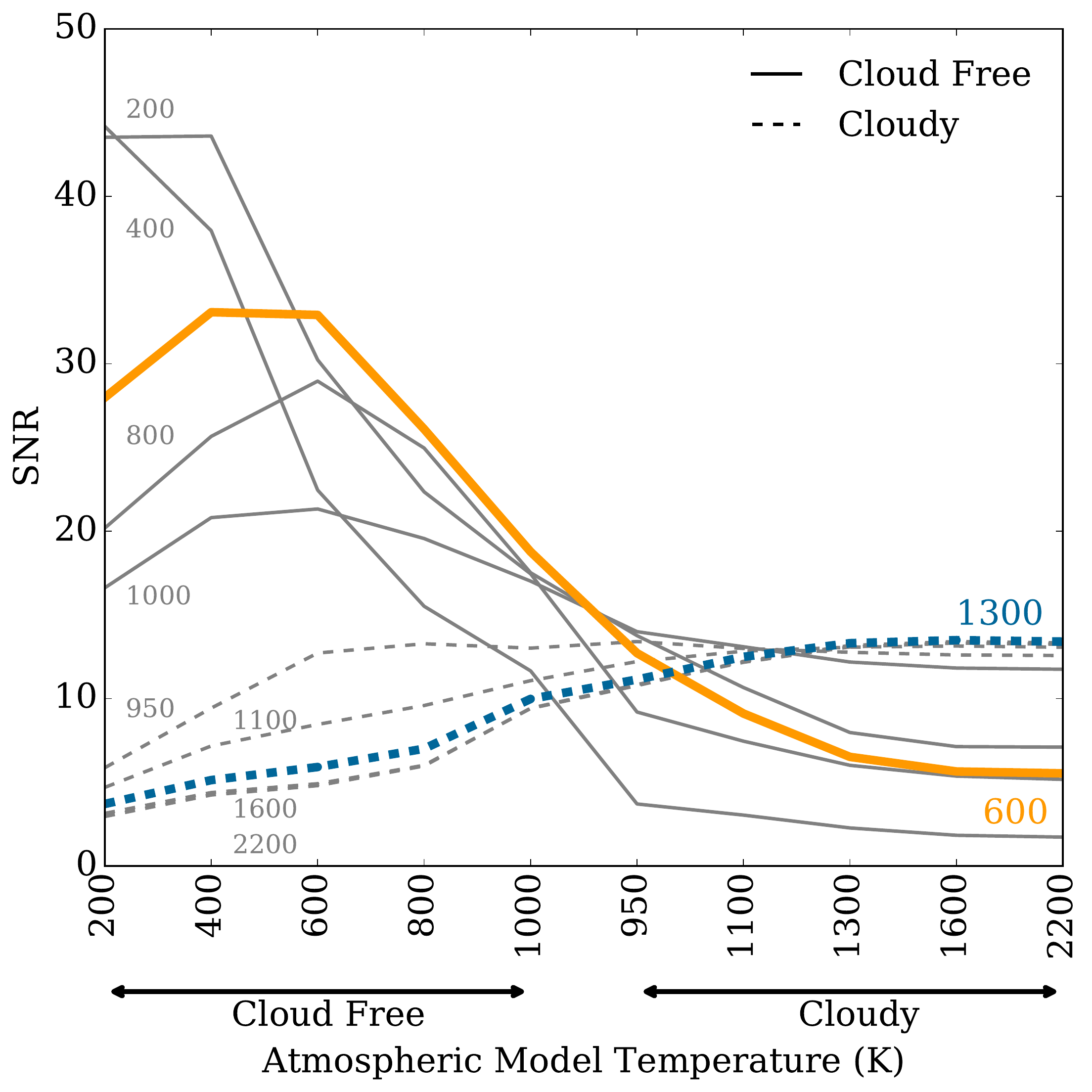}{0.5\textwidth}{(a) Planet Spectrum Vs. Reduction Spectrum}
		\fig{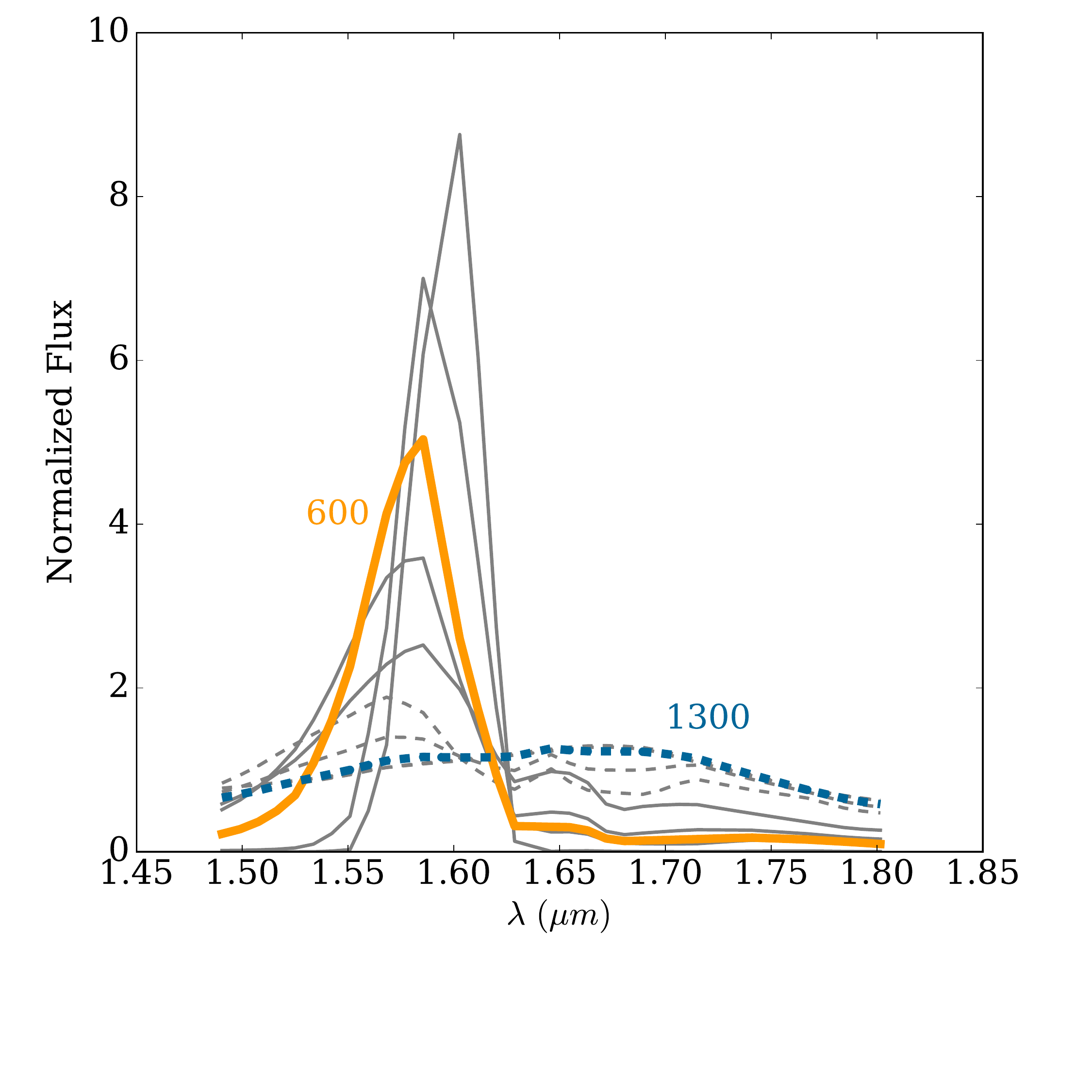}{0.5\textwidth}{(b) H-Band Model Spectra.}
          }
\caption{Effect of a mismatch between the planet spectrum and the reduction spectrum. (a) The median SNR (y-axis) of 16 planets injected in a the HD 131435 dataset is shown as a function of both the planet spectrum (x-axis) and the reduction spectrum (Curves). Figure (b) illustrates the model \textit{H}-band spectra used in (a). The orange and blue spectra have been selected as they allow the recovery of most spectral types without a significant loss of SNR. The spectra are taken from atmospheric models described in Marley et al. (2017, in preparation) and \cite{Saumon2012}. 
\label{fig:specVsSpec}}
\end{figure}

We created ten copies of the HD 131435 GPIES dataset each with 16 simulated planets injected according to the spiral pattern from \autoref{fig:FMMFShowOff}. In each copy of the dataset, the simulated planets were injected with one of ten spectra. The spectra were selected from a list of cloud-free and cloudy atmosphere models such that the most likely spectra from both T-type and L-type are represented. Note that the presence of clouds changes the temperature at which the methane features appear but does not significantly change the shape of the spectra. Then, each simulated dataset was reduced with each of the ten reduction spectra, resulting in 100 different final products. For each of the planet spectra, the best reduction spectrum is defined by the one yielding the best median SNR for the 16 simulated planets. \autoref{fig:specVsSpec} shows the median SNR of simulated planets as a function of their spectrum and the reduction spectrum. Surprisingly, the best reduction spectrum is not the one corresponding to the simulated spectrum.
One possible explanation is that the same reduction spectrum is used for the forward model as for the reference library selection through the exclusion criterion. The spectrum is also a way to weigh the spectral channels differently, which could effectively correct for a biased estimation of the standard deviation where there is planet signal. However, a deeper exploration of this effect is outside the scope of this paper.
We also conclude that only two spectra, the cloud-free $600\,\textrm{K}$ and the cloudy $1300\,\textrm{K}$, are necessary to allow the detection of most giant planets without a significant loss of SNR.
This result is consistent with a similar study in \cite{Johnson-Groh2017} using TLOCI \citep{Marois2014}.
Although the $600\,\textrm{K}$ spectrum is the optimal reduction spectrum for T-type objects, the methane induced peak in H-band is unrealistically sharp. It has indeed not yet been observed in a real directly imaged exoplanet. However, the reduction spectrum and the spectrum of the simulated planets can be different. In order to be more representative of the observations, we use a $1000\,\textrm{K}$ cloud-free model spectrum similar to 51 Eridani b for the T-type injected planet, which has a softer methane induced peak. The L-type simulated planets use the same $1300\,\textrm{K}$ cloudy model spectrum as for the reduction.

For the remainder of the paper, the T-type reduction refers to a cloud-free $600\,\textrm{K}$ reduction spectrum while the L-type reduction refers to a cloudy $1300\,\textrm{K}$ reduction model. 

\section{Noise Characterization}
\label{sec:noiseCharac}

\subsection{Noise Distribution}
\label{sec:noiseDistrib}

\begin{figure}[bth]
\gridline{\fig{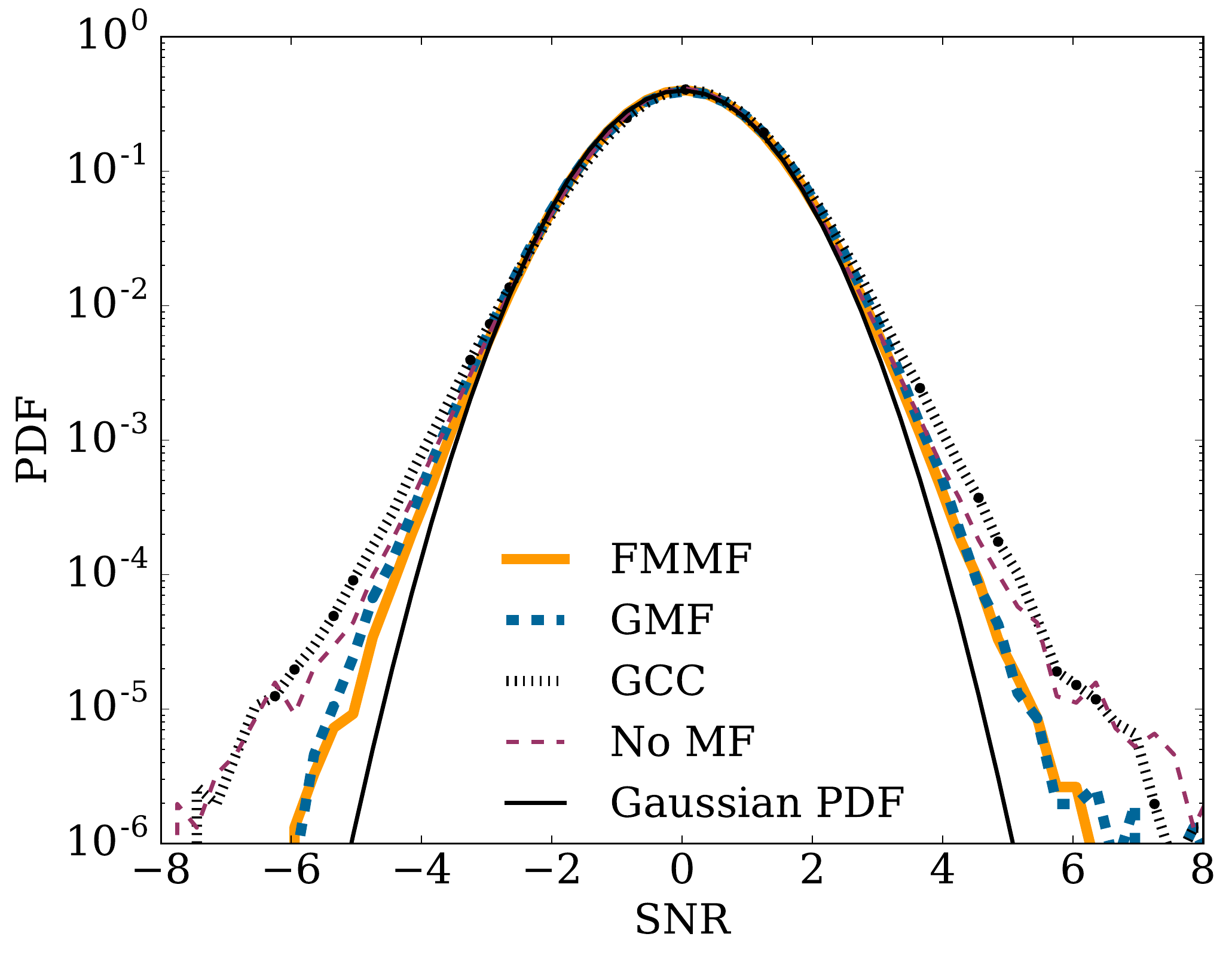}{0.5\textwidth}{(a) T-type}
		\fig{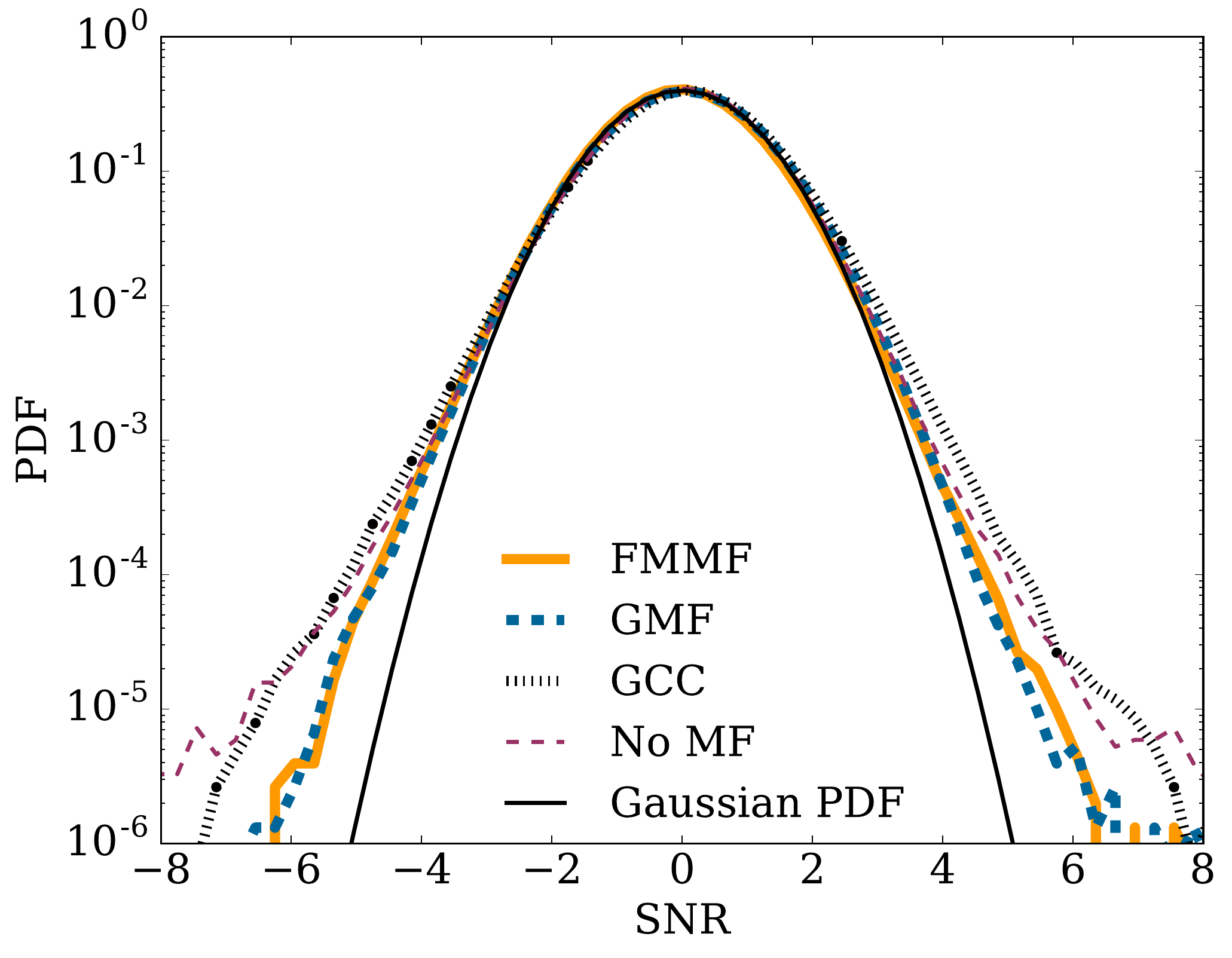}{0.5\textwidth}{(b) L-type}
          }
\caption{Probability Density Function (PDF) of the GPIES SNR maps for the three algorithms: (Solid Orange) Forward Model Matched Filter (FMMF), (Dashed Blue) Gaussian Matched Filter (GMF) and (Dotted Black) Gaussian Cross Correlation (GCC). The matched filter residuals can be compared to the residuals without any matched filter (Dashed Purple) and to an ideal Gaussian PDF (Solid Black). The PDFs are given for both the T-type (a) and the L-type (b) reductions. A total of $330$ GPIES \textit{H}-band datasets were used in which any known astrophysical signal was removed.}
\label{fig:PDF}
\end{figure}

\begin{figure}[bth]
\gridline{\fig{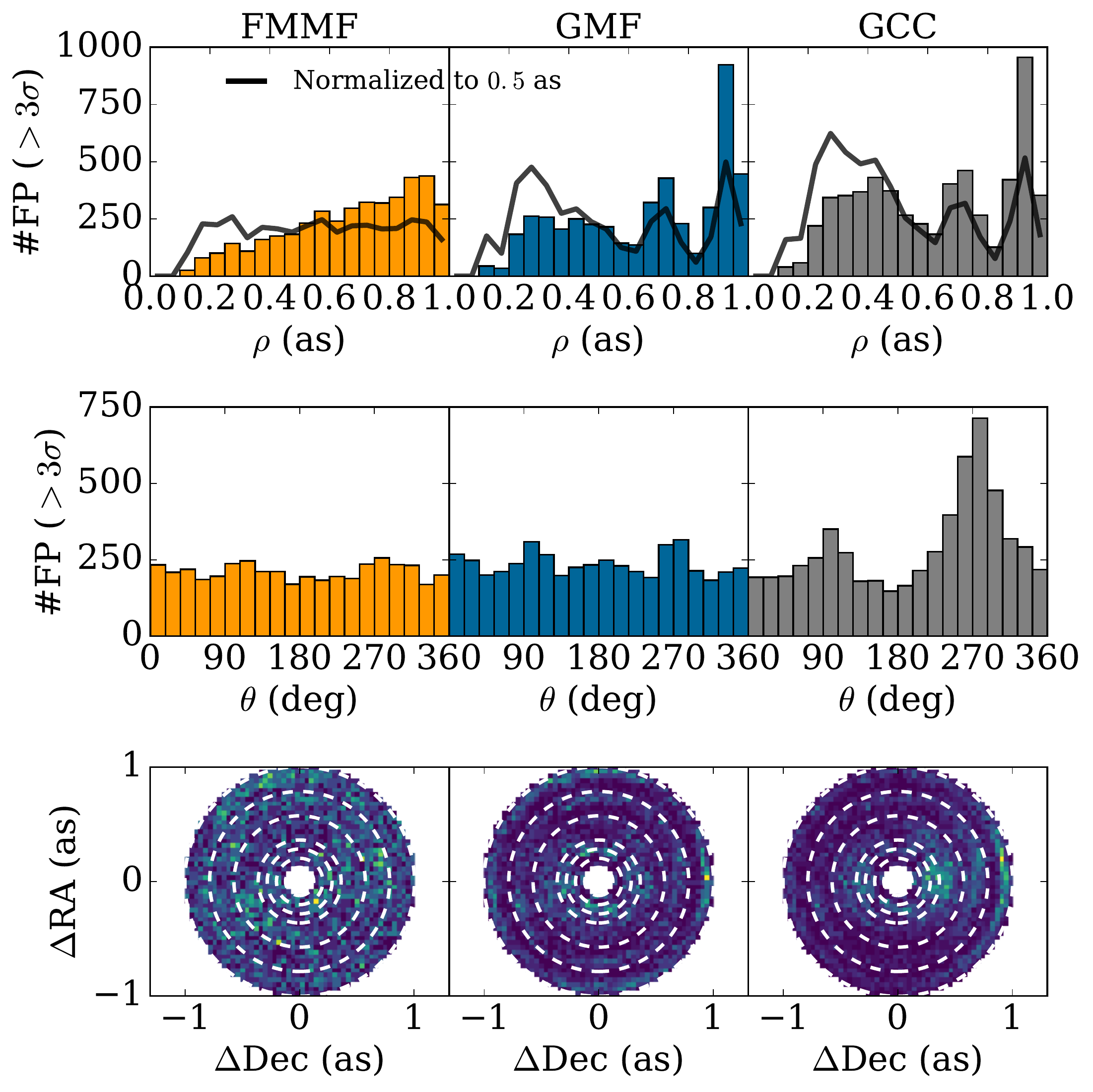}{0.5\textwidth}{(a) T-type}
		\fig{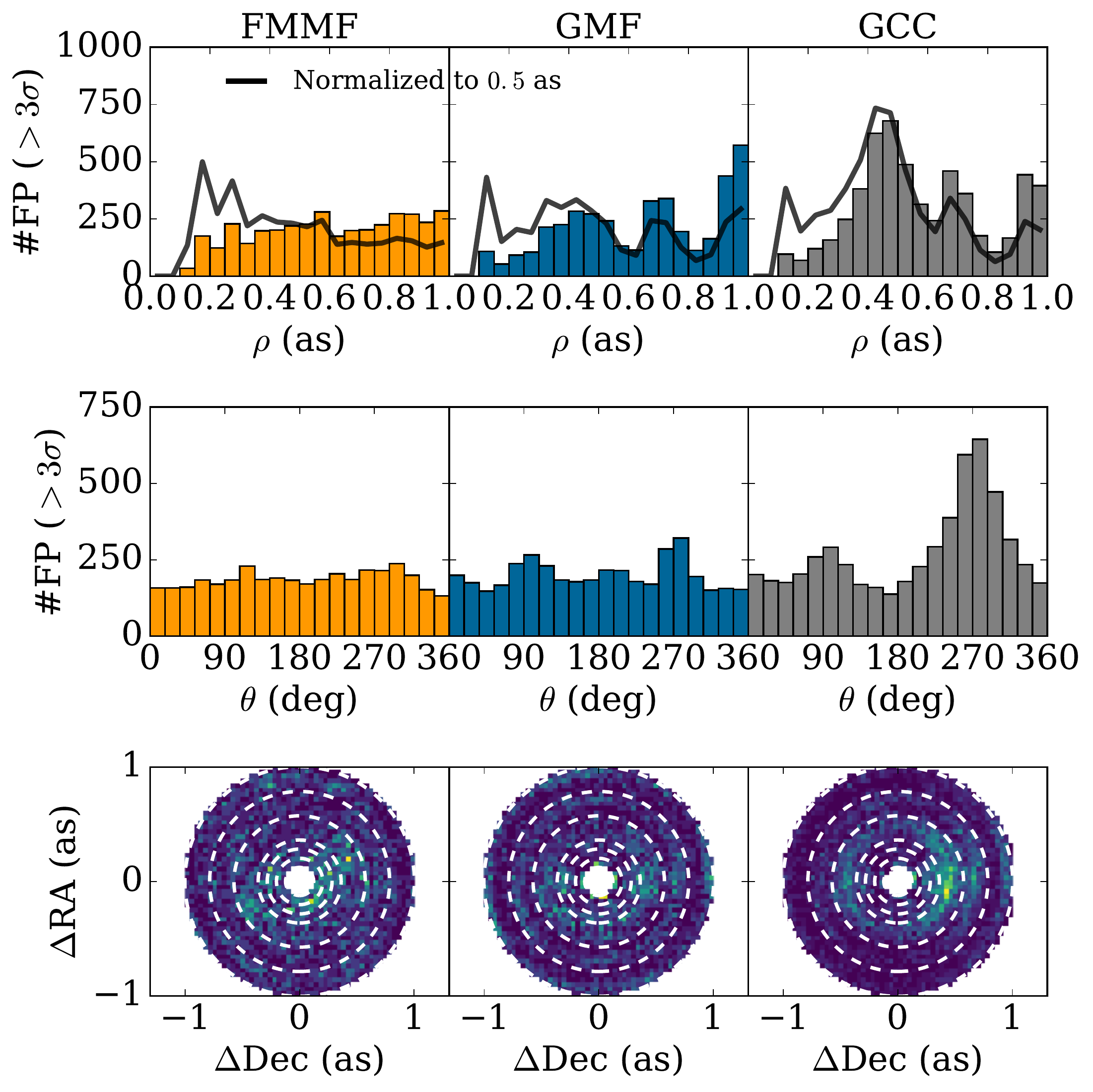}{0.5\textwidth}{(b) L-type}
          }
\caption{Spatial density of False Positives (FP) brighter than $3\sigma$ in GPIES for three algorithms: (First column, Orange) Forward Model Matched Filter (FMMF), (Second column, Blue) Gaussian Matched Filter (GMF) and (Third column, Grey) Gaussian Cross Correlation (GCC). The histograms are given for both the T-type (a) and the L-type (b) reductions. The first and second row feature the number of false positives per bin as a function of separation and position angle respectively.
The density of false positives is expected to increase at larger separation due to the larger area available. The solid black line gives the equivalent number of false positives at $0.5\arcsec$ after normalizing by the area at each separation. The bottom row shows the two-dimensional density of false positives as a function of declination and right ascension, or equivalently $x,y$-axes. A total of $330$ GPIES \textit{H}-band datasets were used in which any known astrophysical signal was removed.}
\label{fig:FPspatialDistrib}
\end{figure}

By combining the $330$ GPIES \textit{H}-band observations after removing any known astrophysical point sources, we have been able to estimate the Probability density Functions (PDF) of the residual noise up to an unprecedented precision in the fully reduced SNR maps. \autoref{fig:PDF} compares the ideal Gaussian PDF with the PDF of the different algorithms calculated from the normalized histograms of the pixel values of all the SNR maps.
The statistics of the noise strongly deviates from the Gaussian distribution at occurrence rates much greater than the frequency of planets \citep{nielsen2013,Bowler2015,Galicher2016,Vigan2017}, demonstrating the high occurrence of false positives with high SNR in direct imaging.
The Gaussian cross correlation has the same PDF as the SNR maps calculated from the speckle subtracted images with no cross correlation or matched filter. The Gaussian matched filter and the FMMF both significantly improve the statistics of the residuals but remain quite remote from an ideal Gaussian distribution. The excess of high SNR occurrences can be explained by either a truly non-Gaussian statistic or by a poor estimation of the standard deviation of the noise when calculating the SNR. For example, an underestimated standard deviation for a pixel will result in more high SNR false positives. The relative improvement between the cross correlation and the matched filter suggests that the latter explanation may still be dominant, which we will discuss in more detail in the next paragraph. 

\autoref{fig:FPspatialDistrib} shows the spatial densities of $3\sigma$ false positives as a function of separation and position angle. Due to the correlation in the final SNR maps, one should count the number of speckles and not the raw number of pixels above a given threshold in order to count the number of false positives. Indeed, for any high SNR false detection, the size of the bump, and therefore the number of pixels above the threshold, will depend on the correlation length of the noise, which also depends on the algorithm used.
The detection of false positives is therefore done recursively as follows. The highest SNR pixel is flagged, and a four pixel area radius is masked around it. This process is repeated until the only false positives left are below a predefined SNR threshold.
Both the Gaussian Cross Correlation and the Gaussian Matched Filter exhibit strong radial variations of their false positive densities at the position of the sector boundaries as seen in the top row plots of \autoref{fig:FPspatialDistrib}. In addition, the Gaussian Cross Correlation has a significant excess of false positives around $90^\circ$ and $270^\circ$ in position angle. This feature can be explained by the excess of speckle noise on both sides of the focal plane mask in the direction of the wind, which we refer to as the wind-butterfly. The pattern is still visible after combining the entire survey because the wind direction overwhelmingly favors the North-East on Cerro Pachon at the Gemini South telescope. The wind-butterfly breaks the assumption of azimuthally identically distributed noise, which we use when estimating the standard deviation in concentric annuli. The wind-butterfly explains why the probability density function of the GCC in \autoref{fig:PDF} is significantly higher than the other matched filters.
The Gaussian Matched Filter doesn't suffer from this effect because the matched filter includes a normalization with respect to the local standard deviation estimated around each pixel. The FMMF features a similar PDF meaning that a similar SNR detection threshold should yield the same number of false positives. The real performance of each algorithm will be studied in the next sections.
While a cross correlation is a common planet detection approach, our analysis suggests that it can be ill-suited if the noise varies azimuthally. One should instead use the expression for the matched filter from \autoref{eq:datasetSNR}. An alternative approach would be to vary the SNR threshold for each dataset and as a function of position but the lack of local independent samples to estimate a position dependent PDF at small false positive rates makes this endeavor very challenging. Indeed, one needs to have probed the PDF of the noise at high SNR in order to evaluate the false positive probability of any detection. For example, $5\sigma$ events are sufficiently rare that their occurrence rate can only be estimated from the data of a entire survey and not individual images.

The FMMF L-type reduction has significantly more false positives near the mask than a T-type reduction. This is likely due to the fact that there is a higher density of speckles near the mask and the L-type spectrum is a better match to them than a sharper spectrum. We have also seen in \autoref{fig:PDF} that for the three algorithms, the L-type PDF has wider tails than the T-type PDF suggesting that the number of L-type false positives will be higher at constant SNR.

\subsection{Receiver Operating Characteristic}
\label{sec:ROC}

\begin{figure}[bth]
\gridline{\fig{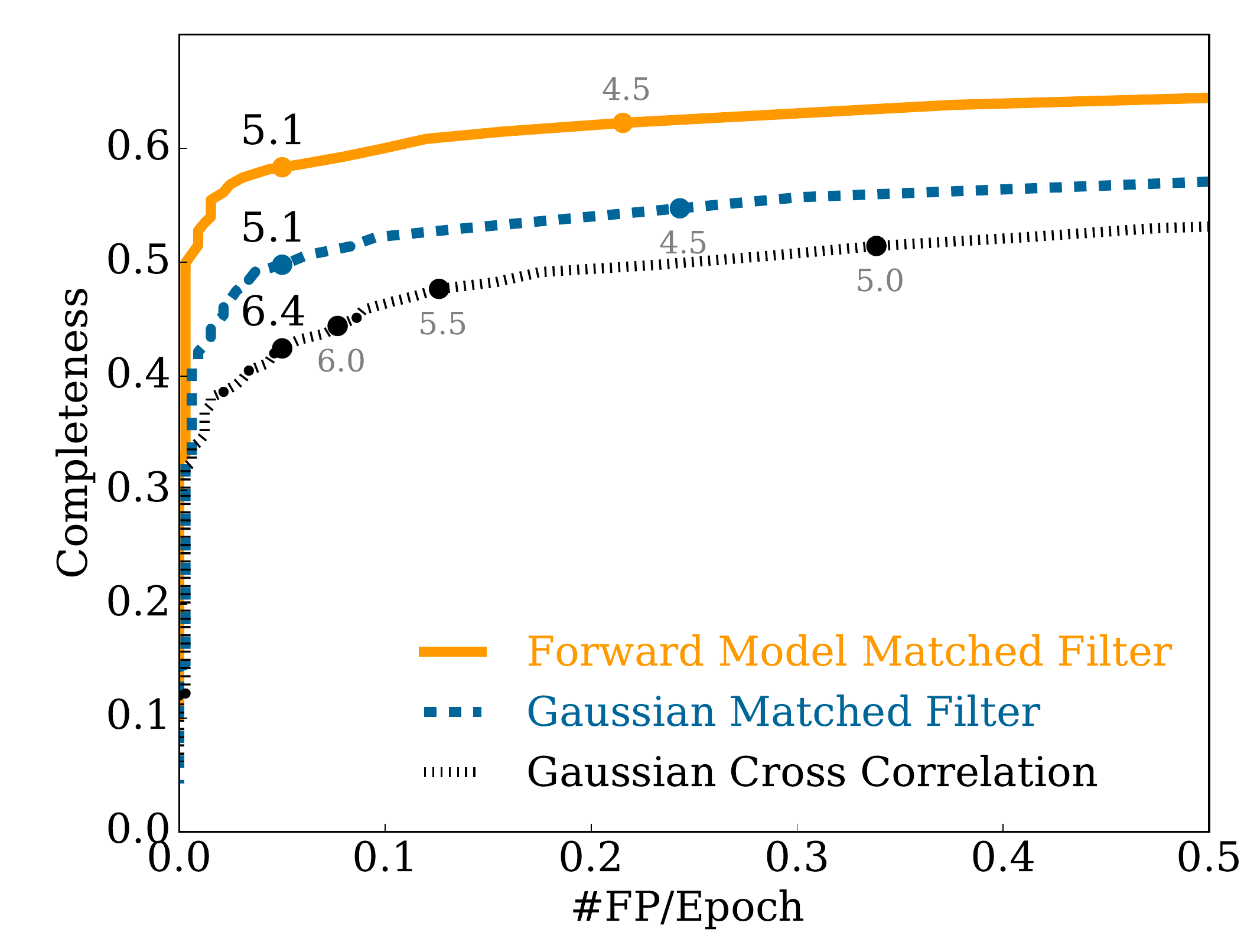}{0.5\textwidth}{(a) T-type}
		\fig{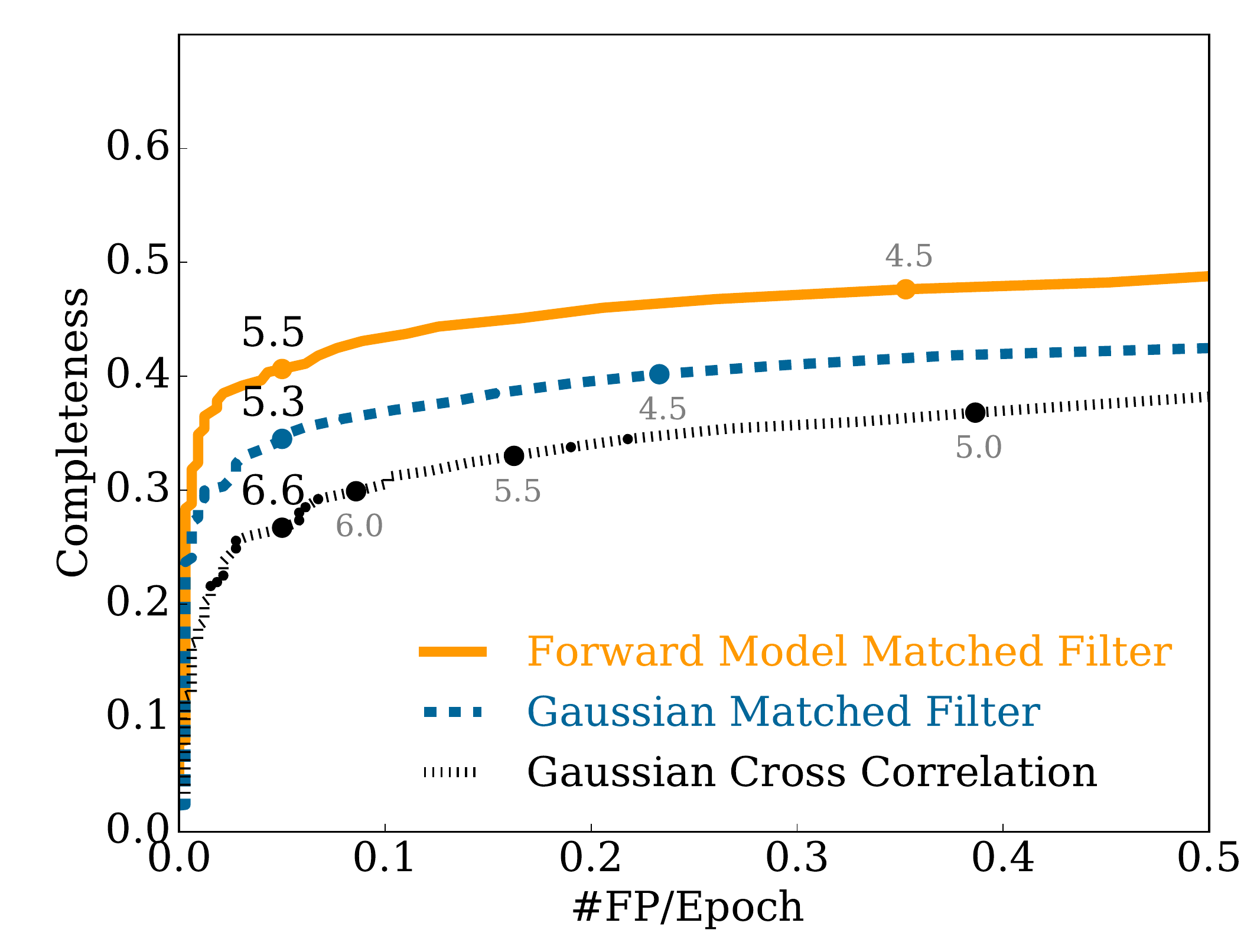}{0.5\textwidth}{(b) L-type}
          }
\caption{Receiver Operating Characteristic (ROC) for three different algorithms: (Orange solid line) Forward Model Matched Filter (FMMF), (Blue dashed line) Gaussian Matched Filter and (Grey dotted line) Gaussian Cross Correlation. The ROC curves feature the current GPIES completeness to T-type (a) and L-type (b) planets with a $4\times10^{-6}$ contrast integrated over all separation and angle as a function of the number of False Positives (FP) per epoch. A few values of the SNR threshold is annotated in grey on each curves. The threshold corresponding to a fraction of false positive per epoch of $0.05$ is written in a larger black font.
A total of $330$ GPIES \textit{H}-band datasets were used in which any known astrophysical signal was removed.}
\label{fig:ROC}
\end{figure}

In general, an improvement in the SNR does not guarantee a better detection efficiency because the false positive rate could increase in the mean time. It is therefore important to compare the number of detected planets to the number of false positives. For example, \autoref{fig:SNRvsMvt} showed that the Gaussian cross correlation tends to have slightly higher SNR than the Gaussian matched filter but we will show in this section that the cross correlation leads to more false positives. For this reason, Receiver Operating Characteristic (ROC) have become increasingly popular in direct imaging to compare different algorithms (\citeauthor{Caucci2007} \citeyear{Caucci2007}; \citeauthor{Choquet2015} \citeyear{Choquet2015}; \citeauthor{Pairet2016} \citeyear{Pairet2016}; \citeauthor{Pueyo2016} \citeyear{Pueyo2016}; Jensen-Clem et al. 2017, submitted).
A ROC curve compares the false positive fraction to the true positive fraction, \textit{i.e.} completeness, as a function of the detection threshold. Alternatively, we have decided to replace the false positive fraction by the number of false detections per epoch integrated over the entire image since it is easily translatable to survey efficiency and astrophysical background occurrence rate. We have also chosen to fix the planet contrast and assume a uniform planet distribution in separation and position angle. Although this approach does not address the dependence of the ROC curve on separation and planet contrast, it is sufficient to evaluate the relative performance of each algorithm.
Multi-dimensional ROC curves could be used but the contrast curves calculated in \Secref{sec:contCurve} already include most of the relevant information.
The ROC curves for the three algorithms are shown in \autoref{fig:ROC}.
Each ROC curve has been built by injecting 16 simulated exoplanets, using either a $1000\,\textrm{K}$ cloud-free T-type spectrum, reduced with a $600\,\textrm{K}$ cloud-free model spectrum, or a $1300\,\textrm{K}$ cloudy L-type spectrum, reduced with a the same cloudy spectrum. All planets were injected at a $4\times10^{-6}$ contrast in 330 GPIES \textit{H}-band datasets using the spiral pattern illustrated in \autoref{fig:FMMFShowOff}. 
The advantage of using a large number of datasets is to marginalize over the conditions in any one particular epoch.
\autoref{fig:ROC} shows that FMMF yields a better completeness at any false positive rate. Also, the SNR threshold corresponding to a given false positive rate is always higher in the case of the cross correlation than for the matched filters due to the larger tail in the PDF. For example, it has a T-type false positive rate roughly $6$ times higher at $5\sigma$ than the two other algorithms. 

The detection threshold should be defined by the number of false positives that can reasonably be followed up during the survey. We set this false positive rate at $0.05$ per epoch corresponding to 30 false positives for the entire survey. The SNR threshold therefore depends on the algorithm as shown in \autoref{fig:ROC}.

\section{Contrast Curve}
\label{sec:contCurve}

\begin{figure}[bth]
\gridline{\fig{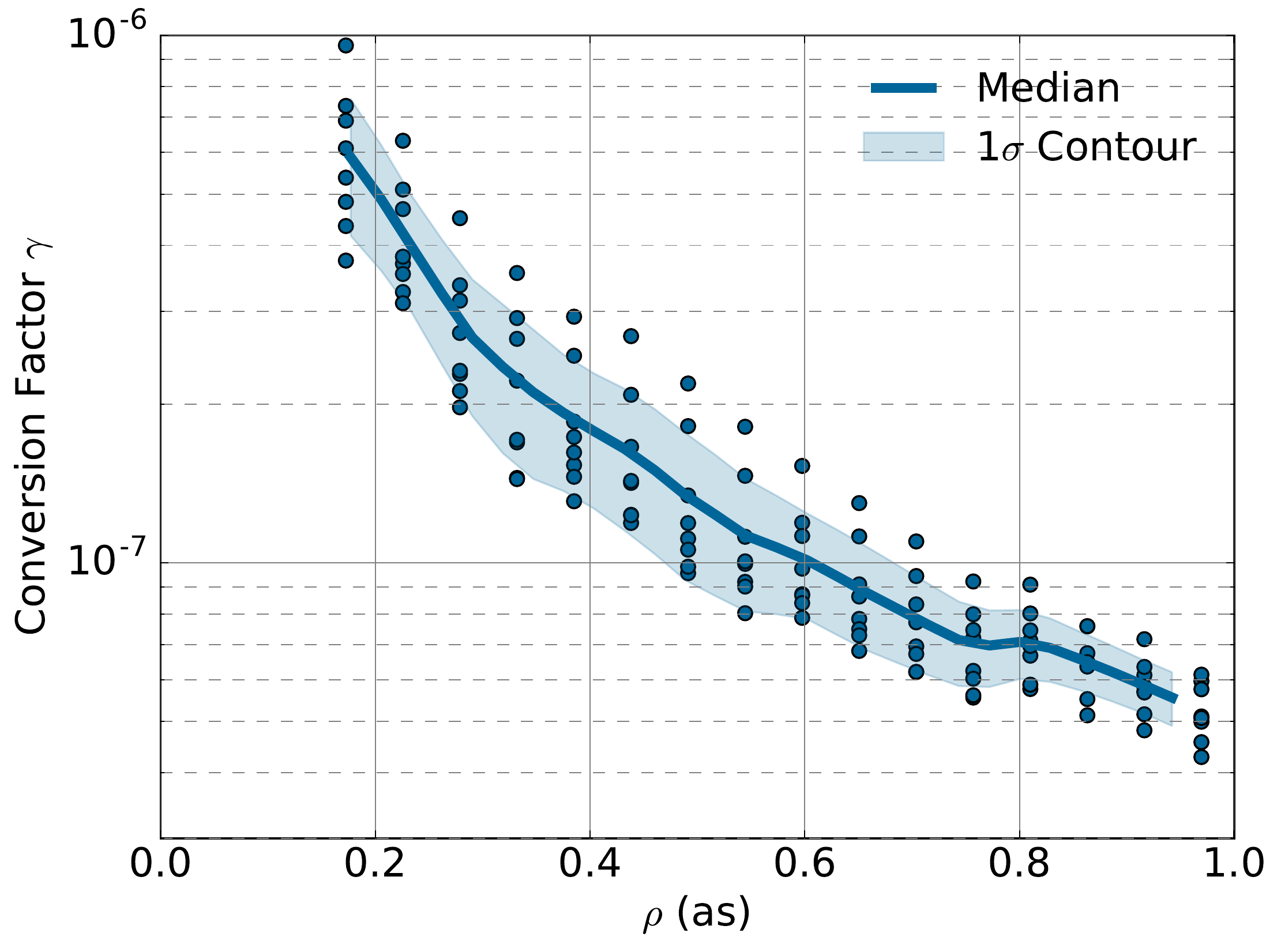}{0.5\textwidth}{(a) Conversion Factor}
		\fig{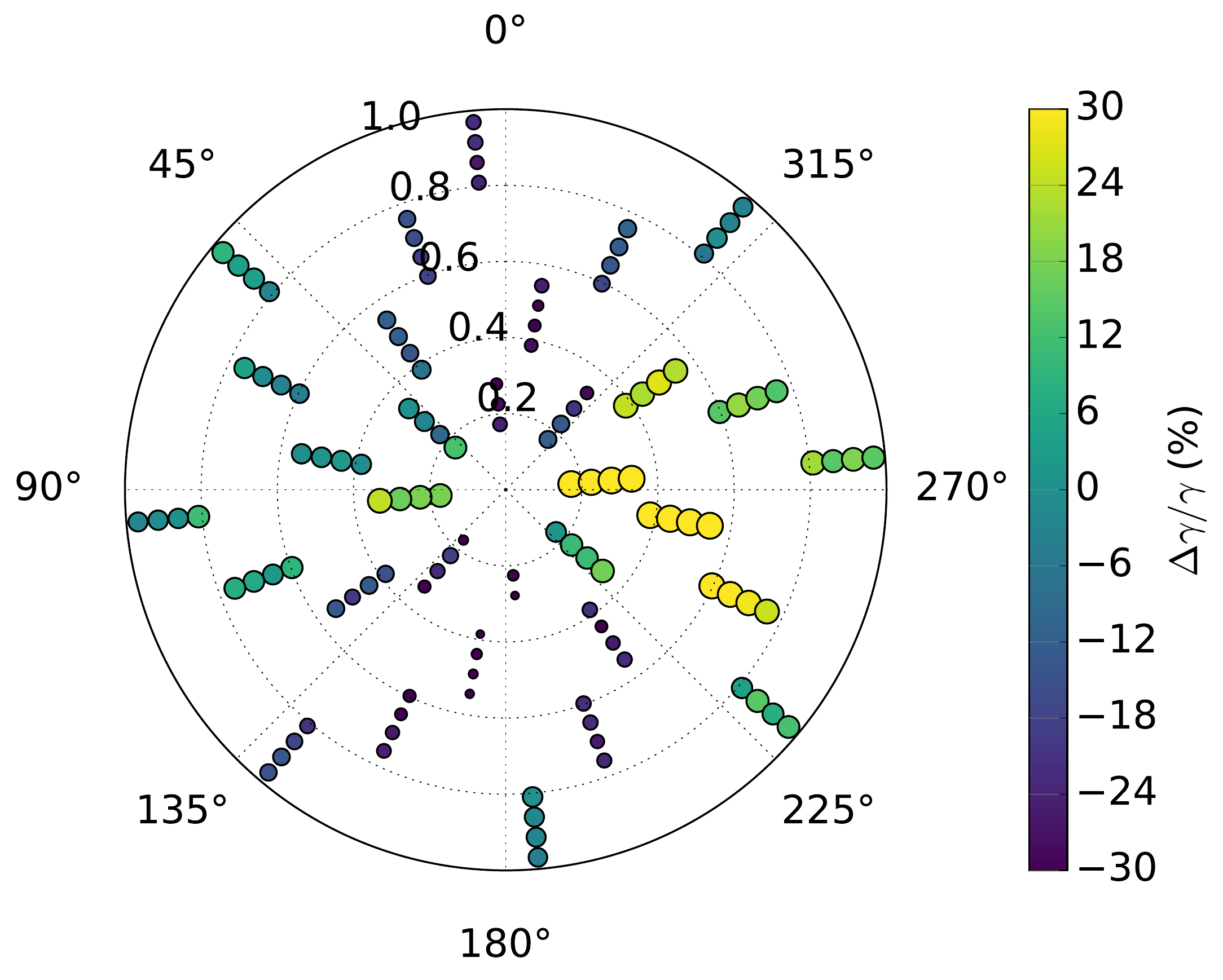}{0.5\textwidth}{(b) Relative Variations}
          }
\caption{Illustration of a T-type conversion factor calibration with simulated injected planets. (a) The conversion factor is calculated from the median over eight simulated planets at each separation. A total of 128 simulated exoplanets injected in eight different copies of the dataset is used to calibrate the conversion factor. (b) The polar plot shows the relative azimuthal variation of the conversion factor (represented by both the color map and the dot scale) as a function of separation in arc-seconds and position angle in degrees. The dataset has been chosen to feature a strong wind-butterfly effect resulting in a higher conversion factor around $90^\circ$ and $270^\circ$.}
\label{fig:convFactor}
\end{figure}

\begin{figure}[bth]
\gridline{\fig{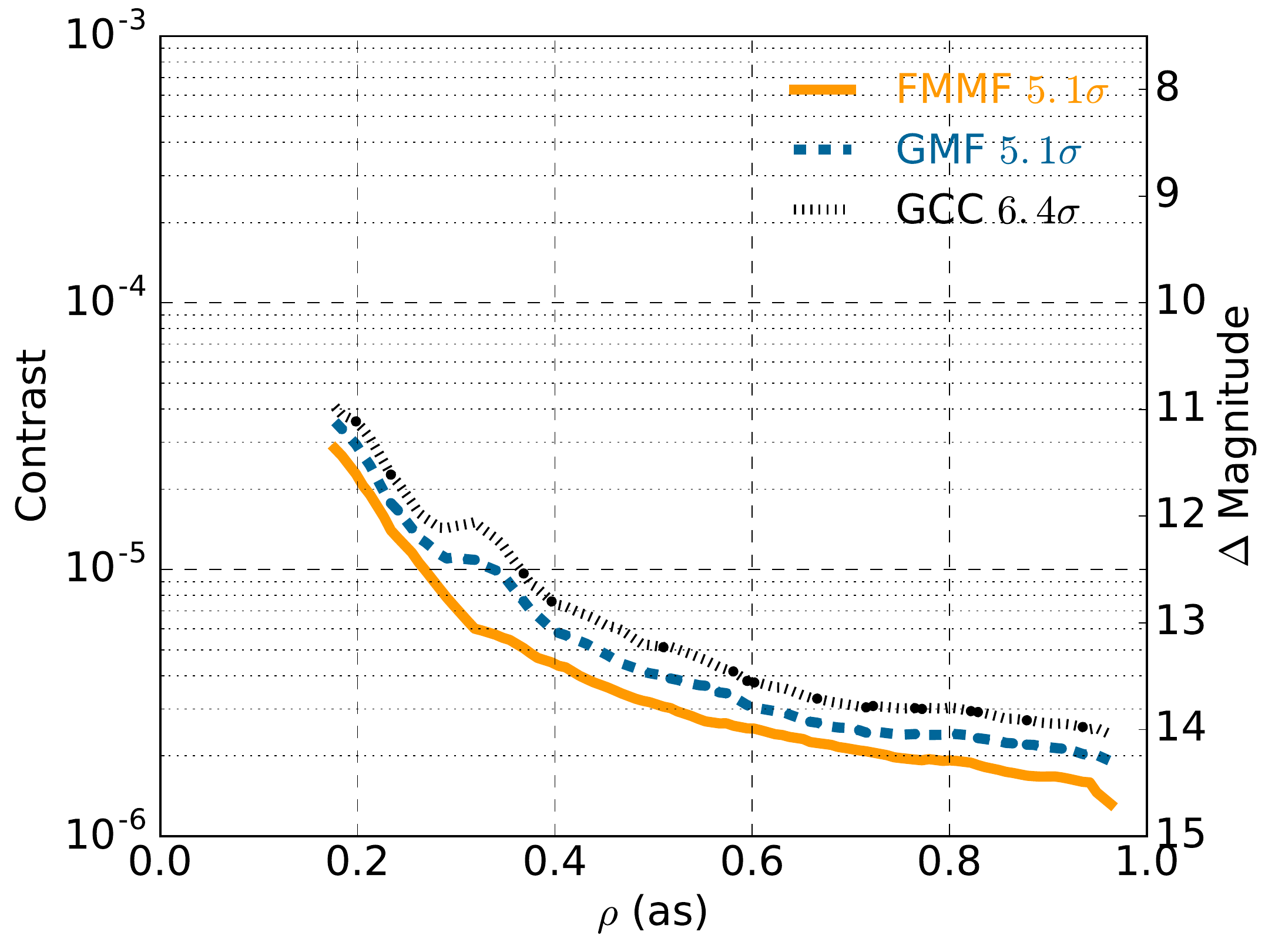}{0.5\textwidth}{(a) T-type}
		\fig{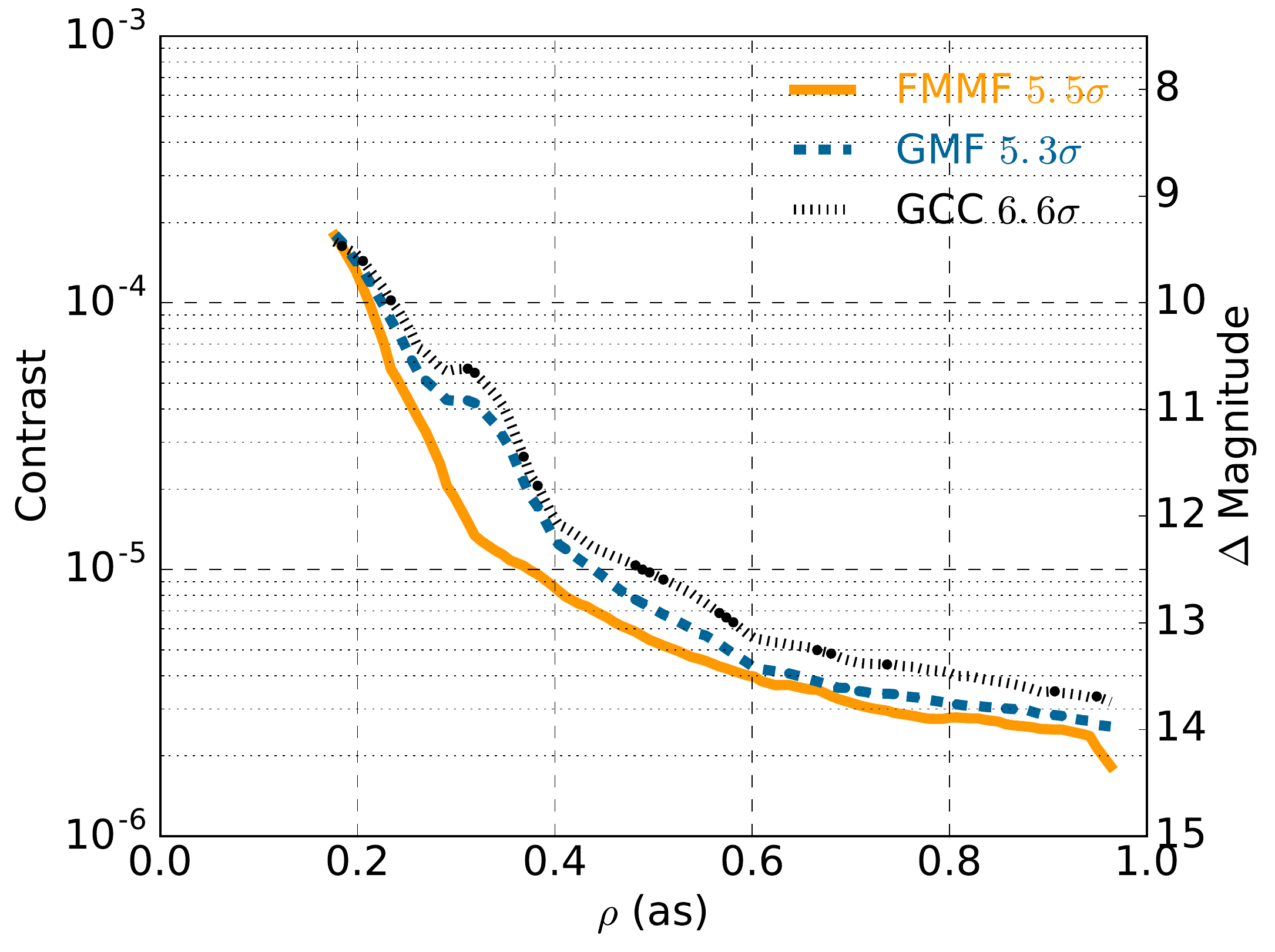}{0.5\textwidth}{(b) L-type}
          }
\caption{Median GPIES contrast curves for T-type (a) and L-type (b) reductions. Three different algorithms are compared: (Orange solid line) Forward Model Matched Filter (FMMF), (Blue dashed line) Gaussian Matched Filter (GMF) and (Grey dotted line) Gaussian Cross Correlation (GCC).  The contrast on the y-axis refers to the companion to host star brightness ratio with a $50\%$ completeness and a false positive rate of 0.05 per epoch. The detection threshold, which is indicated in the legend, varies from one algorithm to the other in order to always yield the same number of false positives. A total of $330$ GPIES \textit{H}-band datasets were used in which any known astrophysical signal was removed. }
\label{fig:medContrast}
\end{figure}

\begin{figure}[bth]
\gridline{\fig{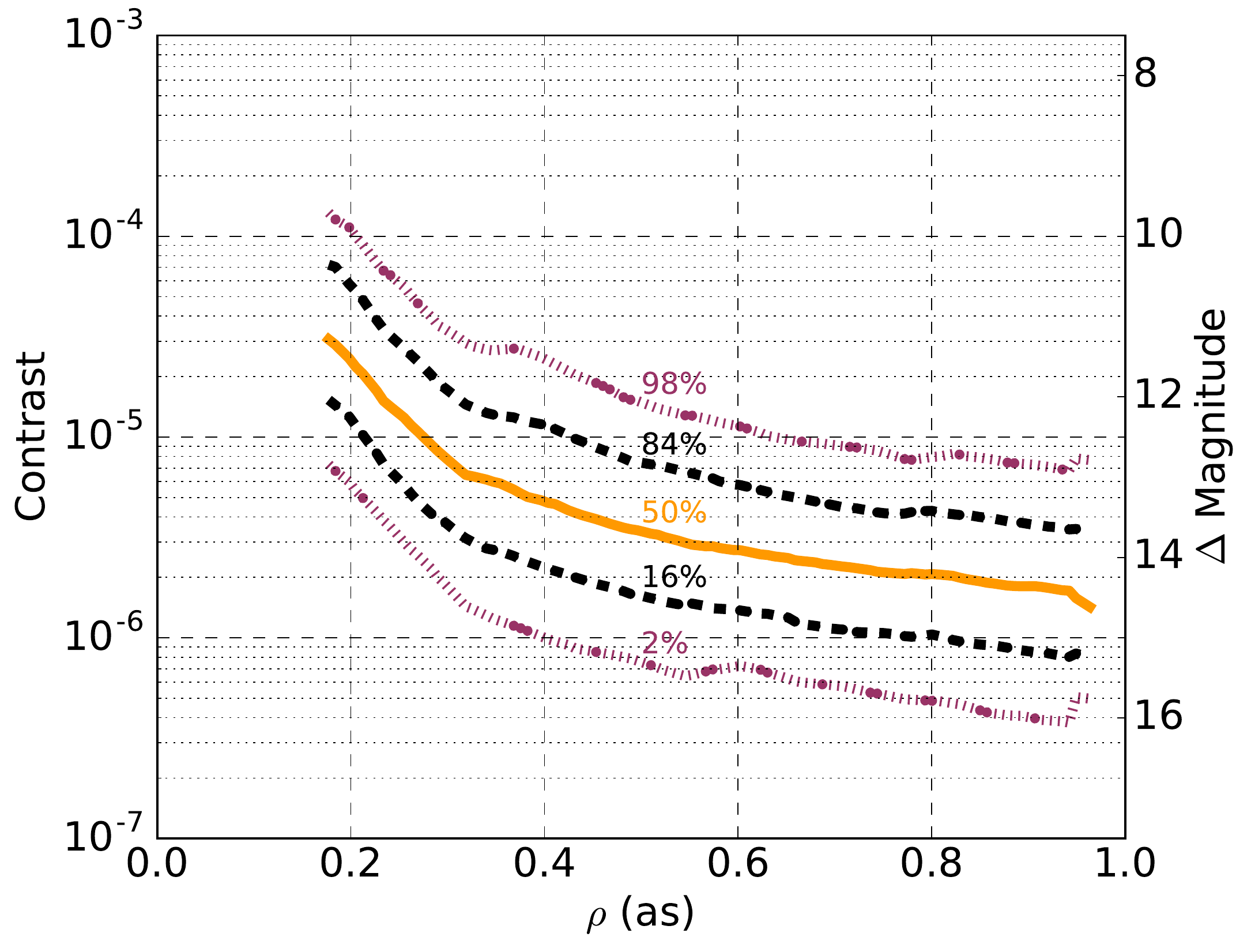}{0.5\textwidth}{(a) T-type}
		\fig{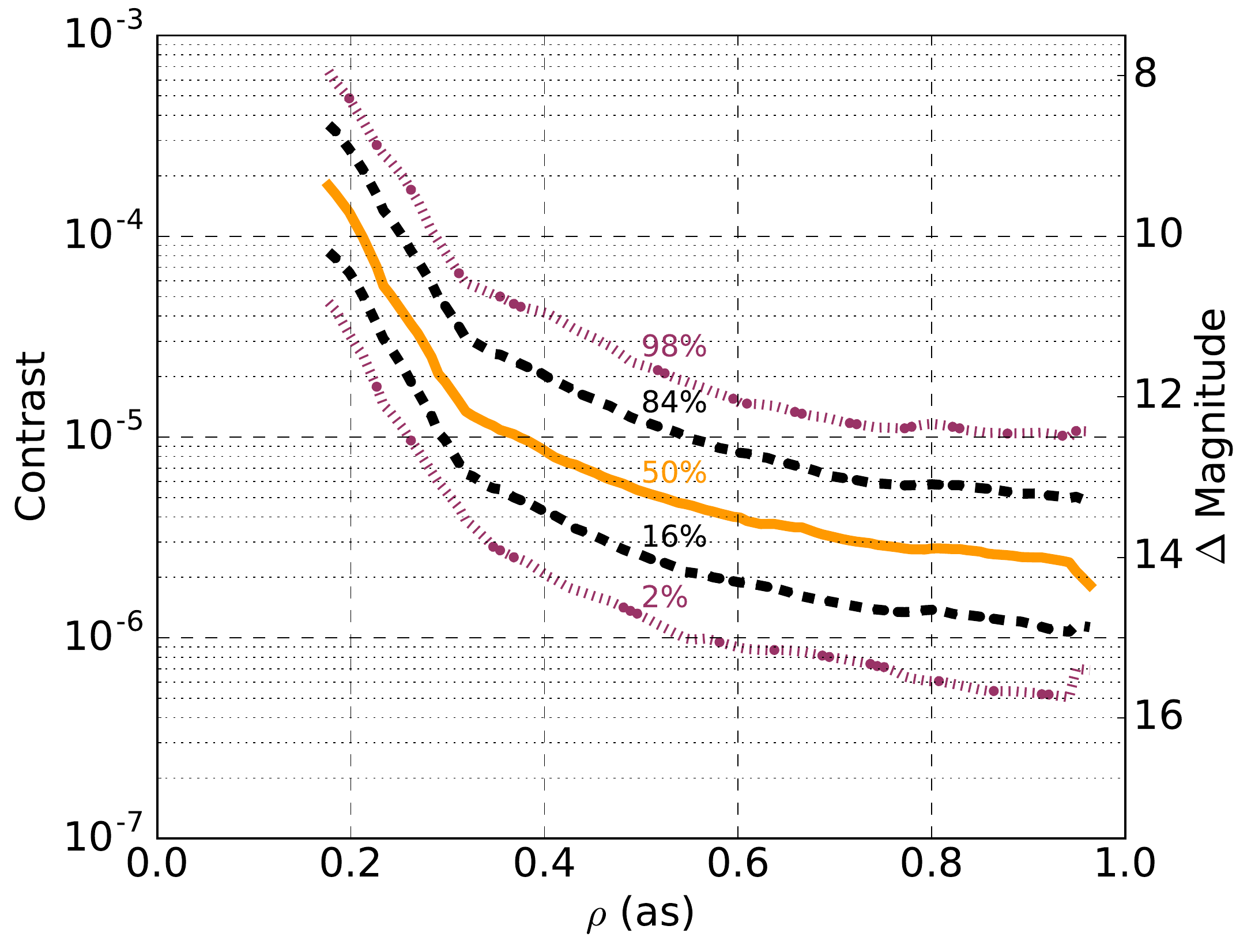}{0.5\textwidth}{(b) L-type}
          }
\caption{Percentiles of GPIES contrast curves with the Forward Model Matched Filter (FMMF) for T-type (a) and L-type (b) reductions. The contrast on the y-axis refers to the companion to host star brightness ratio with a $50\%$ completeness and a false positive rate of 0.05 per epoch. The detection threshold is set to $5.1\sigma$ and $5.5\sigma$ for the T-type and L-type reductions respectively, in order to always yield the same number of false positives.  A total of $330$ GPIES \textit{H}-band datasets were used in which any known astrophysical signal was removed. }
\label{fig:medContrastPercentile}
\end{figure}

\begin{figure}[bth]
\gridline{\fig{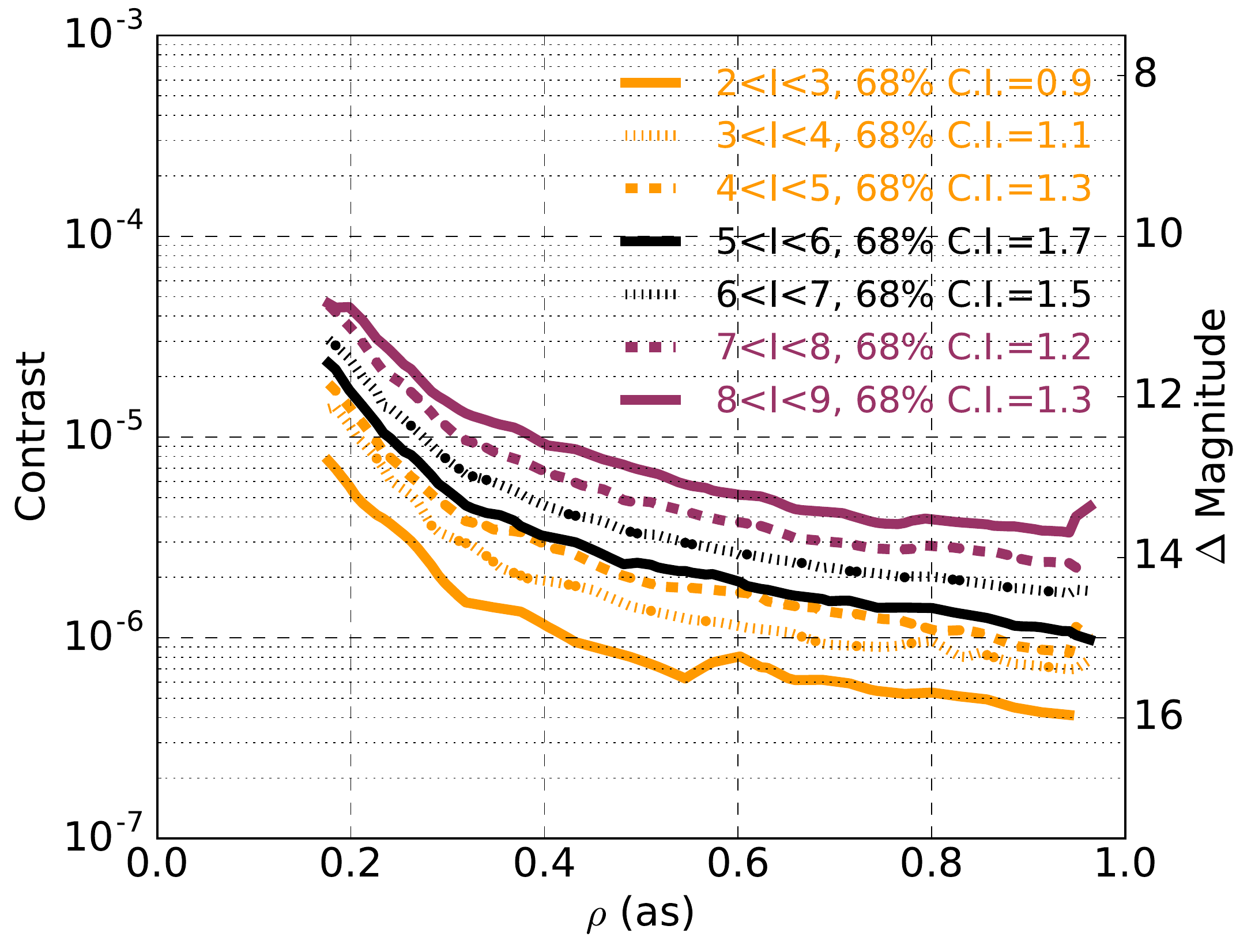}{0.5\textwidth}{(a) T-type}
		\fig{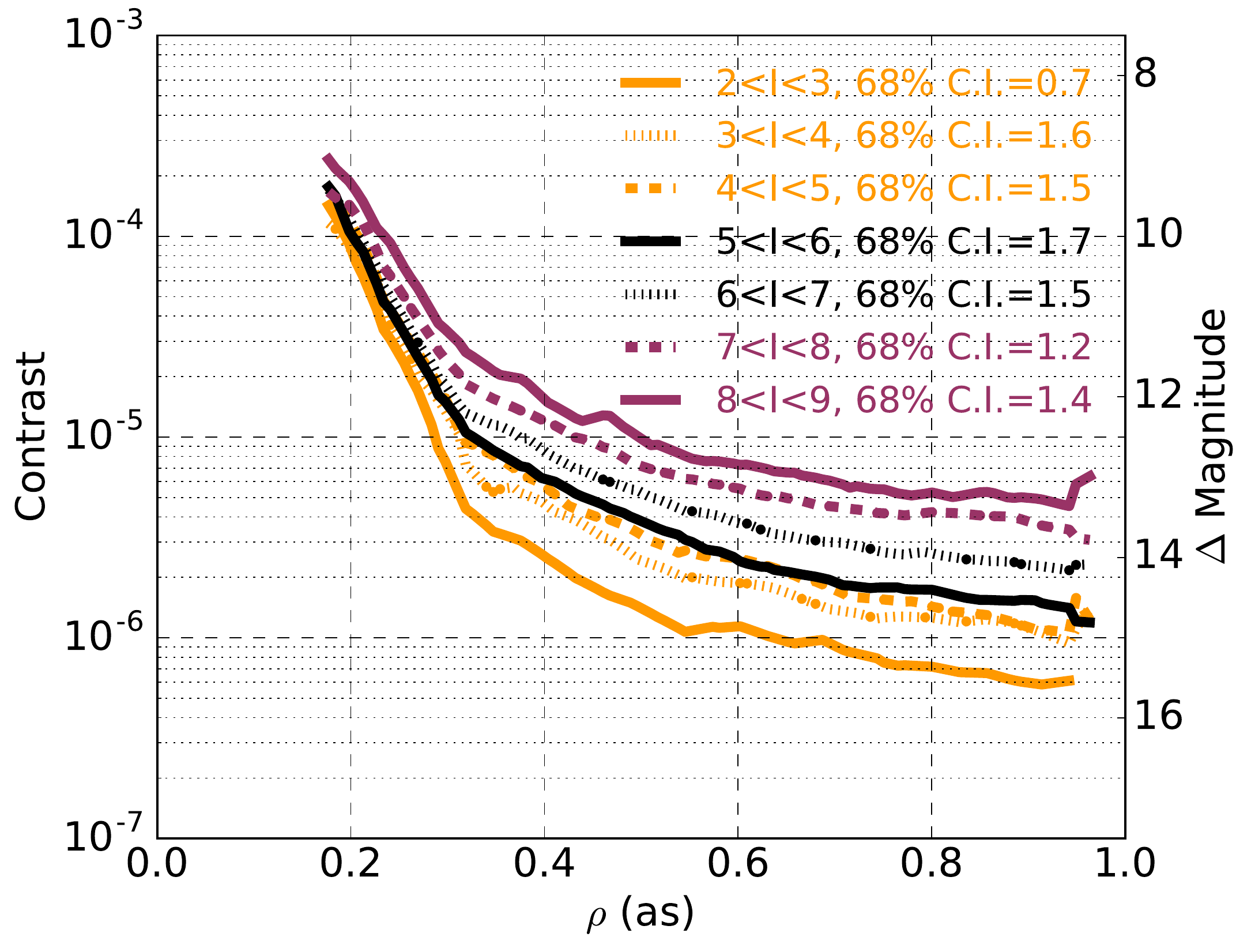}{0.5\textwidth}{(b) L-type}
          }
\caption{Median GPIES contrast curves as a function of the star \textit{I} magnitude with the Forward Model Matched Filter (FMMF) for T-type (a) and L-type (b) reductions. The legend includes the \textit{I} magnitude bin used for each curve as well as the size of the $68\%$ Confidence Interval (CI) at $0.4\arcsec$ in magnitude units. The contrast on the y-axis refers to the companion to host star brightness ratio with a $50\%$ completeness and a false positive rate of 0.05 per epoch. The detection threshold is set to $5.1\sigma$ and $5.5\sigma$ for the T-type and L-type reductions respectively, in order to always yield the same number of false positives. A total of $330$ GPIES \textit{H}-band datasets were used in which any known astrophysical signal was removed.}
\label{fig:medContrastImag}
\end{figure}

In this paper, contrast is defined as the broadband flux ratio between the companion and star. The contrast curve is defined as the $50\%$ detection completeness contour assuming a false positive rate sufficiently low to limit the number of false positives. The false positive rate can be expressed in terms of a SNR threshold, which is not necessarily $5\sigma$. Using a hard $5\sigma$ threshold does not allow for a meaningful comparison of contrast curves, because different algorithms can lead to different numbers of false positives for the same SNR (as demonstrated in \Secref{sec:ROC}). Contrast curve calculations require a calibration step to translate pixel values to planet contrast, which is in some cases referred to as throughput correction. Indeed, speckle subtraction algorithm like KLIP are known to over-subtract the signal of the planet and make it appear fainter than it really is. This effect can be calibrated out by inflating the standard deviations by a certain factor, known as throughput, calculated from simulated planet injection. However, the throughput correction only makes sense if the pixel values, and therefore the standard deviation, are in units of contrast, which is not the case in this paper.
The matched filter maps don't directly estimate the contrast of a planet but rather try to estimate a theoretical SNR as defined in \autoref{eq:datasetSNR}. Therefore, the well known throughput correction is here replaced by a conversion factor to relate the matched filter values to actual contrast estimates, which is also calibrated using simulated planets.

As a consequence, we define a contrast curve as $\eta \gamma(\rho)\sigma_{\mathcal{S}}(\rho)$ where $\eta$ is the SNR threshold, $\gamma(\rho)$ is a median conversion factor between the matched filter map and the true contrast of the planet and $\sigma_{\mathcal{S}}(\rho)$ is the standard deviation of the noise in the matched filter map. 

The different thresholds $\eta$ were determined in \Secref{sec:ROC} for the three algorithms to yield a false positive frequency of $0.05$ per epoch. The standard deviation $\sigma_{\mathcal{S}}(\rho)$ is calculated in concentric annuli according to \Secref{sec:SNR}. The conversion factor $\gamma$ is empirically determined by injecting simulated planets with known contrast in each dataset. The contrast of the simulated planets is chosen to result in a signal somewhere between $5-15\sigma(\rho)$ in the final matched filter map. In total, 128 planets are injected at 16 different separations and position angles in eight different copies of the dataset. The original matched filter map is subtracted from the simulated planet reductions in order to remove the effect of the residual noise. The conversion factor is linearly interpolated from the median of the eight simulated planets at each separation as shown in \autoref{fig:convFactor}. Note that the contrast curve is a function of the spectrum of the simulated planets as well as of the reduction spectrum. Ideally, there should be as many contrast curves as there are possible spectra.
We have limited our study to objects with either a 1000K cloud-free T-type spectrum, reduced with a 600K cloud-free model spectrum, or a 1300K cloudy L-type spectrum, reduced with the same cloudy spectrum. A caveat is that the contrast curves are therefore only truly valid for planets with the same spectrum as the injected planets.

The GPIES median contrast curve for each algorithm as well as the associated detection threshold for both spectral type reduction is given in \autoref{fig:medContrast}. FMMF yields the best median contrast curve at all separations. In the T-type reduction and compared to the Gaussian Matched Filter, the FMMF contrast enhancement ranges from a median $25\%$ up to more than a factor 2 in some cases. The L-type median contrast enhancement drops from $25\%$ below $0.5\arcsec$ to $10\%$ at larger separation.
The median contrast improvement relative to the cross-correlation is around $50\%$, which corresponds to a factor 2.3 gain in exposure time assuming a square root increase of the SNR with time. Tests have shown that this assumption generally holds true to the exception of a slight dependence to observing conditions. The local maximum in the contrast curve for the Gaussian Cross correlation and the Gaussian matched filter is likely due to the non-optimal definition of the reduction sectors for a regular KLIP reduction. We will therefore ignore the FMMF contrast increase in the $0.3\arcsec-0.4\arcsec$ range as we do not consider it representative of the overall performance.

The sensitivity of any observation is highly dependent on the observing conditions and the brightness of the star.
\autoref{fig:medContrastPercentile} shows the percentiles contours of the contrast curves for GPIES using FMMF. \autoref{fig:medContrastImag} shows the FMMF median contrast curves as a function of the \textit{I} magnitude of the star. There is an order of magnitude ratio between the sensitivity around the faintest stars ($I_{\textbf{mag}} \approx 9$) of the survey compared to the brightest stars ($I_{\textbf{mag}} \approx 2$).

\section{Candidate vetting}
\label{sec:candidateDetec}

In order for the contrast curves and the planet completeness to be accurate, any signal above the detection threshold must be properly vetted. This means that any high SNR signal should be confirmed as a true or a false positive. The best approach is always to follow up all the candidates, which can be a significant telescope time investment.
In the case of a real astrophysical signal, the second epoch is generally used to determine proper motion and parallax in order to exclude or confirm the possibility of a background object.
In the case of low SNR candidates, if it cannot be detected in the second epoch, it is usually classified as a false positive. However, it is then necessary to improve the contrast curve in the second epoch in order to exclude the real signal hypothesis with a high enough significance. We discuss the necessary contrast improvements in this section.

First, the detection threshold can be lowered in the follow-up observation compared to the first epoch, because the constraint on the noise is higher.
A spurious signal would need to be in the same position in order for it to be mistakenly considered as a true detection. With a $4\sigma$ detection threshold in the second epoch, there is less than a chance in a thousand to wrongly classify the first detection as a real signal, which seems reasonable. Indeed, there is on the order of $10^3$ independent samples in a $1\arcsec$ field of view in a GPI image assuming a conservative characteristic correlation length of the residuals of 3 pixels and a $4\sigma$ threshold yields less than a false positive per epoch for both matched filters.

\begin{figure}
\gridline{\fig{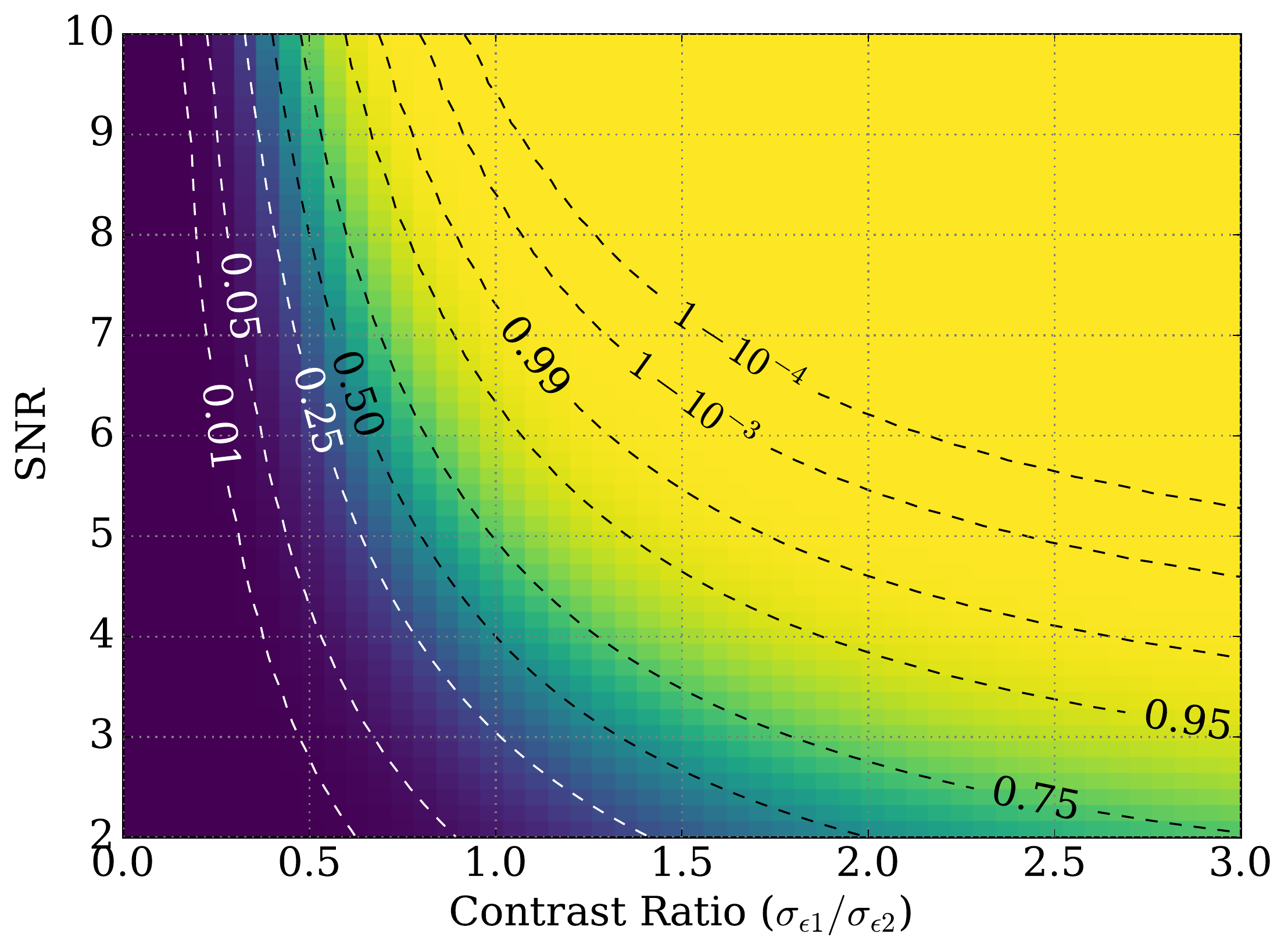}{0.5\textwidth}{(a) $2^{\mathrm{nd}}$ Epoch $4\sigma$ Detection Probability}
		\fig{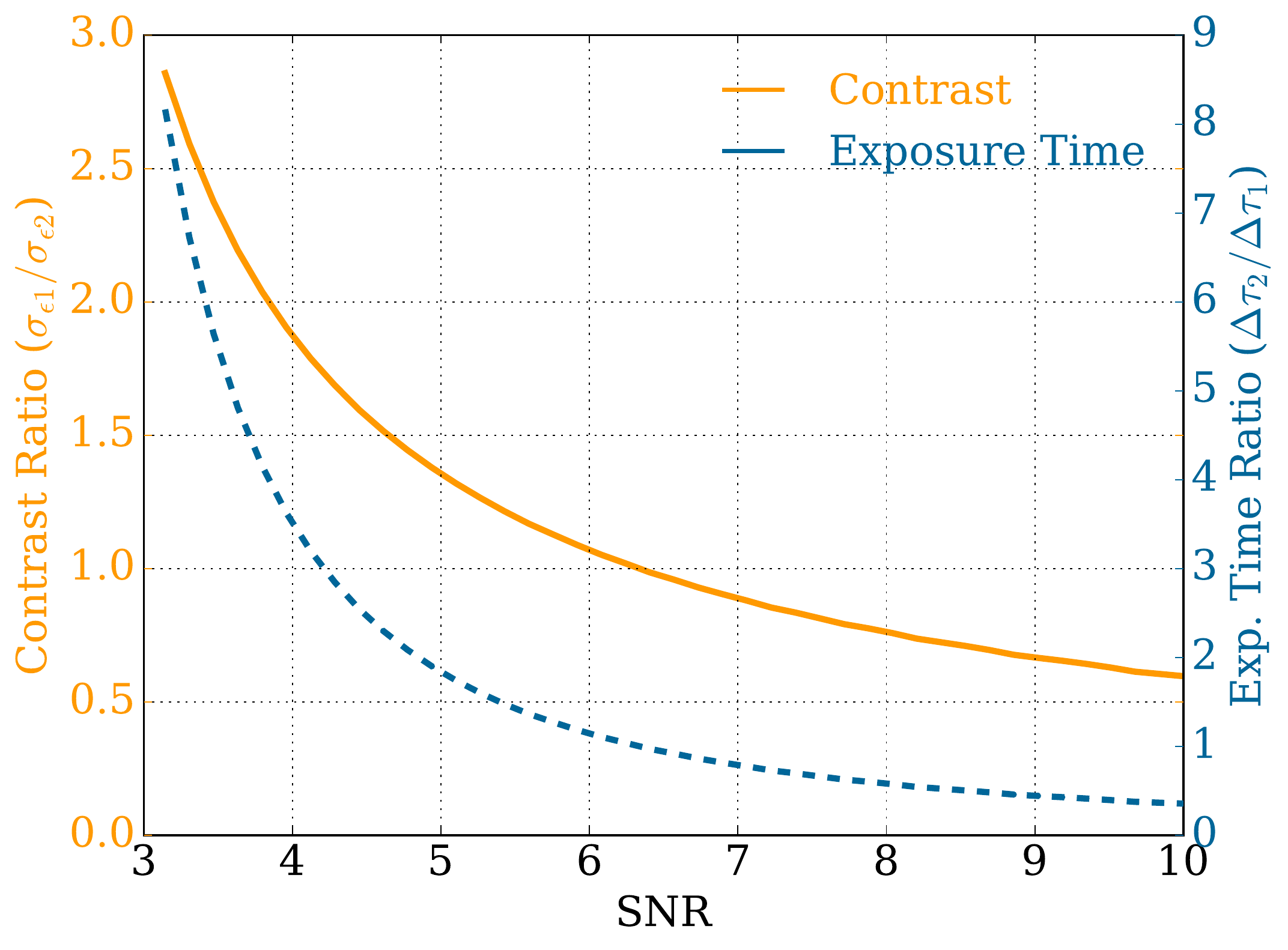}{0.5\textwidth}{(b) $95\%$ Probability}
          }
\caption{Detection probability of a point source candidate in a second epoch. Figure (a) shows the probability of detecting a point source in the follow-up observation (Color map and contours) as a function of the contrast ratio $\sigma_{\epsilon 1}/\sigma_{\epsilon 2}$, where the integer subscript refer to the first or the second epoch and the first epoch SNR. Figure (b) represents the $95\%$ detection probability contour with the corresponding exposure time ratio $\Delta \tau_2/\Delta \tau_1$. The exposure time calculation assumes that the observing conditions are identical for both epochs.
\label{fig:followupSNR}}
\end{figure}

Second, we can estimate the probability of detecting the signal in the new epoch as a function of the original SNR and the ratio of contrast standard deviation $\sigma_{\epsilon 1} / \sigma_{\epsilon 2}$. Assuming a known contrast for a point source, the probability of detection is given by the tail distribution of the planet signal at the detection threshold in the second epoch. However, it needs to be marginalized over the planet contrast due to its uncertainty in the first epoch. The detection probability in the second epoch assuming a detection threshold $\eta_2$ can be written as
\begin{equation}
P_{\mathrm{detec}}(\mathrm{SNR},\sigma_{\epsilon 1} / \sigma_{\epsilon 2}) =
\int_{\mathrm{s}} (1-\mathrm{CDF}(\eta_2 - s\ \sigma_{\epsilon 1} / \sigma_{\epsilon 2}) )\mathrm{PDF}(s - \mathrm{SNR}) \diff s,
\label{eq:followupProbaDetecSNR}
\end{equation}
with $\mathrm{PDF}$ and $\mathrm{CDF}$ respectively the Probability Density Function and the Cumulative Distribution Function of a Gaussian distribution with zero mean and unit standard deviation. \autoref{fig:followupSNR} shows the detection probability contours as a function of the original SNR and the noise ratio.

\begin{figure}
\centering
\includegraphics[width=0.5\linewidth]{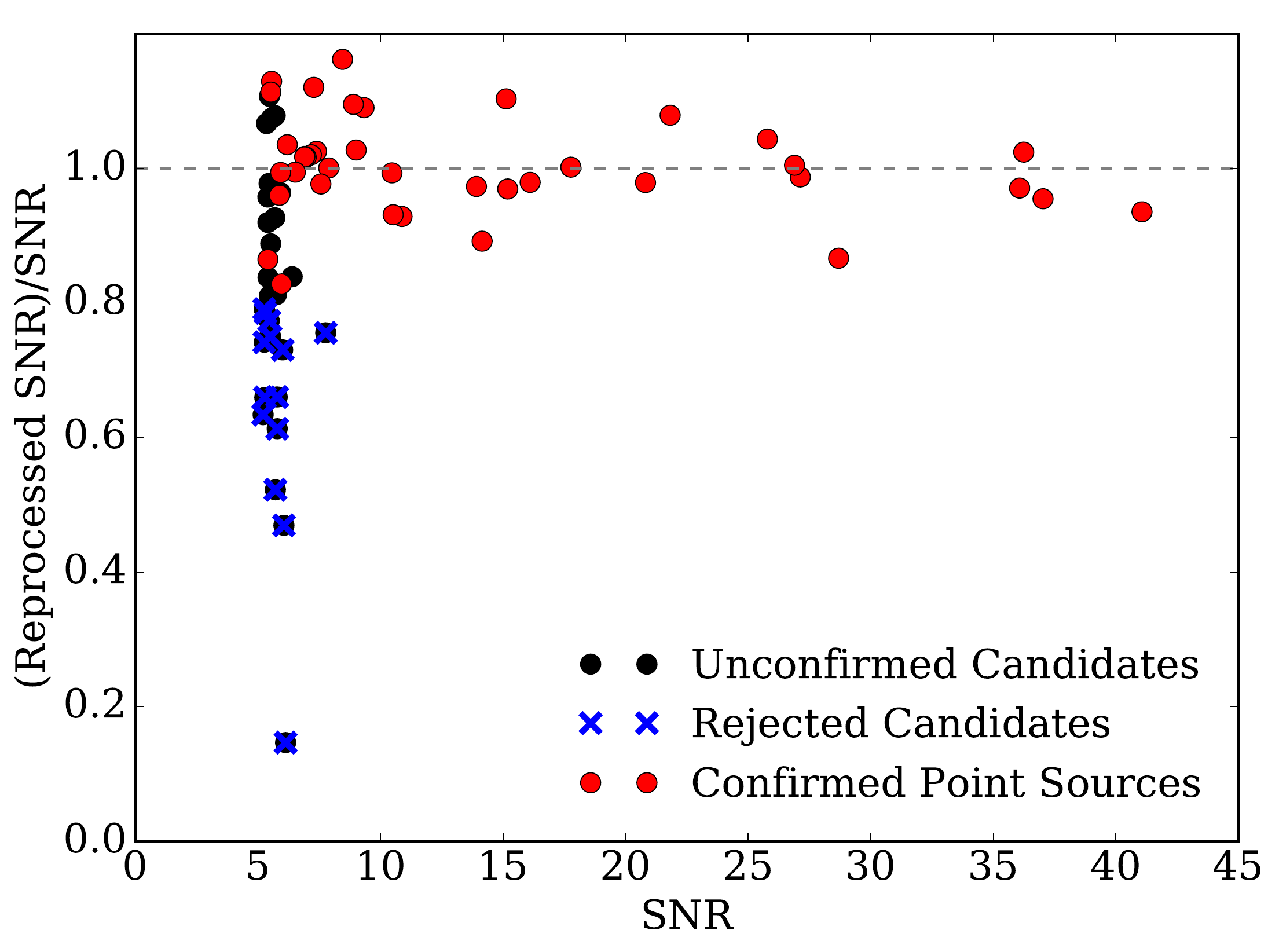}
\caption{Exclusion of detected candidates based on a second reductions of the data with an annulus centered at their separation. The dots represent the L-type candidates above $5.2\sigma$ as a function of their original SNR and the ratio of the SNR between the first and the second reduction. We reject all candidates for which the SNR drops by more than $20\%$ in the new reduction.}
\label{fig:reprocessedSNR}
\end{figure}

The problem is that candidate follow-ups are expensive in telescope time and a significant fraction of the detections are expected to be false positives. 
Indeed, a significant number of candidates appear to be reduction artifacts. Some of these artifacts are very sensitive to the edge of the sectors and can therefore be spotted by running a new reduction with an annulus centered at their separation. Real objects are less sensitive to the definition of the sectors so if the SNR of the candidate drops significantly, it is likely a false positive and should not be reobserved. \autoref{fig:reprocessedSNR} shows the L-type candidates above $5.2\sigma$ as a function of their original SNR and the relative SNR with the new reduction. Because the SNR ratio for the real point sources are always close to unity, we conclude that we can reject all candidates for which this ratio drops under 0.8 without significantly biasing the algorithm completeness. 
This plot cannot be shown for the T-type reduction because there is only one confirmed T-type object in GPIES, which is 51 Eridani b. Most point sources are indeed background stars and the few substellar objects feature a L-type spectrum. The lack of control sample for the T-type reduction does not allow us to define a boundary in this case.
We acknowledge that rejecting candidates based on an ad-hoc criterion is not the best solution and proves that work remains to be done. Ideally, we would want to combine the results of both reductions into a single detection metric rather than applying different cutoffs sequentially.

\section{From Contrast Curve to Completeness}
\label{sec:completeness}

The main goal of a direct imaging exoplanet survey, after discovering new worlds to characterize, is to put constraints on the population of wide orbit sub-stellar objects \citep{nielsen2013,Bowler2015,Galicher2016,Vigan2017}. Deriving the frequency of exoplanets first relies on the completeness of the survey. The most common method to calculate the completeness of an observation is to reduce the same dataset with simulated planets at all separations and contrasts. The completeness at a given point is then the fraction of the fiducial planets detected. This approach requires many reductions for each dataset, which is tractable with classical ADI reductions but computationally out of reach for the FMMF. 
\cite{Lafreniere2007b} showed that, assuming a Gaussian noise, the completeness can be estimated directly from the contrast curve. For example, a planet lying exactly on the contrast curve will have a $50\%$ chance to fall above or below it due to the noise. The probability of detecting a planet of a given contrast is given by the tail distribution of a Gaussian distribution centered on the contrast of the planet evaluated at the contrast curve. We have shown that the Gaussian assumption is not verified in the tail of the noise distribution but this effect would only become important for completeness of planets far from the contrast curve. In the ideal case, the $(\eta+1) \gamma(\rho)\sigma_{\mathcal{S}}(\rho)$ and the $(\eta-1) \gamma(\rho)\sigma_{\mathcal{S}}(\rho)$ curves respectively represents the $84\%$ and $16\%$ completeness contour. One caveat is that the azimuthal variations of the conversion factor (\textit{e.g.} due to the wind-butterfly) add a scatter term to the measured contrast of the planet effectively widening the completeness contours. For example, a planet inside the wind-butterfly suffers from a higher conversion factor than a planet outside of it, resulting in a lower detection probability.

\begin{figure}
\centering
\includegraphics[width=1.0\linewidth]{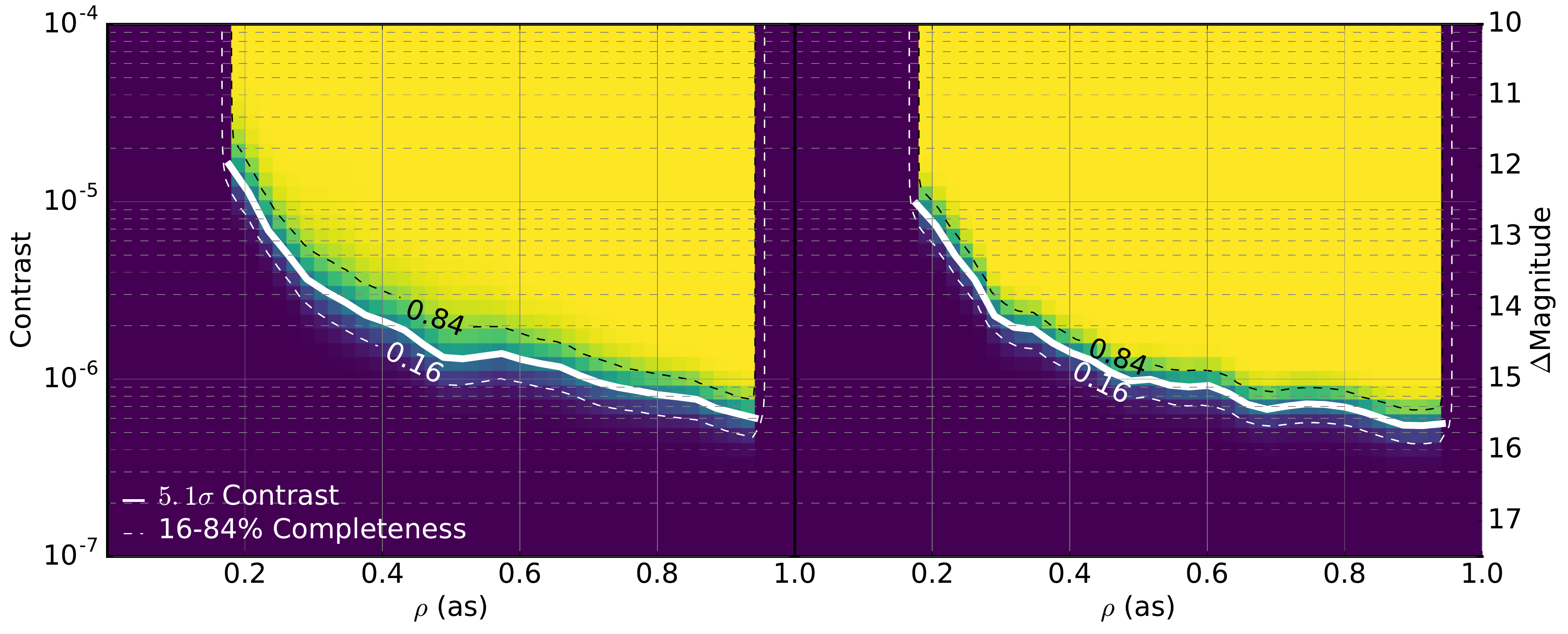}
\caption{Exoplanets completeness as a function of contrast and separation $\rho$. The left panel shows an example of strong wind-butterfly resulting in the conversion factor to vary significantly as a function of the position angle. The right panel shows an example in which the azimuthal scatter of the conversion factor is negligible. (Solid line) Contrast curve at $5.1\sigma$ corresponding to the $50\%$ detection completeness contour for a FMMF T-type reduction. (Dashed line) $16\%$ and $84\%$ completeness contours.}
\label{fig:completeness}
\end{figure}

The azimuthal variations of the conversion factor due to the wind-butterfly or other artifacts can be interpreted as a noise term with a standard deviation $\sigma_{\gamma}$.
Assuming that the residual noise and the conversion factor variations are independent, the standard deviation of the contrast can be written
\begin{equation}
\sigma_\epsilon(\epsilon) = \sqrt{\gamma^2 \sigma_{\mathcal{S}}^2 + \frac{\epsilon^2}{\gamma^2}\sigma_{\gamma}^2},
\end{equation}
which is used to estimate the completeness.
\autoref{fig:completeness} shows two examples of completeness contours, one with a strong wind-butterfly and another where it is negligible.

\section{Conclusion}
\label{sec:conclusion}

In order to improve GPIES planet sensitivity, we implemented a new matched filter based algorithm using KLIP and a forward modeled PSF template. This algorithm includes the speckle subtraction efficiency of KLIP, while mitigating the PSF distortion penalty. We also presented a Gaussian cross correlation and a Gaussian matched filter for comparison. The cross correlation requires an identically distributed noise in order to be valid. This assumption is not verified in GPI images and it therefore leads to an increase of false positives in the part of the image, like the wind-butterfly, where the noise is consistently stronger. An accurate matched filter needs to include a normalization by the local noise standard deviation, which improves significantly the statistics of the residual noise in the final SNR maps.

After optimizing the aggressiveness of the algorithm, we showed that two reduction spectra, a T-type and a L-type, are sufficient to recover sub-stellar companions candidates with their expected spectra without a significant loss in SNR.

The uniform reduction of more 330 datasets from GPIES allowed the unprecedented characterization of the PDF of the residual noise, which significantly deviates from a Gaussian distribution. We showed that the FMMF has a uniform spatial distribution of false positives that does not feature any excesses due to the sector boundaries or the wind-butterfly. The improved performance provided by the FMMF also appeared in the ROC curves and the contrast curves that have been shown for GPIES. We built the contrast curves using a different detection threshold for each spectral type and algorithm such that it yields the same number of false positives in each case. The FMMF allows for example the detection of objects that are $50\%$ fainter than a classical cross correlation.
We reaffirm that a hard $5\sigma$ detection threshold doesn't allow meaningful comparison of algorithms performance and can overestimate the sensitivity of a survey. 
In the future, more diverse algorithms should be compared following a similar recipe to the one presented in this paper.

To finish, we showed how the planet completeness can be derived from each contrast curve, which is simply the $50\%$ detection completeness contour. These completeness contours can be used in a future exoplanet population study.

\acknowledgments
{\bf Acknowledgements:}
This work is based on observations obtained at the Gemini Observatory, which is operated by the Association of Universities for Research in Astronomy under a cooperative agreement with NSF on behalf of the Gemini partnership, whose membership includes: NSF (United States); the National Research Council (Canada); the Comisi\'on Nacional de Investigaci\'on Cient\'ifica y Tecnol\'ogica (Chile); the Australian Research Council (Australia); the Minist\'erio da Ci\^encia, Tecnologia e Inova\c{c}\~ao (Brazil); and Ministerio de Ciencia, Tecnolog\'ia e Innovaci\'on Productiva (Argentina). GPI data are archived at the Gemini Observatory Archive\footnote{\url{http://www.gemini.edu/sciops/data-and-results/gemini-observatory-archive}}: \url{https://archive.gemini.edu/searchform}.
The research was supported by grants from NSF, including AST-1411868 (J.-B.R. B.M., K.F., A.R., J.L.P.) and AST-1518332 (J.J.W., R.J.D, J.R.G., and P.K.).
Support was provided by grants from NASA, including NNX14AJ80G (B.M., E.N., and V.P.B.), NNX15AD95G (J.J.W., R.J.D, J.R.G., and P.K.). This work benefited from NASA’s Nexus for Exoplanet System Science (NExSS) research coordination network sponsored by NASA’s Science Mission Directorate. K.M.M.'s and T.B.'s work is supported by the NASA Exoplanets Research Program (XRP) by cooperative agreement NNX16AD44G. M.A.M.B. was supported for this work by NASA through Hubble Fellowship grant \#51378.01-A awarded by the Space Telescope Science Institute, which is operated by the Association of Universities for Research in Astronomy, Inc., for NASA, under contract NAS5-26555.
J.R and R.D acknowledge support from the Fonds de Recherche du Quebec.


\vspace{5mm}
\facilities{Gemini:South(GPI)}

\software{pyKLIP\footnote{Documentation available at \url{http://pyklip.readthedocs.io/en/latest/}.} \citep{Wang2015}, GPI Data Reduction Pipeline\footnote{Documentation available at \url{http://docs.planetimager.org/pipeline/}.} \citep{Perrin2016}, astropy\footnote{\url{http://www.astropy.org}} \citep{Astropy2013}}



\newpage
\appendix

\section{Method}
\label{sec:appmethod}

In this appendix, we summarize the key theoretical results used in this paper. First, the Karhunen-Lo\`eve Image Processing (KLIP) framework and its effect on the planet PSF, \textit{i.e.} self-subtraction, is introduced in \Appref{sec:appklip}. Then, the principles of a matched filter are presented in \Appref{sec:appmatchedFilter}. Finally, the forward model is defined in \Appref{sec:KLIPFM}.

Throughout the paper, with some rare exceptions, scalars are lower case Greek characters, vectors are bold lower case Latin characters and matrices are bold upper case Latin characters. A summary of the notations can be found in \Appref{sec:notations}.

\subsection{Karhunen-Lo\`eve Image Processing}
\label{sec:appklip}

\subsubsection{Formalism}
\label{sec:appklipForma}

Karhunen-Lo\`eve Image Processing (KLIP) \citep{Soummer2012} is one of the most widely used speckle subtraction frameworks. The general concept of speckle subtraction algorithms is to use a library of reference images from which a model of the speckle pattern is computed and then subtracted for each science image, effectively whitening the noise.
KLIP calculates the principal components of this reference library and filter out the higher order modes that contain more noise and more planet signal.
The reference library is a set of images spatially scaled to the same wavelength and containing random realization of the correlated speckles. We will use a general observing strategy by combining ADI and SDI observations, sometimes referred to as ASDI, which is possible with an IFS like GPI. This choice of strategy implies that the astrophysical signal will be found both in the reference images and the science image. As a consequence, the algorithm is subject to self-subtraction, which is discussed in \Secref{sec:appselfsubtraction}.
A detailed description of the mathematical formalism underlying the KLIP algorithm can be found in \cite{Savransky2015}, which we quickly summarize here. 

The $N_R \times N_{\mathrm{pix}}$ matrix $\boldsymbol{R}$ is defined by concatenating the vectorized mean subtracted reference images $r_{j}$, which are vectors with $N_{\mathrm{pix}}$ elements, such that
\begin{equation}
\boldsymbol{R} = \left[ {\bm r}_{1}, {\bm r}_{2}, \dots, {\bm r}_{N_R} \right]^{\top}.
\end{equation}
Then, the image-to-image sample covariance matrix describing the correlation of the images in the reference library is described, up to a proportional factor, by
\begin{equation}
\boldsymbol{C} = \boldsymbol{R}\boldsymbol{R}^{\top},
\end{equation}
with its eigenvectors and eigenvalues ${\bm v}_{1}, {\bm v}_{2}, \dots, {\bm v}_{N_R}$  and $ \mu_{1}, \mu_{2}, \dots, \mu_{N_R}$, respectively. The Karhunen-Lo\`eve images form the optimal orthonormal basis that best represents any realization of the noise in a least square sense. They are defined as
\begin{equation}
{\bm z}_{k }= \frac{1}{\sqrt{\mu_{k}}} \sum_{j=1}^{N_R} {\bm v}_{k}[j] {\bm r}_{j}.
\end{equation}
The images can be written in a more compact form by defining the matrices $\boldsymbol{V}_{K} = [ {\bm v}_{1}, {\bm v}_{2}, \dots, {\bm v}_{K} ]^{\top}$, $\boldsymbol{Z}_{K} = [ {\bm z}_{1}, {\bm z}_{2}, \dots, {\bm z}_{K} ]^{\top}$ and the diagonal matrix $\boldsymbol{\Lambda} = \textrm{diag}\left( \mu_{1}, \mu_{2}, \dots, \mu_{K} \right)^{\top}$, with $K \leq N_R$ being the number of selected Karhunen-Lo\`eve images kept for the subtraction. Then, the collection of horizontally stacked Karhunen-Lo\`eve images $\boldsymbol{Z}_{K} = [ {\bm z}_{1}, {\bm z}_{2}, \dots, {\bm z}_{K} ]^{\top}$ is equivalently written as
\begin{equation}
\boldsymbol{Z}_{K} = \sqrt{\boldsymbol{\Lambda}^{-1}} \boldsymbol{V}_{K} \boldsymbol{R}.
\end{equation} 
The speckle subtraction consists in subtracting the projection of the science image $i$ onto the Karhunen-Lo\`eve basis from itself , which is given by
\begin{equation}
{\bm p} = {\bm i} - \sum_{k=1}^{K} \langle {\bm i} \vert z_{k} \rangle z_{k}= {\bm i} - \boldsymbol{Z}_{K}^{\top}\boldsymbol{Z}_{K} {\bm i},
\label{eq:klip}
\end{equation}
where $p$ refers to the processed image. The science and the reference images are often part of the same dataset and the science image simply refers to the image of a specific exposure and wavelength from which the speckle pattern is being subtracted.
Typically, speckle subtraction algorithms are performed independently on small sectors of the image to account for spatial variations of the speckle noise, instead of using the full image, however, the formalism is identical.

\subsubsection{Self-Subtraction}
\label{sec:appselfsubtraction}

In the this section, we review the effect of an astrophysical signal as a small perturbation to \autoref{eq:klip}, following \cite{Pueyo2016}. In this section only, variables defined in \Secref{sec:appklipForma} are assumed to be planet free, and the hat indicates the perturbed variable, such that
\begin{equation}
\forall j \in [1,N_R],\quad \hat{{\bm r}}_j = {\bm r}_{j}+\Delta {\bm r}_j ,
\end{equation}
with $\Delta {\bm r}_j = \epsilon {\bm a}_j$, where ${\bm a}_j$ is the normalized planet signal.  The planet signal is defined by its amplitude $\epsilon$, a spectrum and the instrumental PSF before speckle subtraction. It is more convenient to normalize the planet signal to the same brightness as the host star such that $\epsilon$ is in practice the contrast of the planet. Similarly, we write $\hat{{\bm z}}_j = {\bm z}_{j}+\Delta {\bm z}_j$, $\hat{\boldsymbol{Z}}_K = \boldsymbol{Z}_{K}+\Delta \boldsymbol{Z}_K$, $\hat{{\bm i}} = {\bm i}+\Delta {\bm i} = {\bm i} + \epsilon {\bm a}$, and $\hat{{\bm p}} = {\bm p}+\Delta {\bm p} \approx {\bm p} + \epsilon {\bm m}$, where ${\bm m}$ is the planet model that will be used to build the matched filter template ${\bm t}$ introduced in \autoref{eq:datasetSNR}.

Applying KLIP to the perturbed images gives
\begin{align}
\hat{{\bm p}} &= \hat{{\bm i}} - \hat{\boldsymbol{Z}}_K^{\top}\hat{\boldsymbol{Z}}_K \hat{{\bm i}} , \nonumber \\
{\bm p}+\Delta {\bm p} &= \left(  {\bm i}+\epsilon {\bm a} \right) - \left( \boldsymbol{Z}_{K}+\Delta \boldsymbol{Z}_K \right)^{\top} \left( \boldsymbol{Z}_{K}+\Delta \boldsymbol{Z}_K \right) \left(  {\bm i}+\epsilon {\bm a} \right) , \nonumber \\
\label{eq:klipPert}
\end{align}

and expanding \autoref{eq:klipPert}, one identifies ${\bm p}$, which should only contain residual speckles such that
\begin{equation}
{\bm p} = {\bm i} - \boldsymbol{Z}_{K}^{\top}\boldsymbol{Z}_{K} {\bm i} \approx 0 .
\end{equation}
Then, the perturbed processed image is given by
\begin{align}
\Delta {\bm p} =& && \epsilon {\bm a} && \text{Original planet signal,} \nonumber \\
& -&& \epsilon \boldsymbol{Z}_{K}^{\top} \boldsymbol{Z}_{K} {\bm a} && \text{Over-subtraction,} \nonumber \\
& -&& \left( \boldsymbol{Z}_{K}^{\top} \Delta \boldsymbol{Z}_K + \left(  \boldsymbol{Z}_{K}^{\top} \Delta \boldsymbol{Z}_K \right)^{\top} \right) {\bm i}  && \text{Self-subtraction,} \nonumber \\
& +&& \mathcal{O}(\epsilon^2) && \text{Neglected second order terms.} \nonumber \\
\label{eq:selfsubtraction}
\end{align}
The second and third terms of \autoref{eq:selfsubtraction} are the first order distortions introduced by KLIP on the planet signal as shown in \autoref{fig:forwardModelExample} (b).  
The over-subtraction comes from the projection of the planet onto the unperturbed Karhunen-Lo\`eve modes, while the self-subtraction is the result of the projection of the speckles onto the perturbation of the modes. The former is unavoidable in a least squares approach. The characteristic patterns for the self-subtraction are negative lobes in the radial and azimuthal direction due to the movement of the planet in the reference frames with wavelength and parallactic angle. The self-subtraction vanishes when there is no planet signal in the reference library as $\Delta \boldsymbol{Z}_K = 0$, which is the case when using a Reference star Differential Imaging (RDI) approach. Using the Locally Optimized Combination of Images (LOCI) \cite{Lafreniere2007a}, another way to avoid self-subtraction is to use optimization and subtraction region that do not overlap \citep[see,][]{Marois2010}. However, the price of these strategies is a possible less effective speckle subtraction.

\subsubsection{Reference Library Exclusion Criterion}
\label{sec:exclusionCrit}

\begin{figure}[h]
\centering
\includegraphics[width=1.0\linewidth]{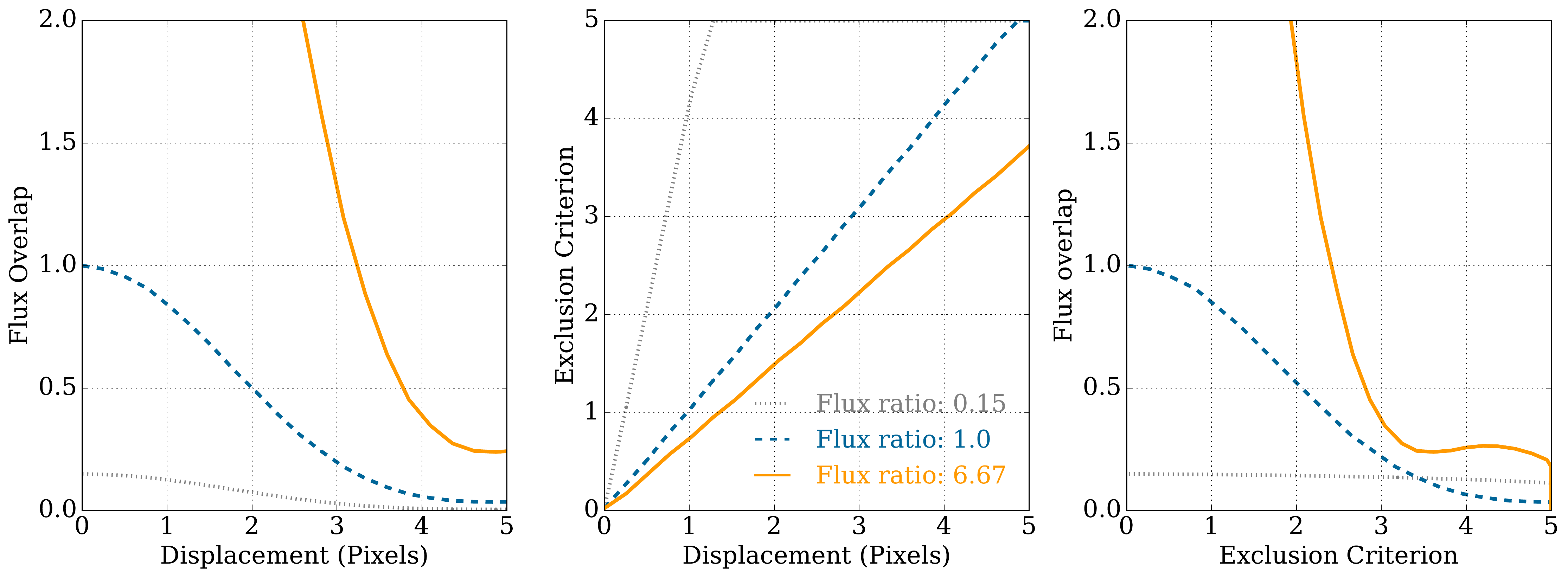}
\caption{Conversion plots between the exclusion criterion, the PSF flux overlap and the PSF displacement between two images. The flux ratio is defined as the ratio of the peak value of the PSF in the reference frame over the peak value in the science image. The ratio of $0.15$ represents a typical value for the peak to bottom ratio of a spectrum with a strong methane signature.}
\label{fig:exculsionCrit}
\end{figure}

Part of the planet signal is lost in the processed image, especially when the planet PSF overlap in the reference images and in the science image spatially. The distortion of the central peak of the PSF is more important when the overlap is big.
A common practice to limit the distortion with ADI and SDI is to only include images in the reference library in which the planet has moved significantly compared to the science image. The importance of the distortion also depends on the relative brightness of the planet in the science image and in the references, which itself depends on the planet spectrum. Therefore, the selection of the reference images can be based on a the relative flux overlap with the planet signal in science image \citep{Marois2014} instead of the sole consideration of the displacement.

\begin{figure}
\gridline{\fig{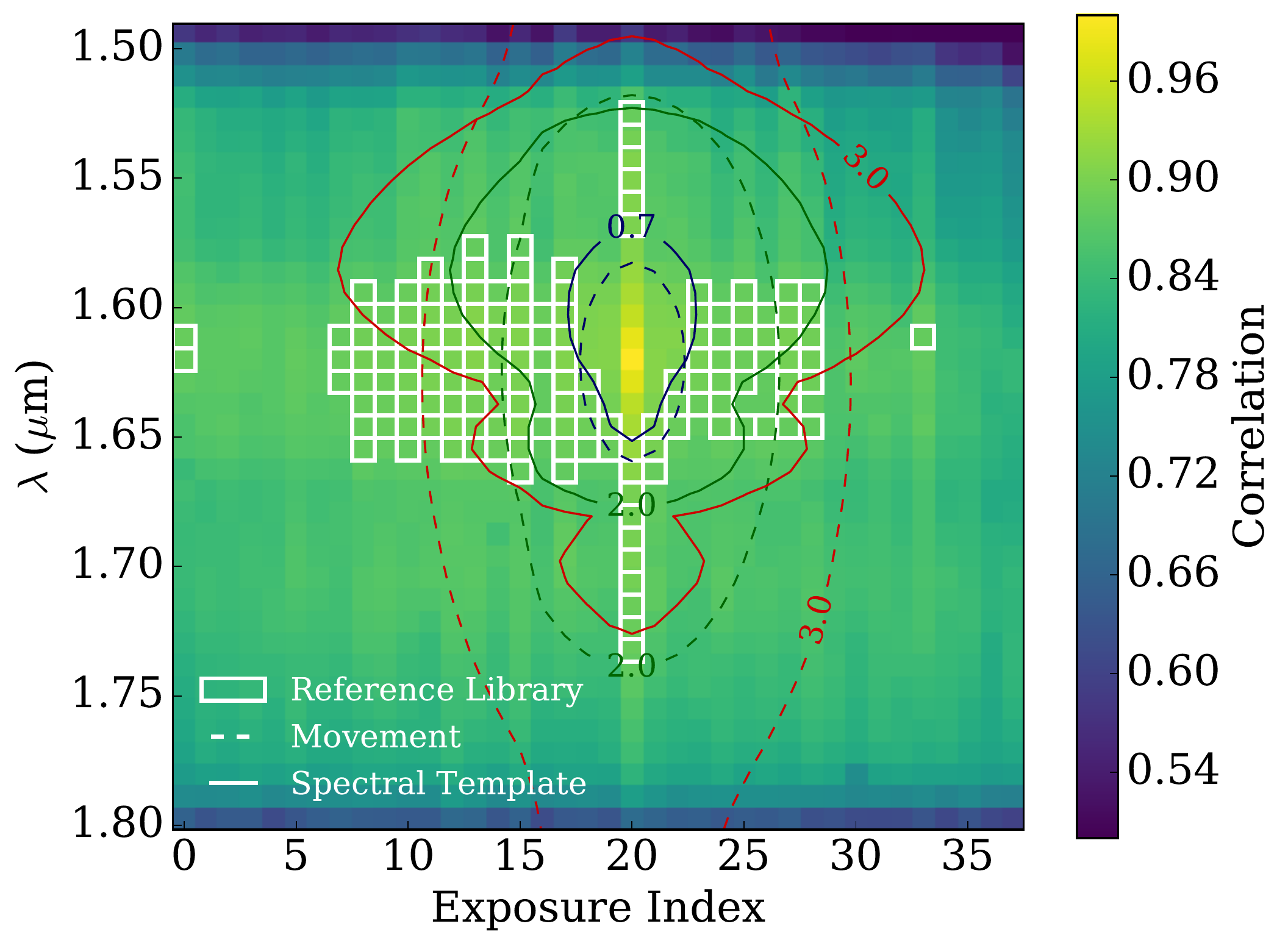}{0.45\textwidth}{(a) $20^{th}$ exposure, $1.60\ \mu m$}
          \fig{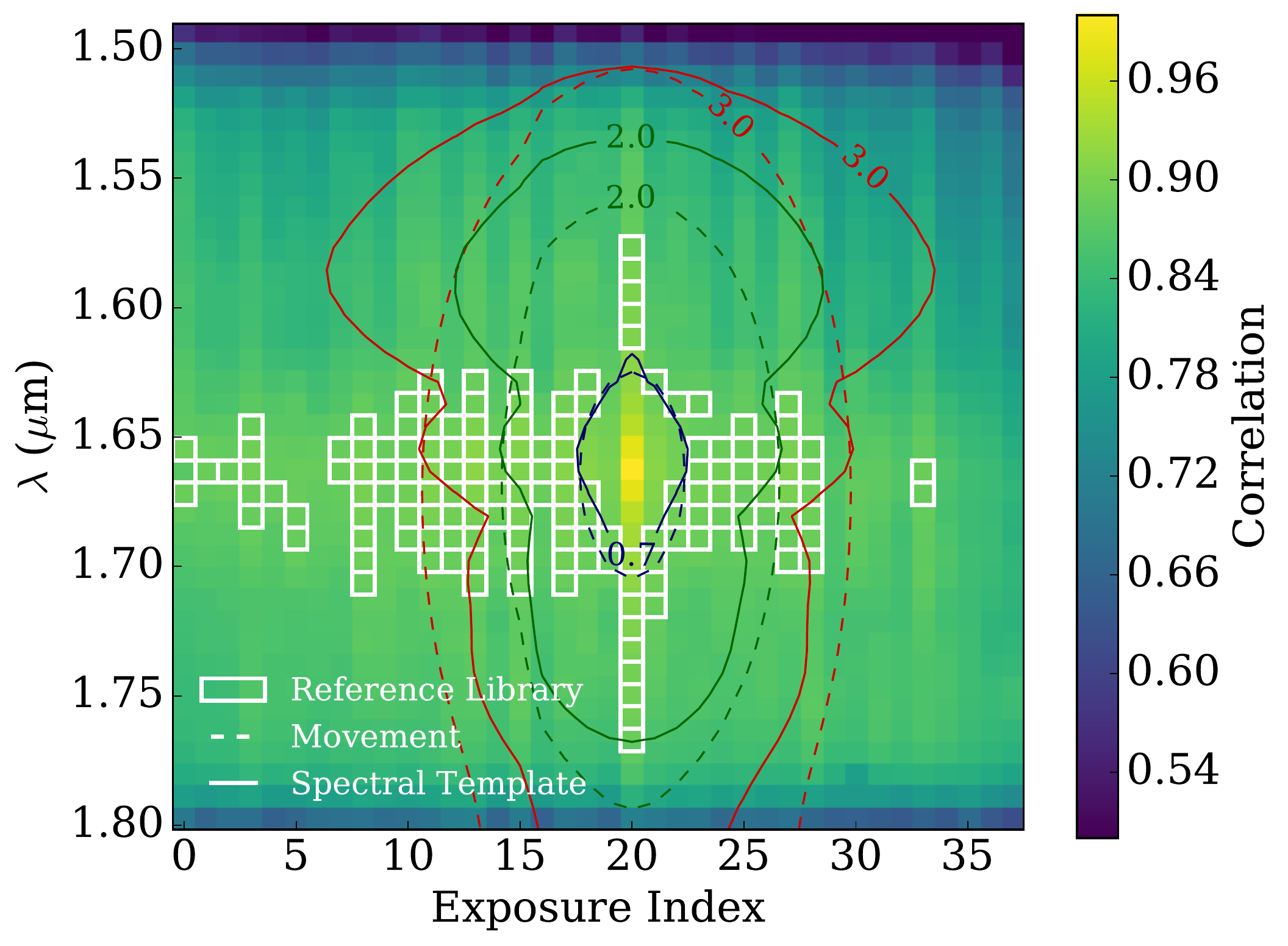}{0.45\textwidth}{(b) $20^{th}$ exposure, $1.65\ \mu m$}
          }
\caption{Illustration of the reference library selection for a given science image in a GPI dataset. Two examples are shown with science images in the same exposure but at different wavelengths in H-band: (a) $1.60$ and (b) $1.65\ \mu \mathrm{m}$. The colormap represents the correlation between each image in the dataset with the science image. The dataset is represented as a function of the exposure (x-axis) and the wavelength (y-axis). The exclusion criterion used in this paper for the reference library selection is represented with solid contours and includes a spectral template. It is a hybrid between a pure displacement in pixels, represented with dashed lines, and a flux overlap criterion. The white squares highlight the $150$ most correlated images satisfying a $0.7$ exclusion criterion, which will form the reference library. More images tends to be exluded around $1.58\ \mu \mathrm{m}$ because we chose a spectrum with a strong methane signature resulting in a peak flux at this wavelength. The dataset used here is the 51 Eridani b GPIES discovery epoch on 2014 December 18. The signal displacement is calculated for an hypothetical planet at 30 pixel ($\approx 0.42$ as) separation and the total rotation during the full sequence is $24^\circ$. \label{fig:refLibrary}}
\end{figure}

The exclusion parameter chosen in this paper is a hybrid between a flux overlap and a displacement parameter.
It corresponds to a pure displacement when comparing images at similar wavelength but it accepts more or less images in the reference library when the wavelengths are different. This implementation of the exclusion criterion has given significantly better result than any of the two other methods. The conversion between the three metrics is given for a specific planet PSF flux ratios between the science and the reference images in \autoref{fig:exculsionCrit}.
The selection of the reference images is illustrated in \autoref{fig:refLibrary} in a typical GPI dataset. Only the $150$ most correlated images with any given science frame are kept for the speckle subtraction in order to limit the size of the covariance matrix $\boldsymbol{C}$ and speed up the reduction.
There is a stronger correlation between images with respect to wavelength than with respect to time, suggesting that SDI will give better results than ADI.

The choice of the exclusion criterion is a trade-off between an effective speckle subtraction and a limited self-subtraction. A small value of the exclusion criterion includes more correlated images in the reference library and results in a more aggressive reduction, due to the larger degree of self-subtraction. The aggressiveness is also affected by the number of Karhunen-Lo\`eve modes used in the subtraction, however we keep this number fixed to $K=30$ throughout this paper. We describe how the optimization of the exclusion parameter is performed in \Secref{sec:aggressOpti}. 

\subsection{Matched Filter}
\label{sec:appmatchedFilter}

In the field of signal processing, a matched filter is the linear filter maximizing the SNR of a known signal in the presence of additive noise \citep{Kasdin2006}. If the noise samples are independent and identically distributed, the matched filter corresponds to the cross-correlation of a template with the noisy data. In the context of high-contrast imaging, the pixels are neither independent nor identically distributed (i.e., heteroskedastic). Note that the matched filter is a very general tool that can be applied regardless of the algorithm used for speckle subtraction.

We define $d$ as the vectorized, speckle-subtracted dataset representing a series of individual exposures of a star. Because GPI is an Integral Field Spectrograph (IFS), the one-dimensional array $d$ typically contains $N_x N_y N_\lambda N_\tau$  elements, where $N_x$ and $N_y$ are the number of pixels in the spatial dimensions (for a square image, $N_x = N_y$), $N_\lambda$ is the number of spectral channels, and $N_\tau$ is the number of exposures. A typical GPIES dataset is the concatenation of all the processed images $\left\lbrace p_{l} \right\rbrace$ from an epoch, with $l=(\tau,\lambda)$, so $N_x N_y N_\lambda N_\tau = 281\times 281\times 37 \times 38$.
If we assume that the dataset contains a single point source, it can be decomposed as
\begin{equation}
{\bm d} = \epsilon {\bm t}(\rho,\theta) + {\bm n},
\label{eq:noisyData}
\end{equation}
With ${\bm t}(\rho,\theta)$ the vectorized template of the planet with separation and position angle $(\rho,\theta)$. In practice we conveniently normalize the template to the same flux as the star such that $\epsilon$ is again the contrast.
It is the concatenation of the planet model PSF for each image $\left\lbrace {\bm m}_{l}\right\rbrace$ and it has the same dimension as $d$. The second term $n$ is for now assumed to be a Gaussian random vector assuming a zero mean and a covariance matrix ${\bm \Sigma}$.
For conciseness, the $(\rho,\theta)$ notation in ${\bm t}(\rho,\theta)$ will be dropped in the following of the paper but the position dependence should be assumed for any point source related variables.

A matched filter can also be interpreted as a maximum likelihood $\mathcal{L}(\epsilon,\rho,\theta)$ estimator of the planet amplitude and position assuming a Gaussian noise:
\begin{equation}
\mathcal{L}(\epsilon,\rho,\theta) = \frac{1}{\sqrt{2\pi|{\bm \Sigma}|}} \mathrm{exp}\left[ \frac{-1}{2} ({\bm d}-\epsilon {\bm t})^{\top} {\bm \Sigma}^{-1} ({\bm d}-\epsilon {\bm t})\right].
\end{equation}

The most likely amplitude $\tilde{\epsilon}$ as a function of position is given by
\begin{equation}
\tilde{\epsilon}(\rho,\theta) = {\bm d}^{\top} {\bm \Sigma}^{-1} {\bm t}  \Big / {\bm t}^{\top} {\bm \Sigma}^{-1} {\bm t}.
\label{eq:MLEa}
\end{equation}

The theoretical SNR of a point source is given as a function of position by,
\begin{align}
\mathcal{S}(\rho,\theta)  &= \tilde{\epsilon} / \sigma_{\tilde{\epsilon}} \nonumber \\
&= {\bm d}^{\top} {\bm \Sigma}^{-1} {\bm t} \Big / \sqrt{{\bm t}^{\top} {\bm \Sigma}^{-1} {\bm t}},
\label{eq:gaussianSNR}
\end{align}
Which is a linear function of the square root of the maximized log-likelihood \citep{Rover2011, Cantalloube2015}, where maximization is carried out over the point source amplitude.

Assuming that the noise is uncorrelated and that its variance is constant in the neighborhood of the planet, the planet amplitude becomes
\begin{equation}
\tilde{\epsilon}(\rho,\theta)  = \sum\limits_{l} {\bm p}_{l}^{\top} {\bm m}_{l}/\sigma_{l}^{2} \Bigg / \sum\limits_{l} {\bm m}_{l}^{\top} {\bm m}_{l}/\sigma_{l}^{2},
\label{eq:appdatasetAmpl}
\end{equation}
where $\sigma_{l}$ is the local standard deviation at the position $(\rho,\theta)$. Note that the planet model ${\bm m}_{l}$ approaches zero rapidly when moving away from its center $(\rho,\theta)$ allowing one to only consider postage-stamp sized images containing the putative planet.

The theoretical SNR of the planet becomes
\begin{equation}
\mathcal{S}(\rho,\theta)  =  \sum\limits_{l} {\bm p}_{l}^{\top} {\bm m}_{l}/\sigma_{l}^{2} \Bigg / \sqrt{ \sum\limits_{l} {\bm m}_{l}^{\top} {\bm m}_{l}/\sigma_{l}^{2}}.
\label{eq:appdatasetSNR}
\end{equation}

A detection can be claimed when the SNR is such that the observation cannot be explained by the null-hypothesis.

\subsection{KLIP Forward Model}
\label{sec:KLIPFM}

It is common practice to perform a matched filter with a simple Gaussian or a box template. Such templates are not optimal because they at best waste planet signal due to the self-subtraction. This paper uses the KLIP forward model from \cite{Pueyo2016} to improve the matched filter planet detection sensitivity. 

The perturbation of the processed image $\frac{\Delta p}{\epsilon}$ from \autoref{eq:selfsubtraction} is what should be used as the template for the matched filter (a.k.a. planet signature in \cite{Cantalloube2015} and forward model in \cite{Pueyo2016}). It can also be thought of as the derivative of the KLIP operator along the direction defined by the planet.

The ideal template $\frac{\Delta p}{\epsilon}$ is a function of the original planet PSF and spectrum. The difficulty is the estimation of the Karhunen-Lo\`eve perturbation term $\Delta \boldsymbol{Z}_K$ because it is a nonlinear function of the planet flux. This is why \cite{Pueyo2016} (E18) derived a first order approximation of this term that can be used to calculate a close estimation of the planet PSF after speckle subtraction:
\begin{equation}
\frac{\Delta  {\bm z}_k}{\epsilon} = -\left(\frac{{\bm v}_{k}^{\top} \boldsymbol{C}_{\boldsymbol{A}\boldsymbol{R}} {\bm v}_{k}}{2\mu_{k}}   \right) {\bm z}_k+\sum_{j=1,j\neq k}^{N_R} \left( \sqrt{\frac{\mu_{j}}{\mu_{k}}}\frac{{\bm v}_{j}^{\top} \boldsymbol{C}_{\boldsymbol{A}\boldsymbol{R}} {\bm v}_{k}}{\mu_{k}-\mu_{j}} \right) {\bm z}_{j} + \frac{1}{\sqrt{\mu_{k}}} \boldsymbol{A}^{\top} {\bm v}_{k} ,
\label{eq:deltaz}
\end{equation} 
With $\boldsymbol{A} = \left[ {\bm a}_1, {\bm a}_2, \dots, {\bm a}_{N_R} \right]^{\top}$ and $\boldsymbol{C}_{\boldsymbol{A}\boldsymbol{R}} = \boldsymbol{A} \boldsymbol{R}^{\top} + \left( \boldsymbol{A} \boldsymbol{R}^{\top} \right)^{\top}$.

The forward model of the planet, a.k.a template, is then given by
\begin{equation}
{\bm m} = {\bm a} - \boldsymbol{Z}_{K}^{\top} \boldsymbol{Z}_{K} {\bm a} - \left( \boldsymbol{Z}_{K}^{\top} \Delta \boldsymbol{Z}_K + \left(  \boldsymbol{Z}_{K}^{\top} \Delta \boldsymbol{Z}_K \right)^{\top} \right) \frac{{\bm i}}{\epsilon} ,
\label{eq:fm}
\end{equation}
with ${\bm m} \approx \frac{\Delta {\bm p}}{\epsilon}$.
As a reminder, we have dropped the $(\rho, \theta)$ upper script at the very beginning of this paper but ${\bm m}$, ${\bm a}$ and $\Delta \boldsymbol{Z}_K$ all depend on the position in the image.

\section{Mathematical Notations}
\label{sec:notations}

We present here a summary of the notations used in this paper. \autoref{tab:scalar}, \autoref{tab:vector}, and \autoref{tab:matrix} respectively contain the definition of the scalars, the vectors and the matrices.

\clearpage

%\floattable
\startlongtable
\begin{deluxetable}{ccp{12cm}}
\tablecaption{Definition of Scalar Variables.}
\tablecolumns{2}
%\tablenum{2}
\tablewidth{0pt}
\tablehead{
\colhead{Symbol} &
\colhead{Type} &
\colhead{Description} \\
}
\startdata
\label{tab:scalar}
$N_x,\ N_y$ & $\mathbb{N}\times\mathbb{N}$ & Dimensions of the image.\\
$N_{\mathrm{pix}}$ & $\mathbb{N}$ &  Number of pixels in a sector. If the sector is the entire image, then $N_{\mathrm{pix}}=N_xN_y$. \\
$N_\lambda$ & $\mathbb{N}$ & Number of wavelength channels in the IFS data cubes.\\
$N_\tau$ & $\mathbb{N}$ & Number of exposures in an epoch.\\
$N_R$ & $\mathbb{N}$ & Number of images in the reference library.\\
$K$ & $\mathbb{N}$ & Number of selected Karhunen-Lo\`eve images to be used in the speckle subtraction.\\
$\rho$ & $\mathbb{R}^{+}$ & Separation to the star. \\
$\theta$ & $\mathbb{R}$ &  Position Angle with the origin at the image North. \\
$\sigma_{l}$ & $\mathbb{R}^{+}$ & Local standard deviation of the noise at the position $(\rho,\theta)$ in the speckle subtracted images. \\
$\mathcal{S}(\rho,\theta)$ & $\mathbb{R}$ & Theoretical Signal-to-Noise Ratio (SNR) at the position $(\rho,\theta)$ assuming independent Gaussian Noise. This is not the true SNR of the signal. \\
$\epsilon$ & $\mathbb{R}^{+}$ & Contrast of the point source relative to the host star. Different units could be used depending on the normalization of the planet model. \\
$\tilde{\epsilon}(\rho,\theta)$ or $\tilde{\epsilon}$ & $\mathbb{R}$ & Maximum likelihood estimation of the amplitude of the planet signal $\epsilon$ at the position $(\rho,\theta)$.  \\
$\sigma_{\epsilon}$ & $\mathbb{R}^{+}$ & Standard deviation of the planet contrast at the position $(\rho,\theta)$. $\sigma_{\epsilon 1}$ and $\sigma_{\epsilon 2}$ respectively refer to the first and follow-up observation of a star.\\
$\mu_{j}$ & $\mathbb{R}$ & Eigenvalues of the matrix $\boldsymbol{C}$. \\
$\gamma$ & $\mathbb{R}$ & Conversion factor from the contrast matched filter values $\mathcal{S}$ to the contrast $\zeta$ such that $\zeta = \gamma\mathcal{S} $. \\
$\eta$ & $\mathbb{R}$ & SNR detection threshold. $\eta_2$ is the detection threshold in the follow-up observation, which can be smaller. \\
$\sigma_{\gamma}$ & $\mathbb{R}^+$ & Standard deviation of $n_{\gamma}$ \\
$\sigma_{\mathcal{S}}$ & $\mathbb{R}^+$ & Standard deviation of the noise in the matched filter map as a function of position. \\
\enddata
\end{deluxetable}

\clearpage

%\floattable
\begin{deluxetable}{ccp{12cm}}
\tablecaption{Definition of Vectors.}
\tablecolumns{2}
%\tablenum{2}
\tablewidth{0pt}
\tablehead{
\colhead{Symbol} &
\colhead{Length} &
\colhead{Description} \\
}
\startdata
\label{tab:vector}
${\bf r}_{j}$ & $N_{\mathrm{pix}}$ & Vectorized images or sectors from the KLIP reference library. \\
$\hat{{\bf r}}_{j}$ & $N_{\mathrm{pix}}$ & Vectorized perturbed images or sectors from the KLIP reference library including some planet signal. \\
$\Delta {\bf r}_{j}$ & $N_{\mathrm{pix}}$ & Vectorized perturbation of the images or sectors from the KLIP reference library. \\
${\bf v}_{j}$ & $N_R$ & Eigenvectors of the matrix $\boldsymbol{C}$. \\
${\bf z}_{k}$ & $N_{\mathrm{pix}}$ & Karhunen-Lo\`eve images or sectors for the reference library matrix $\boldsymbol{R}$. \\
$\hat{{\bf z}}_{k}$ & $N_{\mathrm{pix}}$ & Perturbed Karhunen-Lo\`eve images or sectors for the reference library matrix $\boldsymbol{R}$ including planet signal. \\
$\Delta {\bf z}_{k}$ & $N_{\mathrm{pix}}$ & Perturbation of the Karhunen-Lo\`eve images or sectors for the perturbed reference library. \\
${\bf i}$ & $N_{\mathrm{pix}}$ & Vectorized science image or sector before speckle subtraction. \\
$\hat{{\bf i}}$ & $N_{\mathrm{pix}}$ & Vectorized perturbed science image or sector before speckle subtraction including planet signal. \\
$\Delta {\bf i}$ & $N_{\mathrm{pix}}$ & Vectorized perturbation of the science image or sector before speckle subtraction. \\
${\bf a}_j$ & $N_{\mathrm{pix}}$ & Vectorized planet signal in the reference image or sector $\hat{r}_j$ before speckle subtraction. \\
${\bf a}$ & $N_{\mathrm{pix}}$ & Vectorized planet signal in the science image or sector before speckle subtraction. \\
${\bf p}$ & $N_{\mathrm{pix}}$ & Vectorized processed image or sector after speckle subtraction. \\
$\hat{{\bf p}}$ & $N_{\mathrm{pix}}$ & Vectorized perturbed processed image or sector after speckle subtraction including planet signal. \\
$\Delta {\bf p}$ & $N_{\mathrm{pix}}$ & Vectorized perturbation of the processed image or sector after speckle subtraction. \\
${\bf m}$ & $N_{\mathrm{pix}}$ & Vectorized planet model in the processed image or sector after speckle subtraction. \\
${\bf d}$ & $N_x N_y  N_\lambda   N_\tau$ &  Vectorized speckle subtracted dataset.\\
${\bf t}(\rho,\theta)$ or ${\bf t}$ & $N_x  N_y  N_\lambda   N_\tau$ &  Vectorized planet template for the speckle subtracted dataset.\\
${\bf n}$ & $N_x  N_y  N_\lambda   N_\tau$ &  Multivariate noise distribution of the speckle subtracted dataset. The noise is assumed to be Gaussian for any mathematical derivation.\\
${\bf p}_{l}$ & $N_x  N_y$ &  Vectorized speckle subtracted image with exposure $\tau$ and wavelength $\lambda$. \\
${\bf m}_{l}(\rho,\theta)$ or ${\bf m}_{l}$ & $N_x  N_y$ &  Vectorized planet template for the speckle subtracted image with exposure $\tau$ and wavelength $\lambda$.\\
\enddata
\end{deluxetable}

\clearpage

%\floattable
\begin{deluxetable}{ccp{12cm}}
\tablecaption{Definition of Matrices.}
\tablecolumns{2}
%\tablenum{2}
\tablewidth{0pt}
\tablehead{
\colhead{Symbol} &
\colhead{Dimensions} &
\colhead{Description} \\
}
\startdata
\label{tab:matrix}
${\bf \Sigma}$ & $(N_x  N_y  N_\lambda   N_\tau)^2$ & Covariance matrix of the noise $n$ in the speckle subtracted  \\
$\boldsymbol{R}$ & $ N_R \times N_{\mathrm{pix}}$ & Reference library matrix such that $\boldsymbol{R} = \left[ r_{1}, r_{2}, \dots, r_{N_R} \right]^{\top}$. \\
$\boldsymbol{C}$ & $ N_R \times N_R$ & Matrix proportional to the time covariance matrix defined as $\boldsymbol{C} = \boldsymbol{R}\boldsymbol{R}^{\top}$.\\
$\boldsymbol{V}_K$ & $ K \times N_R$ & Eigenvectors matrix defined as $\boldsymbol{V}_{K} = [ v_{1}, v_{2}, \dots, v_{K} ]^{\top}$. \\
$\boldsymbol{Z}_K$ & $ K \times N_{\mathrm{pix}}$ & Karhunen-Lo\`eve images matrix defined as $\boldsymbol{Z}_{K} = [ z_{1}, z_{2}, \dots, z_{K} ]^{\top}$. \\
$\hat{\boldsymbol{Z}}_K$ & $ K \times N_{\mathrm{pix}}$ & Perturbed Karhunen-Lo\`eve images matrix defined as $\hat{\boldsymbol{Z}}_{K} = [ \hat{z}_{1}, \hat{z}_{2}, \dots, \hat{z}_{K} ]^{\top}$ including planet signal. \\
$\Delta \boldsymbol{Z}_K$ & $ K \times N_{\mathrm{pix}}$ & Perturbation of the Karhunen-Lo\`eve images matrix defined as $\Delta \boldsymbol{Z}_{K} = [ \Delta z_{1}, \Delta z_{2}, \dots, \Delta z_{K} ]^{\top}$. \\
$\boldsymbol{\Lambda}$ & $ K \times N_{\mathrm{pix}}$ & Diagonal matrix defined as $\boldsymbol{\Lambda} = \textrm{diag}\left( \lambda_{1}, \lambda_{2}, \dots, \lambda_{N_R} \right)^{\top}$. \\
$\boldsymbol{A}$ & $ N_R \times N_{\mathrm{pix}}$ & Planet signal component in the reference library $\boldsymbol{R}$ and $\boldsymbol{A} = \left[ a_{1}, a_{2}, \dots, a_{N_R} \right]^{\top}$. \\
$\boldsymbol{C}_{\boldsymbol{A}\boldsymbol{R}}$ & $ N_R \times N_R$ & Matrix used in the forward calculation defined as $\boldsymbol{C}_{\boldsymbol{A}\boldsymbol{R}} = \boldsymbol{A} \boldsymbol{R}^{\top} + \left( \boldsymbol{A} \boldsymbol{R}^{\top} \right)^{\top}$. \\
\enddata
\end{deluxetable}

\clearpage




\newpage
\bibliographystyle{aasjournal}
\bibliography{bibliography}



\end{document}